\documentclass[11pt]{article}

\RequirePackage{silence}
\usepackage[T1]{fontenc}
\usepackage{lmodern,microtype}
\usepackage[margin=1in]{geometry}
\usepackage{amsmath,amssymb,amsthm,mathtools}
\usepackage{booktabs,array}
\usepackage{tikz}
\usetikzlibrary{arrows.meta,positioning,calc,decorations.pathreplacing}
\usepackage{aliascnt}
\usepackage{placeins}
\usepackage[colorlinks=true,linkcolor=blue!45!black,citecolor=blue!45!black,urlcolor=blue!45!black]{hyperref}
\usepackage[nameinlink,noabbrev]{cleveref}

\usepackage[dvipsnames]{xcolor} %

\newtheorem{theorem}{Theorem}[section]
\newaliascnt{lemma}{theorem}
\newtheorem{lemma}[lemma]{Lemma}
\aliascntresetthe{lemma}
\newaliascnt{proposition}{theorem}
\newtheorem{proposition}[proposition]{Proposition}
\aliascntresetthe{proposition}
\newaliascnt{corollary}{theorem}
\newtheorem{corollary}[corollary]{Corollary}
\aliascntresetthe{corollary}
\theoremstyle{definition}
\newaliascnt{definition}{theorem}
\newtheorem{definition}[definition]{Definition}
\aliascntresetthe{definition}
\crefname{lemma}{lemma}{lemmas}
\Crefname{lemma}{Lemma}{Lemmas}
\crefname{proposition}{proposition}{propositions}
\Crefname{proposition}{Proposition}{Propositions}
\crefname{corollary}{corollary}{corollaries}
\Crefname{corollary}{Corollary}{Corollaries}
\crefname{definition}{definition}{definitions}
\Crefname{definition}{Definition}{Definitions}
\crefname{appendix}{appendix}{appendices}
\Crefname{appendix}{Appendix}{Appendices}
\newcommand{\PSPACE}{\mathsf{PSPACE}}
\newcommand{\Kalah}{\mathsf{Kalah}}
\definecolor{ovboardfill}{RGB}{246,237,219}
\definecolor{ovboardline}{RGB}{196,172,132}
\definecolor{ovalice}{RGB}{33,96,168}
\definecolor{ovalicefill}{RGB}{214,230,248}
\definecolor{ovbobc}{RGB}{193,76,50}
\definecolor{ovbobfill}{RGB}{251,222,211}
\definecolor{ovclock}{RGB}{118,78,170}
\definecolor{ovclockfill}{RGB}{233,223,246}
\definecolor{ovweapon}{RGB}{0,122,122}
\definecolor{ovweaponfill}{RGB}{205,236,233}
\definecolor{ovrecord}{RGB}{48,128,72}
\definecolor{ovrecordfill}{RGB}{216,239,220}
\definecolor{ovgold}{RGB}{176,128,12}
\definecolor{ovgoldfill}{RGB}{253,237,192}
\tikzset{
  ovboard/.style={fill=ovboardfill,draw=ovboardline,line width=0.6pt,rounded corners=2.5mm},
  ovdots/.style={font=\scriptsize,text=black!50,inner sep=0pt},
  ovpit/.style={circle,draw=black!55,fill=white,minimum size=5.8mm,inner sep=0pt,font=\scriptsize},
  ovtgt/.style={ovpit,draw=ovalice,fill=ovalicefill,line width=1.1pt},
  ovtimer/.style={ovpit,draw=ovbobc,fill=ovbobfill,line width=1.1pt},
  ovcover/.style={ovpit,draw=ovbobc,fill=ovbobfill},
  ovclockpit/.style={ovpit,draw=ovclock,fill=ovclockfill,line width=0.9pt},
  ovweaponpit/.style={ovpit,draw=ovweapon,fill=ovweaponfill,line width=0.9pt},
  ovrecordpit/.style={ovpit,draw=ovrecord,fill=ovrecordfill,line width=0.9pt},
  ovlab/.style={font=\scriptsize,align=center},
  ovalicelab/.style={ovlab,text=ovalice},
  ovboblab/.style={ovlab,text=ovbobc},
  ovhot/.style={draw=black!55,fill=white,rounded corners=2pt,line width=0.7pt,
                font=\scriptsize,align=center,minimum height=7mm,inner sep=1pt},
  ovsow/.style={->,>={Stealth},line width=0.7pt,draw=ovalice},
  ovbob/.style={->,>={Stealth},line width=0.7pt,densely dashed,draw=ovbobc},
  ovclocksow/.style={ovbob,draw=ovclock},
}

\newcommand{\onepagetoc}{{\small\def\addvspace##1{\vskip0.5em\relax}\tableofcontents}\setcounter{tocdepth}{2}}

\title{Sowing the Seeds of Intractability: Kalah is PSPACE-complete}
\author{Ethan Dickey\\Purdue University\\\texttt{dickeye@purdue.edu} \and Ethan Mader\\Purdue University\\\texttt{emader@purdue.edu}}
\date{}

\begin{document}

\maketitle

\begin{abstract}
Kalah is a classic two-player game in the mancala family of sowing games. Computers have solved several standard-size starting positions, but the complexity of the game on boards of growing size has remained open. We study the natural generalization in which the number of pits per side is part of the input and an arbitrary position is given, with the stone counts given in unary. We show that deciding whether a designated player can force a win from such a position is $\PSPACE$-complete. The hard positions have Alice to move first and both stores initially empty. The same positions are hard under either common convention for capture opposite an empty pit, and with or without the bonus rule, which gives a player another turn when the last stone is sown into their own store.

We introduce Line Kalah as an intermediate problem, a game on a line in which stones that leave the line effectively disappear and Alice must save each of her target pits before Bob captures it, and we reduce quantified Boolean formulas to it. The reduction uses a set of sequential timer piles whose last stones approach a group of targets, forcing Alice to choose between two captures in order to set the truth value of a variable, and a final capture gadget to test a clause chosen by Bob (only one is needed adversarially).

Since either player may move any pile at any time, the stages are placed so that premature timer firings protect earlier potential targets. For a true formula, Alice's strategy against arbitrary opposing play either gains enough time to save the remaining targets (after Bob deviates from the intended play) or reaches the final test with a correctly encoded, protected input. Finally, the line positions are embedded in Kalah with both stores empty. A separate Bob pit holds the score adjustment, so the winner takes a strict majority of the stones.
\end{abstract}
\thispagestyle{empty}
\clearpage
\onepagetoc
\clearpage
\section{Introduction} \label{sec:intro}

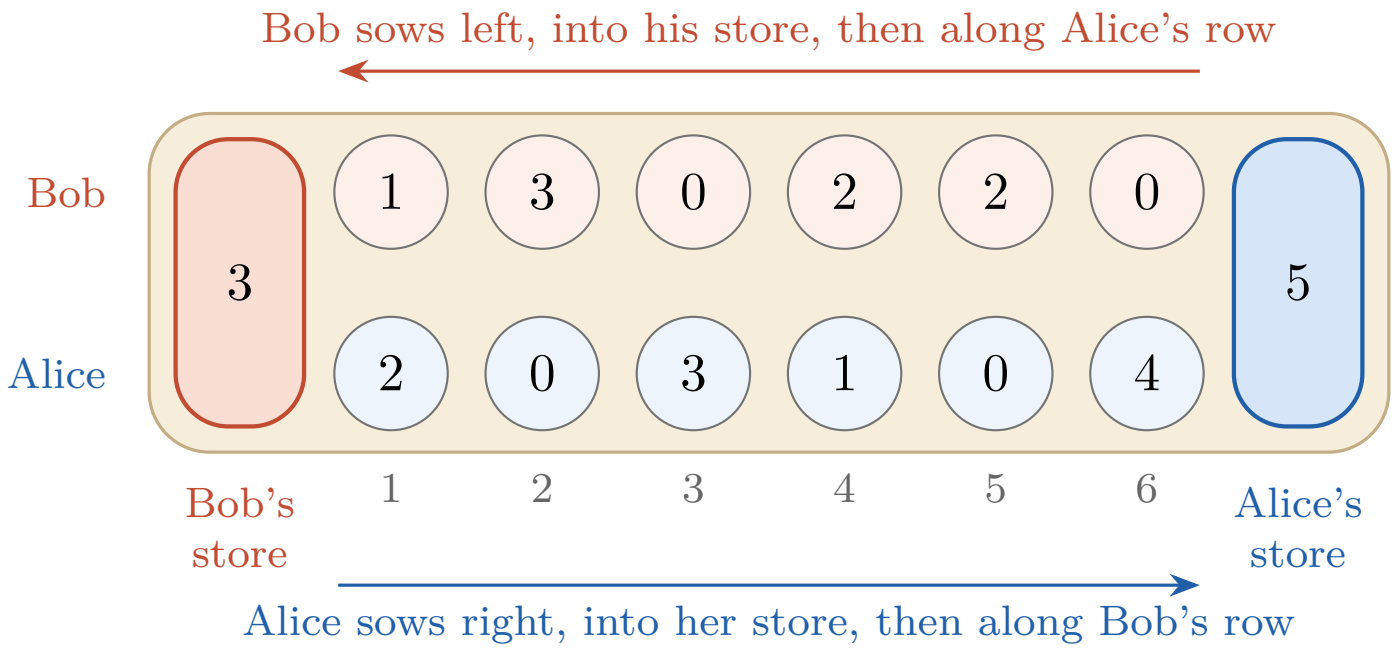
\begin{figure}[h]
    \centering
    \begin{tikzpicture}[
      >={Stealth},
      pit/.style={circle,draw=black!55,fill=white,minimum size=7.5mm,inner sep=0pt,font=\small},
      store/.style={rounded corners=3.5mm,minimum width=8.5mm,line width=0.8pt,
                    minimum height=19mm,font=\small},
      lab/.style={font=\scriptsize},
      dir/.style={->,line width=0.6pt}]
      \draw[ovboard,rounded corners=4mm] (-0.6,-0.52) rectangle (7.6,1.72);
      \foreach \x/\a/\b in {1/2/1,2/0/3,3/3/0,4/1/2,5/0/2,6/4/0}{
        \node[pit,fill=ovalicefill!45] (A\x) at (\x,0) {$\a$};
        \node[pit,fill=ovbobfill!45] (B\x) at (\x,1.2) {$\b$};
        \node[lab,black!60] at (\x,-0.76) {$\x$};
      }
      \node[store,draw=ovbobc,fill=ovbobfill] (SB) at (0,0.6) {$3$};
      \node[store,draw=ovalice,fill=ovalicefill] (SA) at (7,0.6) {$5$};
      \node[lab,align=center,anchor=north,text=ovbobc] at (0,-0.62) {Bob's\\store};
      \node[lab,align=center,anchor=north,text=ovalice] at (7,-0.62) {Alice's\\store};
      \node[lab,anchor=east,text=ovbobc] at (-0.75,1.2) {Bob};
      \node[lab,anchor=east,text=ovalice] at (-0.75,0) {Alice};
      \draw[dir,ovalice] (0.65,-1.4) -- (6.35,-1.4)
        node[midway,below,lab] {Alice sows right, into her store, then along Bob's row};
      \draw[dir,ovbobc] (6.35,2.0) -- (0.65,2.0)
        node[midway,above,lab] {Bob sows left, into his store, then along Alice's row};
    \end{tikzpicture}
    \caption{A Kalah position of width $N=6$ with Alice to move. The small numbers are the columns. Pits in the same column are opposite. Both players sow counterclockwise, and each skips the opponent's store.}
    \label{fig:kalah-board}
\end{figure}

Kalah is a two-player game from the mancala family of sowing games~\cite{IrvingDonkersUiterwijk2000} with stones (or seeds). Each player has a row of pits and a store at the end of that row (\Cref{fig:kalah-board}). On a turn, a player picks up every stone in one of their pits and sows the stones one at a time into the following pits, counterclockwise, through their own store but skipping the opponent's. A last stone that lands in an empty pit on the mover's side can capture the opposite pile. When one row is empty, each player collects the stones left on their own side, and the player with more stones in their store wins. Mancala games are played in many variants around the world, and some, such as Oware, are played in national tournaments. Oware never sows into a store and captures from the opponent's row.

Kalah and other mancala games also have a long history in computer science. A Kalah program by Roland Silver was among the first games written for the PDP-1 computer in 1961~\cite{Silver1961}, and Kalah was the test domain of early studies of game-tree search, such as the comparison of minimax and alpha-beta search with new ordering methods by Slagle and Dixon~\cite{SlagleDixon1969}. Later work solved Kalah by computer for boards with six pits per side and up to six stones in each pit~\cite{IrvingDonkersUiterwijk2000,Carstensen2011}, and Romein and Bal solved Awari, a version of Oware, by parallel retrograde analysis and showed that it is a draw under perfect play~\cite{RomeinBal2003}.

These computer solutions concern fixed boards. Computational complexity instead asks how the difficulty of a game grows with its size and has classified many classical two-player games. Generalized to an $n\times n$ board, Chess~\cite{FraenkelLichtenstein1981}, Checkers~\cite{Robson1984}, and Go with the Japanese ko rule~\cite{Robson1983Go} are $\mathsf{EXPTIME}$-complete, while Hex~\cite{Reisch1981} and Othello~\cite{IwataKasai1994}, in which every game ends after polynomially many moves, are $\PSPACE$-complete. In contrast, the complexity of deciding whether a player can force a win from a given position of generalized Kalah has remained open~\cite{Guy1996,Nowakowski2025}. The computer solutions above concern fixed boards and do not resolve the complexity of the game on boards of growing size.

We study the natural generalization in which the number of pits per side is part of the input and an arbitrary position is given, with the stone counts written in unary. Every game then lasts a number of moves polynomial in the input length (\Cref{sec:membership}), so the winner can be decided in polynomial space. With counts written in binary, a position can hold exponentially many stones, so this bound on the length of play is no longer necessarily polynomial in the input length; for that version, we only show membership in $\mathsf{EXPTIME}$ (\Cref{app:extensions}) and leave the exact complexity open.

To establish hardness for unary counts, we exploit two features of Kalah. The column where a sow ends is determined by the size of the pile, so every stone that falls on a pile moves the end of its next sow, and a capture (conventionally) removes an entire opposing pile. We show that these features suffice to simulate quantified Boolean formulas, so that deciding the winner of a given Kalah position is $\PSPACE$-complete (\Cref{thm:main}). The hard positions have both stores empty and Alice to move, and they are hard under both common capture conventions, with or without the bonus rule. At the center of the proof is Line Kalah, a game on a line that we introduce. It separates captures from the rest of the rules, and we hope that this approach helps to classify other sowing games (\Cref{sec:consequences}).

\subsection{The problem and our results}

\begin{definition}[Generalized Kalah] \label{def:kalah}
    A Kalah board of width $N\ge1$ has two rows of $N$ pits, one per player, and a store for each player. The columns are numbered $1,\ldots,N$; pits in the same column are \emph{opposite}. Alice's store is to the right of her row and Bob's is to the left of his (\Cref{fig:kalah-board}). All stone counts are nonnegative integers.
    
    On a turn, a player empties one of their nonempty pits and sows its stones one at a time counterclockwise, skipping the opponent's store. If the last stone lands in a pit on the mover's side that was empty immediately before it arrived, it captures the opposite pile: the last stone and all stones in the opposite pit move to the mover's store. When the opposite pit is empty, we consider two conventions: under the \emph{nonempty-opposite capture rule}, the last stone stays in its pit; under the \emph{empty capture rule}, it moves to the mover's store.
    
    If either row is empty initially or after a move and any resulting capture, play ends and each player adds the stones remaining on their side to their store. Otherwise, the opponent moves next, except that under the optional \emph{bonus rule}, a last stone sown directly into the mover's store gives the mover another turn. A capture into the store does not give a bonus, and the end-of-game rule takes priority. We consider both capture conventions, independently with and without the bonus rule.
\end{definition}

An instance lists $N$, the $2N$ pit counts, the two store counts, and the player to move. In the \emph{unary} version, all counts are written in unary. The question is whether a designated player, Alice, has a strategy that ends the game with strictly more stones in her store than in Bob's. A tie is not a win for Alice.

\begin{theorem}
\label{thm:main}
Deciding whether Alice wins a given unary Kalah position, whose width $N$ is part of the input, is $\PSPACE$-complete under either capture convention, with or without the bonus rule, even when both stores are empty and Alice is to move. The same positions show hardness in all four cases.
\end{theorem}

\paragraph{Our results.} \Cref{thm:main} is our main result. Its proof also gives the following.
\begin{itemize}
\item \emph{Line Kalah.} We introduce Line Kalah, a game on a line in which stones that leave the line disappear and the players interact only through captures, and show that it is $\PSPACE$-complete (\Cref{thm:formula-to-line} and \Cref{sec:consequences}). The hardness of Kalah follows by embedding its positions in Kalah boards with empty stores (\Cref{thm:empty-transfer}).
\item \emph{Robustness.} The same positions are hard under both capture conventions, with and without the bonus rule, with both stores empty and Alice to move. Arbitrary initial stores and a supplied score target give equivalent problems (\Cref{app:extensions}).
\item \emph{Upper bounds.} Every Kalah play has at most $(M+1)(NM+1)$ moves, where $M$ is the number of stones in pits (\Cref{lem:polynomial-play}). This gives membership in $\PSPACE$ and a polynomial-time algorithm for each constant width. With counts written in binary the problem is in $\mathsf{EXPTIME}$ and $\PSPACE$-hard (\Cref{prop:binary-counts}).
\end{itemize}

\subsection{Standard Kalah and prior work}

Kalah is a commercially marketed game in the mancala family of sowing games. Following Irving, Donkers and Uiterwijk \cite{IrvingDonkersUiterwijk2000}, write $\Kalah(N,c)$ for the game with $N$ pits per side, $c$ stones in every pit, and empty stores. The standard game is $\Kalah(6,4)$. Irving et al.\ solved Kalah for several starting configurations with up to six pits and five stones per pit, using full-game databases and game-tree search, and Carstensen later solved $\Kalah(6,6)$~\cite{Carstensen2011}. Their complexity measures are the game-tree and state-space sizes of a fixed board.

\Cref{def:kalah} is the natural generalization of $\Kalah(N,c)$ in which the width is part of the input and the starting position is arbitrary. The $2N$ pit counts need not be equal. \Cref{thm:main} concerns this family of positions and already holds with empty initial stores. It does not settle the complexity of the uniform openings $\Kalah(N,c)$ themselves. Rule sets also differ in one detail of the capture. In the rules of~\cite{IrvingDonkersUiterwijk2000}, a last stone in an empty own pit is captured even when the opposite pit is empty. Our theorem covers this convention as well as the nonempty-opposite capture rule.

FairKalah studies supplied starting positions that give a draw under optimal play~\cite{NellerNeller2023}. Some of its positions include stones already in a store, illustrating why both the pit distribution and the initial scores can be part of a supplied-position question. Maei, Kiya, Hanaka and Ono~\cite{MaeiKiyaHanakaOno2018} study Simple-Kalah, which removes captures and bonuses and continues until both rows are empty. They give sufficient winning conditions and positions with exponentially many distinct reachable configurations. Such a counting result obstructs direct enumeration, but does not by itself exclude a polynomial-time winner algorithm.

Sowing games have also been studied combinatorially. Erickson~\cite{Erickson1996} analyzed two simple sowing games with combinatorial game theory. Broline and Loeb~\cite{BrolineLoeb1995} related endgames of Ayo to Tchoukaillon, the solitaire in which every move must end in the store and so earn a bonus turn, and showed that the number of stones in a winning position on $N$ pits is asymptotically at most $N^2/\pi$. Musesti, Paolini and Reale~\cite{MusestiPaoliniReale2015} proved an optimal bound on the number of moves before open mancala becomes periodic.

Our lower bound reduces from the truth of quantified Boolean formulas, which is $\PSPACE$-complete~\cite{StockmeyerMeyer1973}. For $\PSPACE$-completeness of two-player games more broadly, see Schaefer~\cite{Schaefer1978} and the book of Hearn and Demaine~\cite{HearnDemaine2009}. No hardness result was known for Kalah with a variable number of pits, not even $\mathsf{NP}$-hardness. \Cref{thm:main} settles the complexity of the unary problem.

\section{Proof overview}
\label{sec:overview}

We first prove membership, then describe the reduction. \Cref{tab:glossary} collects the terms that the proof uses.


\begin{figure}[tbp]
\centering
\begin{tikzpicture}[x=0.95cm,y=1cm]
  \draw[ovboard] (0.5,-0.45) rectangle (10.45,1.9);
  \foreach \x/\a in {1/1,2/0,4/0,6/2,7/0,8/0,9/1}{
    \node[ovpit] (A\x) at (\x,0) {$\a$};
  }
  \node[ovtgt] (A3) at (3,0) {$D$};
  \node[ovtgt] (A5) at (5,0) {$D$};
  \foreach \x/\b in {1/0,2/0,3/0,4/0,6/0,7/0,9/0}{
    \node[ovpit] (B\x) at (\x,1.45) {$\b$};
  }
  \node[ovcover] (B5) at (5,1.45) {$1$};
  \node[ovtimer] (B8) at (8,1.45) {$5$};
  \node[ovdots] at (9.85,0) {$\cdots$};
  \node[ovdots] at (9.85,1.45) {$\cdots$};
  \foreach \x in {1,...,9}{\node[ovlab,text=black!50] at (\x,-0.68) {$\x$};}
  \node[ovboblab,anchor=east] at (0.35,1.45) {Bob};
  \node[ovalicelab,anchor=east] at (0.35,0) {Alice};
  \draw[ovbob] (B8.north) to[out=140,in=40]
     node[ovlab,above] {a timer, its last stone lands at $8-5=3$} (B3.north);
  \draw[ovsow] (A3.north east) to[out=45,in=180] (4.2,0.72) -- (11.1,0.72)
     node[ovlab,right,text=black] {leaves\\the line,\\gone forever};
  \node[ovalicelab,above] at (7.3,0.72) {save};
  \node[ovlab] (thr) at (2.2,-1.45) {pending target,\\threatened};
  \draw[->,black!50] (thr) -- (A3.south west);
  \node[ovlab] (cov) at (6.4,-1.45) {pending target, covered\\by Bob's stone opposite};
  \draw[->,black!50] (cov) -- (A5.south east);
  \node[font=\small\bfseries,anchor=north west] at (current bounding box.north west) {(a)};
\end{tikzpicture}

\bigskip
\begin{tikzpicture}[x=1cm,y=1cm]
  \draw[ovboard] (0.55,-0.45) rectangle (10.45,2.35);
  \foreach \x/\l in {1.0/a-2,1.8/a-1,2.6/a}{
    \node[ovtgt] at (\x,0) {};
    \node[ovlab,text=black!55] at (\x,-0.62) {$\l$};
  }
  \foreach \x/\l in {3.4/a+1,4.2/a+2,5.6/g}{
    \node[ovpit] at (\x,0) {};
    \node[ovlab,text=black!55] at (\x,-0.62) {$\l$};
  }
  \foreach \x in {7.2,9.4}{\node[ovpit] at (\x,0) {};}
  \foreach \x in {1.0,1.8,2.6,3.4,4.2}{\node[ovpit] at (\x,1.9) {};}
  \node[ovpit] (G) at (5.6,1.9) {};
  \node[ovtimer] (T) at (7.2,1.9) {};
  \node[ovclockpit] (C) at (9.4,1.9) {};
  \foreach \x in {4.9,6.4,8.3,10.1}{
    \node[ovdots] at (\x,0) {$\cdots$};
    \node[ovdots] at (\x,1.9) {$\cdots$};
  }
  \node[ovboblab,anchor=south west] at (7.32,2.38) {timer};
  \node[ovlab,text=ovclock,anchor=south] at (9.4,2.38) {clock};
  \draw[ovbob,draw=ovbobc!35] (7.2,2.19) .. controls (7.1,3.72) and (2.7,3.72) .. (2.6,2.19);
  \draw[ovbob,draw=ovbobc!65] (7.2,2.19) .. controls (7.1,3.26) and (3.5,3.26) .. (3.4,2.19);
  \draw[ovbob] (7.2,2.19) .. controls (7.1,2.79) and (4.3,2.79) .. (4.2,2.19);
  \node[ovlab,anchor=north] at (4.2,1.56) {now};
  \node[ovlab,anchor=north,text=black!65] at (3.4,1.56) {$+1$};
  \node[ovlab,anchor=north,text=black!50] at (2.6,1.56) {$+2$};
  \draw[ovclocksow] (C.south) .. controls (9.3,1.01) and (5.7,1.01) .. (G.south);
  \node[ovlab,anchor=west] at (10.65,1.0) {a {\color{ovclock}clock} firing\\crosses the timer\\and lands at $g$};
  \node[ovalicelab,anchor=east] at (0.4,-1.05) {save order};
  \node[ovalicelab] at (2.6,-1.05) {1st};
  \node[ovalicelab] at (1.8,-1.05) {2nd};
  \node[ovalicelab] at (1.0,-1.05) {3rd};
  \node[ovboblab,anchor=east] at (0.4,1.9) {Bob};
  \node[ovalicelab,anchor=east] at (0.4,0) {Alice};
  \node[font=\small\bfseries,anchor=north west] at (current bounding box.north west) {(b)};
\end{tikzpicture}
\caption{(a) Line Kalah near the left end of a line. The blue pits are pending targets. Each holds $D$ stones, so a save sows past the right end of the line, where its stones vanish. Bob's timer at column $8$ would land in his empty pit at column $3$ and capture the target opposite, so Alice must save that target now. The target at column $5$ is covered by Bob's stone opposite, which blocks a capture there. (b) A deadline, for a left triple of targets and one of its timers. Each Bob move adds a stone to the timer, here by a clock firing that crosses it, so its landing column moves from $a+2$ to $a+1$ and then to $a$. Alice saves the highest, that is rightmost, pending target first, so each target is saved before the landing column reaches it.}
\label{fig:ov-basic}
\end{figure}
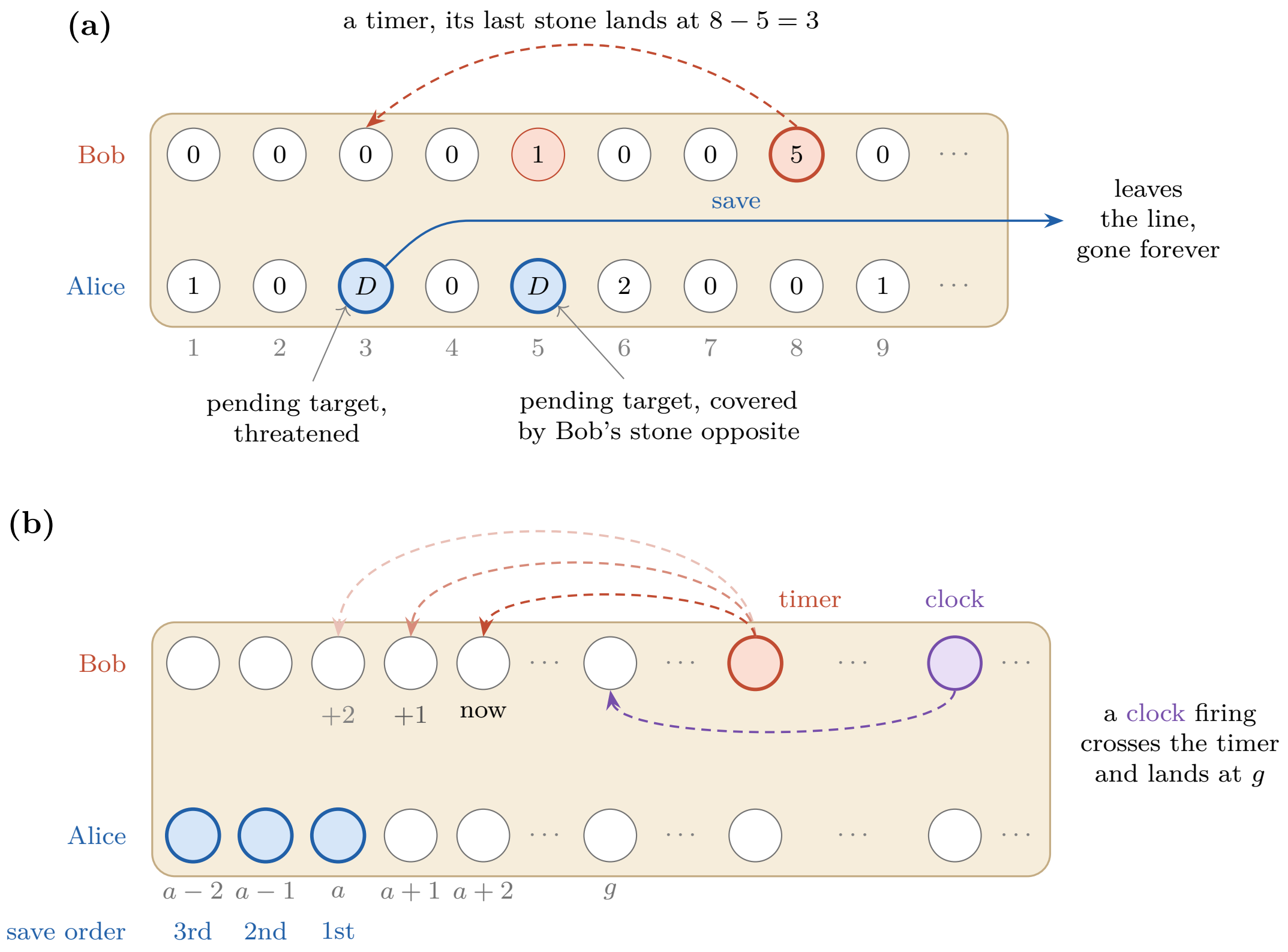

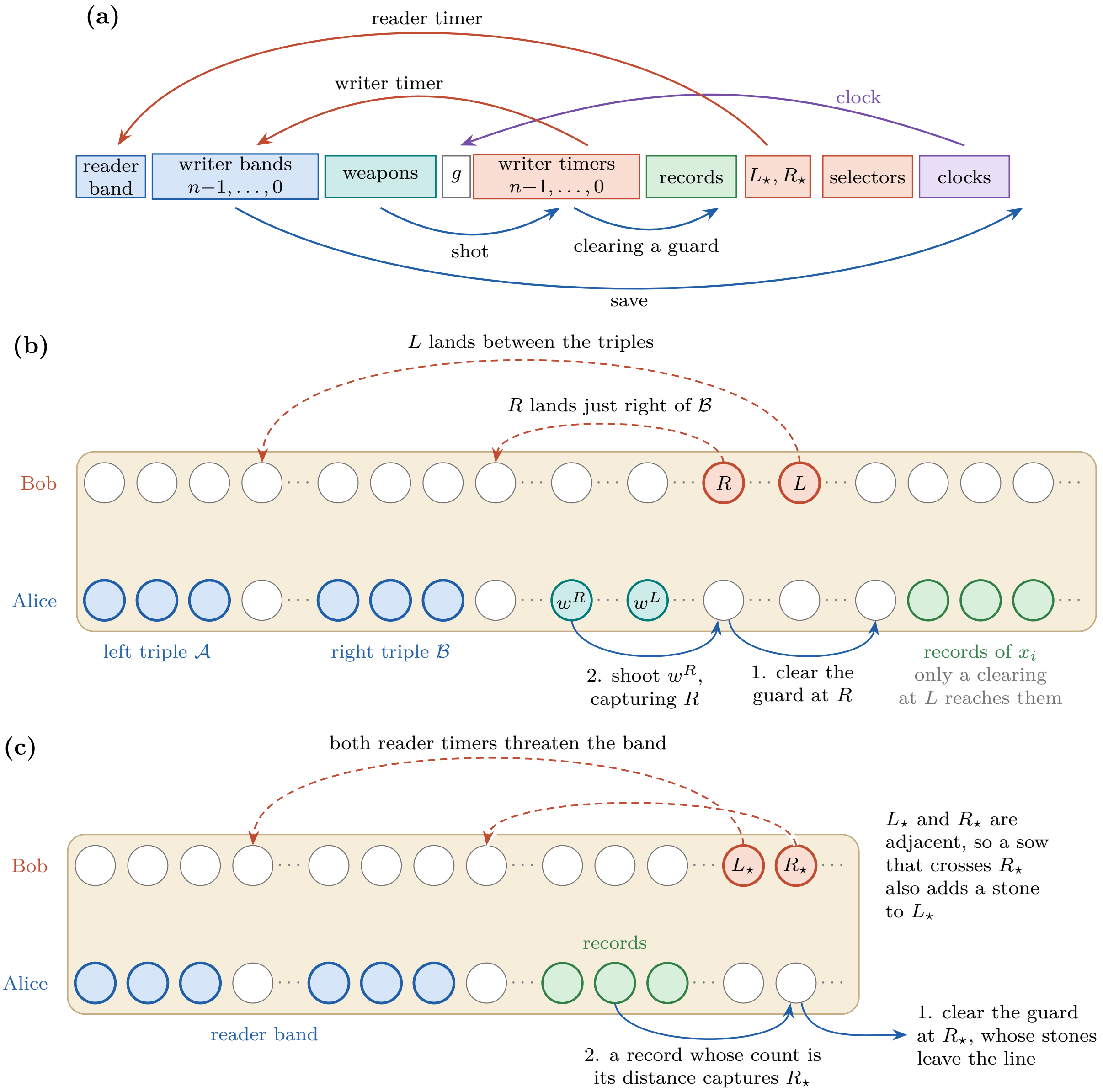
\begin{figure}[tbp]
\centering
\begin{tikzpicture}[x=1cm,y=1cm,>={Stealth},
  seg/.style={draw,minimum height=6mm,inner sep=1pt,line width=0.6pt,
              font=\scriptsize,align=center},
  tgtseg/.style={seg,draw=ovalice,fill=ovalicefill},
  bobseg/.style={seg,draw=ovbobc,fill=ovbobfill},
  lab/.style={font=\scriptsize,align=center}]
\node[tgtseg,minimum width=1.0cm] (RB) at (0.5,0) {reader\\band};
\node[tgtseg,minimum width=2.4cm] (WB) at (2.3,0) {writer bands\\$n{-}1,\ldots,0$};
\node[seg,draw=ovweapon,fill=ovweaponfill,minimum width=1.6cm] (WP) at (4.4,0) {weapons};
\node[seg,draw=black!55,fill=white,minimum width=0.4cm] (G) at (5.5,0) {$g$};
\node[bobseg,minimum width=2.4cm] (TB) at (6.95,0) {writer timers\\$n{-}1,\ldots,0$};
\node[seg,draw=ovrecord,fill=ovrecordfill,minimum width=1.3cm] (RC) at (8.9,0) {records};
\node[bobseg,minimum width=0.9cm] (RT) at (10.15,0) {$L_\star,R_\star$};
\node[bobseg,minimum width=1.3cm] (SE) at (11.45,0) {selectors};
\node[seg,draw=ovclock,fill=ovclockfill,minimum width=1.3cm] (CL) at (12.85,0) {clocks};
\draw[->,thick,ovclock] (12.85,0.45) to[out=160,in=20] node[lab,above,pos=0.2] {clock} (5.55,0.45);
\draw[->,thick,ovbobc] (7.4,0.45) to[out=150,in=30] node[lab,above,pos=0.6,text=black] {writer timer} (2.6,0.45);
\draw[->,thick,ovbobc] (10.0,0.45) .. controls (8,2.6) and (2,2.6) .. node[lab,above,text=black] {reader timer} (0.6,0.45);
\draw[->,thick,ovalice] (7.2,-0.45) to[out=-30,in=-150] node[lab,below,text=black] {clearing a guard} (9.3,-0.45);
\draw[->,thick,ovalice] (4.4,-0.45) to[out=-30,in=-150] node[lab,below,text=black] {shot} (7.0,-0.45);
\draw[->,thick,ovalice] (2.3,-0.45) .. controls (5,-2.0) and (11,-2.0) .. node[lab,below,text=black] {save} (13.7,-0.45);
\node[font=\small\bfseries,anchor=north west] at (current bounding box.north west) {(a)};
\end{tikzpicture}

\smallskip
\begin{tikzpicture}[x=1cm,y=1cm]
  \draw[ovboard] (-0.4,-0.45) rectangle (14.35,2.15);
  \foreach \x in {0,0.76,1.52,3.38,4.14,4.90}{\node[ovtgt] at (\x,0) {};}
  \foreach \x in {2.28,5.66,8.96,10.06,11.16}{\node[ovpit] at (\x,0) {};}
  \foreach \x in {11.92,12.68,13.44}{\node[ovrecordpit] at (\x,0) {};}
  \node[ovweaponpit] at (6.76,0) {$w^R$};
  \node[ovweaponpit] at (7.86,0) {$w^L$};
  \foreach \x in {0,0.76,1.52,2.28,3.38,4.14,4.90,5.66,6.76,7.86,11.16,11.92,12.68,13.44}{
    \node[ovpit] at (\x,1.7) {};}
  \node[ovtimer] at (8.96,1.7) {$R$};
  \node[ovtimer] at (10.06,1.7) {$L$};
  \foreach \x in {2.83,6.21,7.31,8.41,9.51,10.61,13.99}{
    \node[ovdots] at (\x,0) {$\cdots$};
    \node[ovdots] at (\x,1.7) {$\cdots$};
  }
  \node[ovboblab,anchor=east] at (-0.55,1.7) {Bob};
  \node[ovalicelab,anchor=east] at (-0.55,0) {Alice};
  \draw[ovbob] (10.06,1.99) .. controls (9.96,3.95) and (2.38,3.95) ..
     node[ovlab,above] {$L$ lands between the triples} (2.28,1.99);
  \draw[ovbob] (8.96,1.99) .. controls (8.86,2.72) and (5.76,2.72) ..
     node[ovlab,above] {$R$ lands just right of $\mathcal B$} (5.66,1.99);
  \draw[ovsow] (6.76,-0.29) .. controls (6.86,-0.96) and (8.78,-0.96) ..
     node[ovlab,below] {2. shoot $w^R$,\\capturing $R$} (8.88,-0.29);
  \draw[ovsow] (9.04,-0.29) .. controls (9.14,-0.96) and (11.06,-0.96) ..
     node[ovlab,below] {1. clear the\\guard at $R$} (11.16,-0.29);
  \node[ovalicelab,anchor=north] at (0.76,-0.5) {left triple $\mathcal A$};
  \node[ovalicelab,anchor=north] at (4.14,-0.5) {right triple $\mathcal B$};
  \node[ovlab,anchor=north] at (12.68,-0.5) {{\color{ovrecord}records of $x_i$}\\{\color{black!55}only a clearing}\\{\color{black!55}at $L$ reaches them}};
  \node[font=\small\bfseries,anchor=north west] at (current bounding box.north west) {(b)};
\end{tikzpicture}

\smallskip
\begin{tikzpicture}[x=1cm,y=1cm]
  \draw[ovboard] (-0.4,-0.45) rectangle (11.05,2.15);
  \foreach \x in {0,0.76,1.52,3.38,4.14,4.90}{\node[ovtgt] at (\x,0) {};}
  \foreach \x in {2.28,5.66,9.38,10.14}{\node[ovpit] at (\x,0) {};}
  \foreach \x in {6.76,7.52,8.28}{\node[ovrecordpit] at (\x,0) {};}
  \foreach \x in {0,0.76,1.52,2.28,3.38,4.14,4.90,5.66,6.76,7.52,8.28}{
    \node[ovpit] at (\x,1.7) {};}
  \node[ovtimer] at (9.38,1.7) {$L_\star$};
  \node[ovtimer] at (10.14,1.7) {$R_\star$};
  \foreach \x in {2.83,6.21,8.83,10.69}{
    \node[ovdots] at (\x,0) {$\cdots$};
    \node[ovdots] at (\x,1.7) {$\cdots$};
  }
  \node[ovboblab,anchor=east] at (-0.55,1.7) {Bob};
  \node[ovalicelab,anchor=east] at (-0.55,0) {Alice};
  \draw[ovbob] (9.38,1.99) .. controls (9.28,3.66) and (2.38,3.66) ..
     node[ovlab,above] {both reader timers threaten the band} (2.28,1.99);
  \draw[ovbob,preaction={draw=white,line width=2.6pt,solid,-}]
     (10.14,1.99) .. controls (10.04,2.79) and (5.76,2.79) .. (5.66,1.99);
  \node[ovlab,anchor=west,align=left] at (11.3,1.7) {$L_\star$ and $R_\star$ are\\adjacent, so a sow\\that crosses $R_\star$\\also adds a stone\\to $L_\star$};
  \node[ovlab,text=ovrecord,anchor=south] at (7.52,0.36) {records};
  \draw[ovsow] (7.52,-0.29) .. controls (7.62,-0.96) and (9.96,-0.96) ..
     node[ovlab,below] {2. a record whose count is\\its distance captures $R_\star$} (10.06,-0.29);
  \draw[ovsow] (10.22,-0.29) .. controls (10.32,-0.85) and (11.1,-0.75) .. (11.75,-0.75)
     node[ovlab,right,align=left] {1. clear the guard\\at $R_\star$, whose stones\\leave the line};
  \node[ovalicelab,anchor=north] at (2.45,-0.5) {reader band};
  \node[font=\small\bfseries,anchor=north west] at (current bounding box.north west) {(c)};
\end{tikzpicture}
\caption{(a) The order of the line, from low to high columns, not to scale. Bob sows to the left (arrows above) and Alice sows to the right (arrows below). A clock firing crosses every timer and every record, and in the intended play it lands at $g$. A writer timer crosses the timers of later writers, the weapons and the bands of earlier writers, and lands near its own band. A reader timer crosses everything. (b) A writer, with its two backup timers not shown. The left timer $L$ threatens the left triple and the right timer $R$ the right triple, although $L$ lies to the right of $R$ (\Cref{sec:construction}). To capture $R$, Alice first clears her guard opposite $R$, whose stones stop short of the records, and then shoots the weapon $w^R$, whose last stone lands in the emptied guard pit. Capturing $L$ instead uses the other guard and weapon, and that clearing adds one stone to each record of the variable. (c) The reader. Its two timers lie in adjacent columns and both threaten the reader band, so Bob cannot shift their deadlines apart. Alice can capture $R_\star$ in time only with a pile whose count is exactly its distance to $R_\star$. A dummy record has that count when Bob's counter is odd, and otherwise a record has it exactly when the clause named by the counter is satisfied.}
\label{fig:ov-stages}
\end{figure}

\begin{figure}[tbp]
\centering
\begin{tikzpicture}[x=1cm,y=1cm]
  \draw[ovboard] (-0.4,-0.45) rectangle (13.0,2.15);
  \foreach \x in {0,0.76,1.52,3.04,3.80,4.56,6.42,7.18,7.94,9.46,10.22,10.98}{
    \node[ovtgt] at (\x,0) {};}
  \foreach \x in {2.28,5.32,8.70,12.08}{\node[ovpit] at (\x,0) {};}
  \foreach \x in {0,0.76,1.52,2.28,3.04,3.80,4.56,5.32}{\node[ovpit] at (\x,1.7) {};}
  \foreach \x in {6.42,7.18,7.94,8.70,9.46,10.22,10.98}{
    \node[ovcover] at (\x,1.7) {$1$};}
  \node[ovtimer] (T) at (12.08,1.7) {};
  \foreach \x in {5.87,11.53,12.63}{
    \node[ovdots] at (\x,0) {$\cdots$};
    \node[ovdots] at (\x,1.7) {$\cdots$};
  }
  \node[ovboblab,anchor=east] at (-0.55,1.7) {Bob};
  \node[ovalicelab,anchor=east] at (-0.55,0) {Alice};
  \node[ovboblab,anchor=west,align=left] at (13.15,1.7) {timer of\\writer $j$};
  \draw[ovbob] (12.08,1.99) .. controls (11.98,3.4) and (5.42,3.4) ..
     node[ovlab,above] {an early firing crosses band $i$ and lands near band $j$} (5.32,1.99);
  \draw[decorate,decoration={brace,amplitude=4pt,mirror},ovbobc] (6.2,1.3) -- (11.2,1.3);
  \node[ovlab] at (8.7,0.9) {new stones opposite cover these targets};
  \node[ovalicelab,anchor=north] at (2.28,-0.5) {band of a later writer $j$};
  \node[ovalicelab,anchor=north] at (8.70,-0.5) {band of an earlier writer $i<j$};
  \node[font=\small\bfseries,anchor=north west] at (current bounding box.north west) {(a)};
\end{tikzpicture}

\bigskip
\begin{tikzpicture}[x=1cm,y=1cm]
  \draw[ovboard] (-1.5,-0.6) rectangle (14.9,1.75);
  \draw[rounded corners=3mm,draw=ovbobc,fill=ovbobfill,line width=0.7pt] (-1.35,-0.45) rectangle (-0.2,1.6);
  \node[ovlab] at (-0.775,0.58) {Bob's\\store};
  \draw[rounded corners=3mm,draw=ovalice,fill=ovalicefill,line width=0.7pt] (13.6,-0.45) rectangle (14.75,1.6);
  \node[ovlab] at (14.175,0.58) {Alice's\\store};
  \node[ovhot,minimum width=4.0cm,minimum height=8mm] at (2.0,1.15) {};
  \node[ovhot,draw=ovbobc,fill=ovbobfill,minimum width=4.6cm,minimum height=8mm] at (6.3,1.15) {Bob's line row};
  \node[ovhot,minimum width=4.8cm,minimum height=8mm] at (11.0,1.15) {empty};
  \node[ovhot,minimum width=4.0cm,minimum height=8mm] at (2.0,0) {empty};
  \node[ovhot,draw=ovalice,fill=ovalicefill,minimum width=4.6cm,minimum height=8mm] at (6.3,0) {Alice's line row, payloads $\Pi$};
  \node[ovhot,minimum width=4.8cm,minimum height=8mm] at (11.0,0) {empty};
  \node[draw=ovbobc,fill=ovbobfill,line width=1.1pt,rounded corners=2.6mm,minimum height=5.8mm,
        inner xsep=3pt,inner ysep=1.5pt,font=\scriptsize] (Q) at (0.95,1.15) {$(p{-}1)\Pi$};
  \node[ovlab,anchor=west] at (1.75,1.15) {padding};
  \node[ovlab,text=black!55] at (0.1,-0.8) {$1$};
  \node[ovlab,text=black!55] at (0.95,-0.8) {$q$};
  \node[ovlab,text=black!55] at (4.0,-0.8) {$h$};
  \node[ovlab,text=black!55] at (8.6,-0.8) {$h+D$};
  \node[ovlab,text=black!55] at (13.3,-0.8) {$N$};
  \node[ovlab] at (2.0,-1.2) {buffer};
  \node[ovlab] at (6.3,-1.2) {line segment};
  \node[ovlab] at (11.0,-1.2) {long empty stretch};
  \draw[ovsow] (0.2,-1.6) -- (13.4,-1.6) node[ovalicelab,midway,below] {Alice sows right, toward her store};
  \draw[ovsow,draw=ovbobc] (13.4,2.3) -- (0.2,2.3) node[ovboblab,midway,above] {Bob sows left, toward his store};
  \node[font=\small\bfseries,anchor=north west] at (current bounding box.north west) {(b)};
\end{tikzpicture}
\caption{(a) Protection by covering. The band of a later writer $j$ lies to the left of the band of an earlier writer $i<j$. If Bob fires a timer of writer $j$ early, its sow puts a stone in every Bob pit opposite a target of writer $i$, and a covered target cannot be captured. Bob can remove a cover only by selecting the covering pit, and Alice then saves the exposed target on her next move. (b) The Kalah board of \Cref{sec:empty-stores}, not to scale. The shaded middle part is the line position, with a payload of $\Pi$ extra stones on each target. To its left lies a buffer that is empty apart from Bob's padding pile of $(p-1)\Pi$ stones at column $q$, and to its right lies a long empty stretch. Bob collects the padding after any capture of a payload, so Alice wins exactly when she keeps every payload. Stones that Alice sows off the line stay in that stretch, and stones that Bob sows past his store land only in Alice's buffer, so no stone returns to the line.}
\label{fig:ov-cover-kalah}
\end{figure}
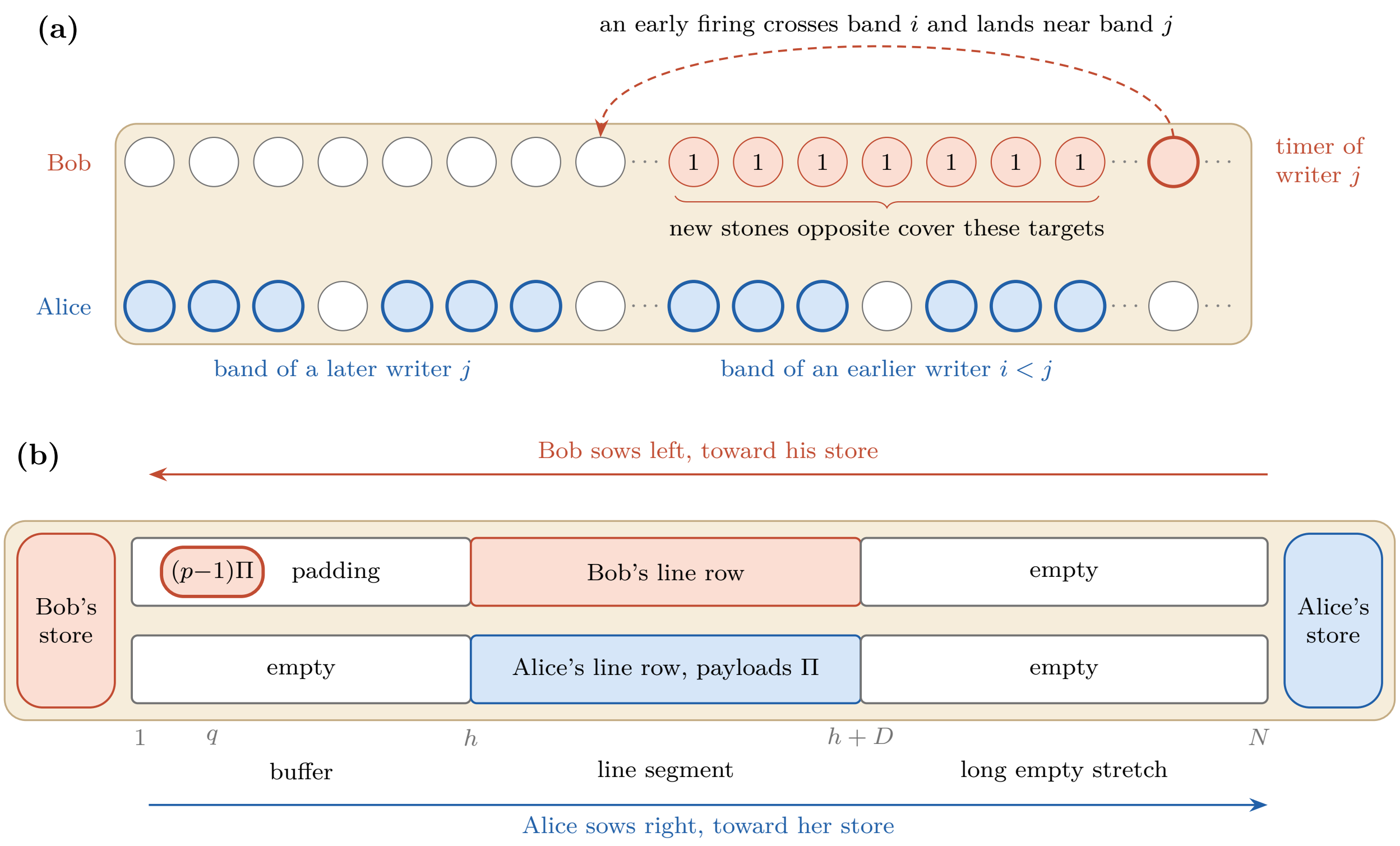

\paragraph{Membership.} A Kalah move either puts a stone into a store or moves all of its stones closer to the mover's store within the mover's own row. Hence every game lasts a number of moves polynomial in the width and the number of stones (\Cref{sec:membership}), and the game tree can be searched in polynomial space.

\paragraph{Line Kalah.} Kalah is played on a cycle, and every stone eventually reaches a store. Most of our construction is easier to describe on a line. In \emph{Line Kalah} (\Cref{sec:linegame}) Alice sows to the right, Bob sows to the left, and stones that leave the line disappear. We use the following terminology. Some of Alice's pits are marked as \emph{targets}. A target is \emph{pending} until Alice selects it, which \emph{saves} it. Alice must avoid captures of pending targets until Bob's row is empty (\Cref{fig:ov-basic}(a)).

Line Kalah isolates the source of hardness in our reduction. There are no stores, and neither player ever sows into the other's row, so captures are the only interaction between the players. In particular, the analysis never has to follow stones around the board. The last paragraph of this overview explains how the line positions of the construction become Kalah positions in which stones still never cross the board.

\paragraph{Deadlines.} On the line, a Bob pile at column $x$ with $c$ stones lands at column $x-c$. Every stone that later falls on the pile moves this landing column one step to the left. A pile whose landing column is approaching a group of targets is therefore a \emph{timer}, because Alice must save each target before the landing column reaches it. Bob also has \emph{clocks}, piles far to the right whose firing adds one stone to every timer, so in the intended play every Bob move brings every deadline one step closer (\Cref{fig:ov-basic}(b)).

\paragraph{Writers.} The hardness construction (\Cref{sec:construction}) consists of one \emph{writer} for each quantified variable, followed by a \emph{reader} (\Cref{fig:ov-stages}(a)). Each of these stages has six targets, its \emph{band}, consisting of a left and a right group of three. Timers threaten both groups. The deadlines leave Alice time for exactly two extra moves, which she must use to remove one of the two threatening timers by capturing it (\Cref{fig:ov-stages}(b)). She first \emph{clears} her pit opposite the timer, called its \emph{guard}, and then \emph{shoots} a \emph{weapon}, a pile whose count is exactly its distance to the timer. Her choice of timer is a bit. At a universal variable Bob chooses the bit by destroying one of Alice's two weapons first. Alice owns one \emph{record} pile for each occurrence of each variable in the formula. One of the two clearings of a writer adds a stone to every record of its variable, so the counts of the records encode the assignment.

\paragraph{The reader.} After all variables, Bob sets a counter of extra universal bits, and an even value of the counter names a clause. The reader then has a deadline that Alice can meet only if some record has exactly the count needed to capture the reader's right timer, which happens exactly when the named clause is satisfied (\Cref{fig:ov-stages}(c)). An odd value names no clause, and then a dummy record has the right count.

\paragraph{Unrestricted moves.} The main difficulty is that either player may select any nonempty pit at any time, not only the moves of the intended play. If the formula is false, Bob wins with the intended moves, because the deadlines force Alice to follow the intended computation and any other move by her loses a target (\Cref{sec:false}). If the formula is true, Alice needs a strategy that defeats every Bob strategy, including those that leave the intended play (\Cref{sec:true}). Two features of the layout make this possible.

First, later stages lie to the left of earlier stages. If Bob fires a timer of a later stage early, its sow crosses every target of the earlier stages and protects them, because a pit opposite a target blocks captures there while it is nonempty (\Cref{fig:ov-cover-kalah}(a)). Alice can then move on and save the skipped targets whenever Bob removes that protection. Other moves are handled by counting the increments received by the timers and the saves already made. The result is enough extra time to save the targets or a protected input for the reader. An already started preparation may still need to be completed.

Second, the reader's two timers occupy adjacent columns. While both remain unfired, every Bob move that adds a stone to the left timer also adds one to the right. Thus, if the left timer has gained no extra time, neither has the right (\Cref{fig:ov-stages}(c)). In that case the records still encode an assignment consistent with Alice's winning formula strategy.

\paragraph{Alice's two modes.} These two features give Alice's strategy for true formulas its shape. She starts in \emph{computing mode}, in which she plays the intended preparation and saves at each writer and chooses her bits by a winning strategy for the formula. At Bob's first move that leaves the intended play, she switches permanently to \emph{escaping mode}. Such a move is a pass, a move of a leftover Bob pile, the early firing of a timer of a later stage (a \emph{jump}), or the firing of the timer whose capture Alice has just prepared. In escaping mode she no longer computes. At each stage she only saves targets, choosing which group of three to finish first, and the counting shows that the deviation has given her enough time to save each band without a preparation. After a jump she moves straight to the later stage and saves the skipped targets whenever Bob removes their protection. At the reader, either Bob's deviations have gained her time, or the reader's deadlines are intact. In the second case every writer was played in computing mode, and the records encode an assignment under which the clause named by Bob's counter is satisfied.


\paragraph{Back to Kalah.} \Cref{sec:empty-stores} embeds the line positions of the construction in a Kalah board with both stores empty (\Cref{fig:ov-cover-kalah}(b)). Long empty stretches on both sides of the line ensure that stones leaving it never return, and moves outside the line act as passes. Alice may pass in Line Kalah, and her strategy for true formulas still works when Bob may pass (\Cref{cor:bob-passes}). The score is carried by the targets, the Alice pits that Bob must not capture. Each of the $p$ targets carries a payload of $\Pi$ stones, more than all other line stones together, and Bob has a separate pile of $(p-1)\Pi$ stones near his store. The total mass is $2p\Pi-1$, so Alice wins by keeping all $p$ payloads, while Bob wins by capturing one and collecting his separate pile. A nearest-store collection lemma shows that he can secure that pile after the capture.

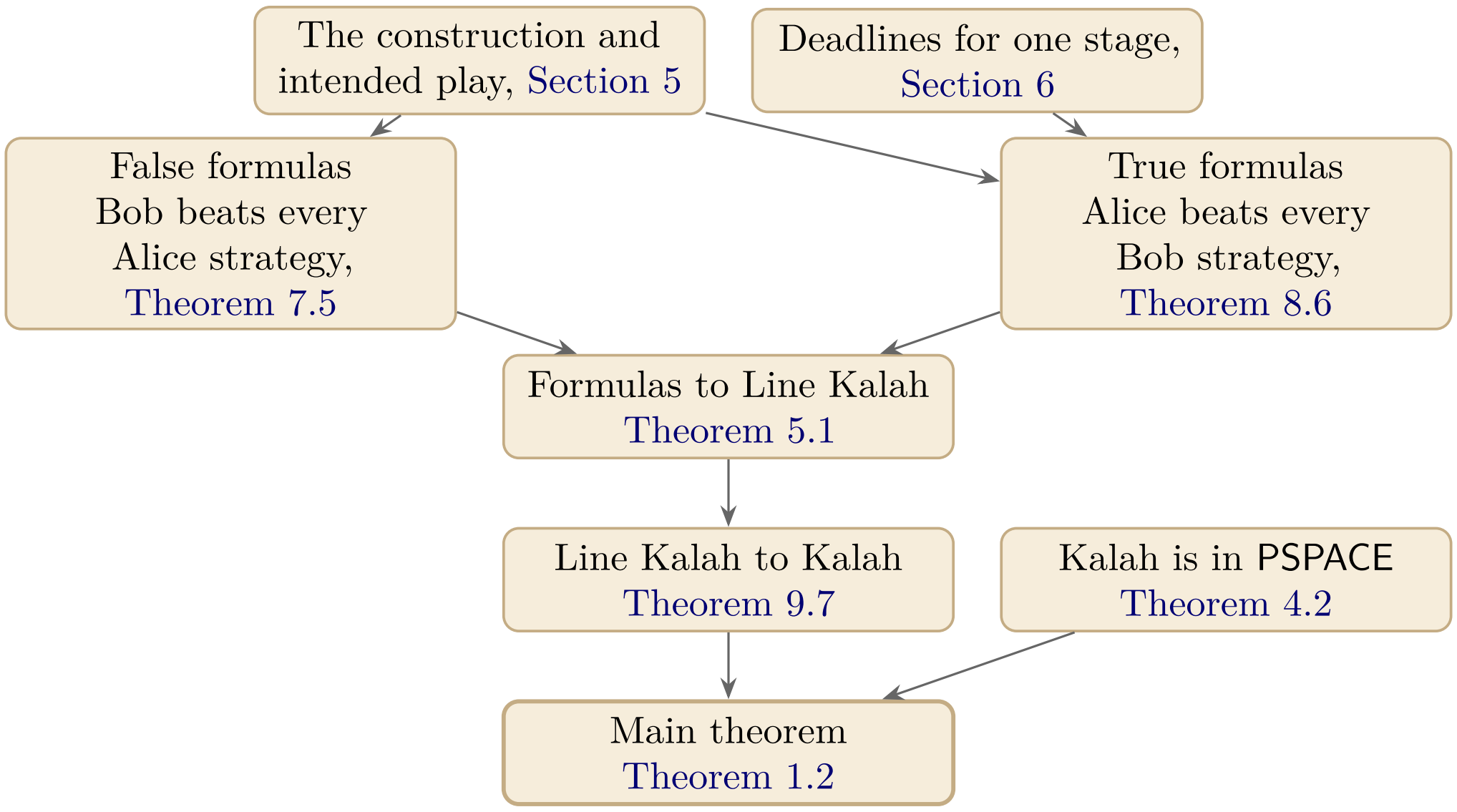
\begin{figure}[tbp]
\centering
\begin{tikzpicture}[
  box/.style={draw=ovboardline,fill=ovboardfill,rounded corners,align=center,font=\small,line width=0.7pt,
              text width=3.9cm,minimum height=0.95cm},
  dep/.style={->,line width=0.6pt,black!60},
  >={Stealth}]
\node[box] (constr) at (-2.3,0)
  {The construction and\\intended play, \Cref{sec:construction}};
\node[box] (bands) at (2.3,0)
  {Deadlines for one stage,\\\Cref{sec:bands}};
\node[box] (false) at (-4.6,-1.6) {False formulas\\Bob beats every\\Alice strategy, \Cref{thm:false}};
\node[box] (true) at (4.6,-1.6) {True formulas\\Alice beats every\\Bob strategy, \Cref{thm:true}};
\node[box] (line) at (0,-3.2)
  {Formulas to Line Kalah\\\Cref{thm:formula-to-line}};
\node[box] (transfer) at (0,-4.8)
  {Line Kalah to Kalah\\\Cref{thm:empty-transfer}};
\node[box] (length) at (4.6,-4.8)
  {Kalah is in $\PSPACE$\\\Cref{thm:membership}};
\node[box,line width=1.1pt] (main) at (0,-6.4) {Main theorem\\\Cref{thm:main}};
\draw[dep] (constr)--(false);
\draw[dep] (constr)--(true);
\draw[dep] (bands)--(true);
\draw[dep] (false)--(line);
\draw[dep] (true)--(line);
\draw[dep] (line)--(transfer);
\draw[dep] (transfer)--(main);
\draw[dep] (length)--(main);
\end{tikzpicture}
\caption{How the main results depend on each other. The two strategy theorems handle arbitrary play by the opponent, not only the intended play. Every part of the hardness proof also uses the facts about landing columns and covers in \Cref{sec:toolkit}.}
\label{fig:proof-tree}
\end{figure}

\paragraph{Organization.} \Cref{fig:proof-tree} shows how the main results depend on each other. \Cref{sec:prelim} defines Line Kalah and collects facts about single moves, and \Cref{sec:membership} proves membership. \Cref{sec:construction,sec:bands,sec:false,sec:true} reduce quantified formulas to Line Kalah, \Cref{sec:empty-stores} reduces Line Kalah to Kalah, and \Cref{sec:consequences} discusses open questions. The appendices contain the coordinates of the construction with the routine calculations about them (\Cref{app:coordinates}), the proofs of the lemmas of \Cref{sec:bands,sec:false,sec:true,sec:empty-stores} (\Cref{app:deadline-proofs,app:false-proofs,app:true-proofs,app:kalah-proofs}), and score targets and binary counts (\Cref{app:extensions}).

\begin{table}[tbp]
\centering
\small
\begin{tabular}{@{}p{0.24\textwidth} p{0.75\textwidth}@{}}
\toprule
Term & Meaning\\
\midrule
pile & a nonempty pit\\
target, pending, save & an Alice pit that Bob must not capture. It is pending until Alice first selects it, which saves it\\
saturated & a target whose selection sows past the end of the line\\
landing column $e(x)$ & the column $x-B_x$ where the last stone of Bob's pile at $x$ lands\\
cross & a move crosses the columns strictly between its source and its last stone, and puts one stone on each\\
structural, fire, unfired & Bob's designated initial piles. One fires when Bob first selects it\\
residue & every Bob pile other than an unfired structural pile\\
frontier $E_\Omega$ & the smallest landing column of a Bob pile outside the set $\Omega$\\
covered, exposed & an Alice pit is covered while Bob's opposite pit is nonempty, which blocks captures there. A pending target is exposed when Bob's last move removed its cover\\
quiet move & a residue move far above a band, which changes nothing that the band depends on\\
band, triple & the six targets of a stage, a left and a right triple\\
slack & how far the frontier lies above the highest pending target of a triple\\
timer, backup, clock, selector & Bob's structural piles. Timers threaten bands, a clock firing adds a stone to every timer, and selectors capture weapons at universal writers\\
guard, clear, weapon, shot & a guard is Alice's pit opposite a timer, which she clears by selecting it. A weapon is a pile whose count is its distance to the timer, and a shot fires it to capture the timer\\
record & an Alice pile whose count stores the bit of one occurrence of a variable\\
stage, jump & a writer or the reader. A jump is Bob's firing of a timer of a stage after the active one\\
intended play, branch, order & the play the construction is designed for. An intended branch is Alice's clearing, shot and six saves at a writer, and the intended order is the order of saves that Bob's intended strategy forces\\
computing, escaping mode & Alice's strategy for true formulas before and after Bob first leaves the intended play\\
\bottomrule
\end{tabular}
\caption{Terms used in the proof, in order of first use.}
\label{tab:glossary}
\end{table}

\section{Preliminaries}
\label{sec:prelim}

This section defines Line Kalah, the game on a line used by the construction, and collects the facts about single moves that every later argument uses. Kalah itself was defined in \Cref{def:kalah}.

\subsection{Line Kalah}
\label{sec:linegame}

Line Kalah removes the stores and the cyclic board, and replaces the score by a safety condition. \Cref{fig:ov-basic}(a) shows part of a line position, and \Cref{fig:kalah-board} shows a Kalah position for comparison.

\begin{definition}[Line Kalah]
\label{def:linegame}
A \emph{line position} consists of a width $D\ge1$, nonnegative integers $A_x$ and $B_x$ for $x\in\{1,\ldots,D\}$, and a set $P\subseteq \{1,\ldots,D\}$. The number $A_x$ is the count of \emph{Alice's pit} at column $x$ and $B_x$ is the count of \emph{Bob's pit} at column $x$. These two pits are \emph{opposite}. The columns in $P$ are Alice's \emph{targets}.

A move selects a nonempty pit of the player to move and empties it. Alice sows to the right. Selecting her pit at $x$ adds one stone to each of her pits $x+1,x+2,\ldots$ in turn, and stones that would go beyond column $D$ are discarded. Bob sows to the left in the same way, through his pits $x-1,x-2,\ldots$, and stones that would go below column~$1$ are discarded. If the last stone lands in one of the mover's own pits at a column $y$ with $1\le y\le D$, that pit was empty before the stone arrived, and the opposite pit is nonempty, then the last stone and the opposite pile are removed from the board (a \emph{capture} at column $y$). We use the same two capture conventions as in \Cref{def:kalah}. With empty capture, the last stone is also removed when its landing pit was empty and the opposite pit is empty. Under the nonempty-opposite rule, that stone stays in its pit. There are no stores in this game, so all captured stones are discarded.

Alice moves first and the players alternate. Alice may also pass. Bob must move while his row is nonempty. Each target is \emph{pending} until Alice first selects it, and that selection \emph{saves} the target. Bob wins as soon as he captures a pending target. If Bob's row is empty at the start, or becomes empty after a move without such a capture, the game ends and Alice wins.

A target $x$ is \emph{saturated} if $A_x\ge D-x+1$. The \emph{Line Kalah problem} asks, for a line position whose targets are all saturated and whose counts are written in unary, whether Alice has a strategy that wins. The capture convention is fixed for the whole game. Unless a statement specifies otherwise, the results below hold separately under either convention.
\end{definition}

A \emph{pile} is a nonempty pit. An Alice pile at column $x$ with at least $D-x+1$ stones adds one stone to every Alice pit to its right, whatever its exact count, and its last stone leaves the line, so it captures nothing. Likewise a Bob pile at column $x$ with at least $x$ stones adds one stone to every Bob pit to its left and captures nothing. In particular, a save never captures and never changes Bob's row. A pending target only gains stones until it is saved or captured, so it stays saturated while it is pending. Every nonempty pit may be selected at any time. The construction of \Cref{sec:construction} assigns roles to pits, but these roles never restrict a legal move. We say that a column lies below or above another when it is smaller or larger. The figures draw columns increasing to the right, so below means to the left.

\subsection{Landing columns and covers}
\label{sec:toolkit}

This subsection collects the facts about line positions that every later argument uses. Throughout, we fix a line position, call the start of the game time~$0$, and assume that every target is saturated. All statements in this subsection and the next hold under either capture convention.

\begin{definition}[Landing columns, residue and covers]
\label{def:landing}
For a Bob pile at column $x$ its \emph{landing column} is $e(x)=x-B_x$. If $e(x)\ge1$, selecting it puts one stone on each of Bob's pits $x-1,x-2,\ldots,e(x)$, and the last stone lands at $e(x)$. Otherwise the last stone leaves the line. A move \emph{crosses} the columns strictly between the selected column and the column of its last stone, or every column beyond the selected one in the direction of sowing if its last stone leaves the line. So a Bob move from $x$ crosses the columns $y$ with $\max\{e(x),0\}<y<x$, and an Alice move from $x$ with $A_x=c$ crosses the columns $x+1,\ldots,\min\{x+c-1,D\}$.

The construction of \Cref{sec:construction} designates some of Bob's initial piles as \emph{structural}. A structural pile is \emph{unfired} until Bob first selects its pit, which \emph{fires} it, or Alice captures it. After that its pit is an ordinary pit, and any stones that later fall into it form a new pile. A Bob pile that is not an unfired structural pile is called \emph{residue}, and a Bob move that selects a residue pile is a \emph{residue move}.

For a set $\Omega$ of unfired structural piles, the \emph{frontier} $E_\Omega$ is the smallest landing column of a Bob pile not in $\Omega$, or $+\infty$ if there is none. A pile of $\Omega$ that fires or is captured leaves $\Omega$. We write $E$ for $E_\emptyset$.

Alice's pit at column $z$ is \emph{covered} if $B_z>0$. A pending target is \emph{exposed} at a time when Alice is to move if its pit was covered before Bob's last move and is not covered now.
\end{definition}

\begin{lemma}[Single moves]
\label{lem:moves}
\leavevmode
\begin{enumerate}
\item\label{mv:capture} \textbf{Captures of opposing piles.} A Bob move from column $x$ captures a nonempty Alice pile only at column $e(x)$, and does so exactly when $e(x)\ge1$, $B_{e(x)}=0$ before the move, and $A_{e(x)}>0$. An Alice move from column $x$ with $A_x=c$ captures a nonempty Bob pile only at column $x+c$, and does so exactly when $x+c\le D$, $A_{x+c}=0$ before the move, and $B_{x+c}>0$.
\item\label{mv:covered} \textbf{Covers persist.} If Alice's pit at $z$ is covered when Bob is to move, then Bob's move does not capture at $z$, and afterwards the pit is still covered unless the move selected Bob's pit at $z$.
\item\label{mv:sow} \textbf{A sow covers what it crosses.} After a Bob move from column $x$, every Bob pit at a column $y$ with $\max\{e(x),0\}<y<x$ is nonempty, so Alice's pit at $y$ is covered.
\item\label{mv:frontier} \textbf{The frontier moves slowly.} Let $\Omega$ be a set of unfired structural piles. A Bob move that fires no pile of $\Omega$ lowers $E_\Omega$ by at most one, and an Alice move never lowers $E_\Omega$. The landing column of an unfired structural pile never increases.
\item\label{mv:empty} \textbf{Empty capture.} From the same position and selected pit, the two capture conventions give different positions only if the last stone lands on the line in a previously empty own pit whose opposite pit is empty. The empty-capture convention then removes that last singleton. This removes no pile that was nonempty before the move, changes no deposit strictly between the source and landing columns, and does not change the opposing row. For a Bob move it cannot occur opposite a pending target.
\end{enumerate}
\end{lemma}
\begin{proof}
A sow on the line never reaches the same pit twice, so the pit where the last stone lands receives exactly one stone from the move. It was empty before that stone arrived if and only if it was empty before the move, and item~\labelcref{mv:capture} follows from either capture rule of \Cref{def:linegame}.

A Bob move removes its own stones only from the pit it selects and, when it captures, from the pit where its last stone lands, and the latter pit was empty before the move. This gives item~\labelcref{mv:covered}. For item~\labelcref{mv:sow}, the move puts one stone on each of Bob's pits $x-1,\ldots,\max\{e(x),1\}$. If $e(x)\ge1$, the last stone lands at $e(x)<y$. Otherwise it leaves the line. Either way the capture rule can remove a deposited Bob stone only at column $e(x)$.

For item~\labelcref{mv:frontier}, a Bob move from $x\notin\Omega$ adds at most one stone to every other pile, which lowers its landing column by at most one. The selected pile, whose landing column was $e(x)\ge E_\Omega$, disappears. Any new pile that survives the move holds one stone and lies at a column $y\ge e(x)$, so its landing column $y-1$ is at least $E_\Omega-1$. A capture only removes piles. Alice changes Bob's row only by capturing a pile, and a captured pile of $\Omega$ leaves $\Omega$. Finally, an unfired structural pile only gains stones. The two conventions differ only in the case described in item~\labelcref{mv:empty}. Its last assertion holds because the Alice pile at a pending target is nonempty until Bob captures it.
\end{proof}

\subsection{Quiet moves}
\label{sec:quiet}

Some Bob moves happen far away from the targets that a local argument protects, and change nothing that the argument depends on. The bound $H$ below will be the parameter of the construction.

\begin{definition}[Quiet move]
\label{def:quiet}
Fix a positive integer $H$ and an integer $\lambda$. A Bob move is \emph{$\lambda$-quiet} if it selects a residue pile at a column $z$ with $z-2H>\lambda$, and no unfired structural pile lies at a column in $[z-H,\,z-1]$.
\end{definition}

\begin{lemma}[Quiet moves]
\label{lem:quiet}
Suppose every residue pile holds at most $H$ stones before and after a $\lambda$-quiet Bob move from column $z$. Then the following hold.
\begin{enumerate}
\item Among Bob pits, the move changes only those at columns in $[z-H,z]$, and it changes no unfired structural pile.
\item Every Bob pile that the move removes, changes or creates has landing column greater than $\lambda$ both before and after the move.
\item The move captures a nonempty Alice pile only at a column in $[z-H,z-1]$.
\item For every set $\Omega$ of unfired structural piles, the quantity $\min\{E_\Omega,\lambda+1\}$ is unchanged.
\item Every covered Alice pit other than the one at $z$ stays covered.
\end{enumerate}
\end{lemma}
\begin{proof}
The selected pile holds at most $H$ stones, so it sows into the pits $z-1,\ldots,z-H$ at most, and its landing column is at least $z-H>\lambda$. The pits in $[z-H,z-1]$ hold no unfired structural pile, so they hold residue, with at most $H$ stones before and after the move, so their landing columns are at least $z-2H>\lambda$. This proves the first three items. For the fourth, piles with landing column at most $\lambda$ are not affected, and every affected pile has landing column at least $\lambda+1$ before and after the move. The fifth item is \Cref{lem:moves}(\labelcref{mv:covered}).
\end{proof}

\paragraph{Passes.} In \Cref{sec:true} Bob may also \emph{pass}, that is, use a turn without selecting a pit. We count a pass as a Bob move that selects no pit and changes nothing, and we call it $\lambda$-quiet for every $\lambda$. Every statement of \Cref{lem:moves,lem:quiet} about a Bob move then holds trivially for a pass. The proofs in \Cref{sec:bands} consider one Bob move at a time and handle quiet moves through these lemmas, so all results of that section hold, with the same proofs, for plays in which Bob may pass.

\section{Kalah is in PSPACE}
\label{sec:membership}

This section proves the upper bound of \Cref{thm:main}. Every Kalah game lasts a number of moves polynomial in the width and the number of stones, so the game tree can be searched in polynomial space. The bound holds for every Kalah position, including sows that go around the board several times, captures under either convention, and bonus turns.

\begin{lemma}[Polynomial play length]
\label{lem:polynomial-play}
Every play from a Kalah position of width $N$ whose pits hold $M$ stones in total has at most $(M+1)(NM+1)$ moves under either capture convention, with or without the bonus rule. The bound also holds if captures are omitted.
\end{lemma}
\begin{proof}
Let $\Sigma$ be the total number of stones in the two stores, and let $\Psi$ be the sum, over all stones in pits, of the distance from the stone's pit to its owner's store. A pit at distance $d\in\{1,\ldots,N\}$ before its owner's store is the pit from which a sown stone reaches that store after $d$ steps. Stones never leave a store, so $\Sigma$ never decreases and increases at most $M$ times in total. Also $0\le\Psi\le NM$ at every time.

Call a move a \emph{store move} if it puts at least one stone into a store, by sowing, by a capture or by the final sweep. A store move increases $\Sigma$, so there are at most $M$ store moves. Consider any other move. Its mover selects a pit at some distance $d$ holding $c$ stones. If $c\ge d$, the sow would reach the mover's store, so $c<d$, and the stones land in the mover's own pits at distances $d-1,\ldots,d-c$. No capture occurs, since a capture would put stones into the store. So the move lowers $\Psi$ by $1+2+\cdots+c\ge1$ and changes nothing else. Between two consecutive store moves, and before the first and after the last, there are therefore at most $NM$ other moves. Hence the play has at most $M+(M+1)NM\le(M+1)(NM+1)$ moves. The argument does not depend on the capture convention, on bonus turns, or on whether captures are allowed.
\end{proof}

The proof also gives an order in which to evaluate positions. Every move either increases $\Sigma$, or leaves $\Sigma$ unchanged and lowers $\Psi$. So every move increases the pair $(\Sigma,-\Psi)$ in the lexicographic order.

\begin{theorem}[Membership]
\label{thm:membership}
Deciding whether Alice wins a given unary Kalah position belongs to $\PSPACE$ under either capture convention, with or without the bonus rule.
\end{theorem}
\begin{proof}
With unary counts, $(M+1)(NM+1)$ is polynomial in the input length, so by \Cref{lem:polynomial-play} the game tree has polynomial depth. A position, including both store counts and the player to move, has an encoding of polynomial length, and its successors can be listed one at a time. A depth-first evaluation of the game tree therefore uses polynomial space. Alice wins exactly when her final store exceeds Bob's.
\end{proof}

For each constant width $N$, the winner can even be computed in polynomial time under either capture convention, with or without the bonus rule. There are at most $2\binom{M+2N+1}{2N+1}=M^{O(1)}$ ways to distribute the $M$ pit stones over the pits and stores, with either player to move, and they can be evaluated in decreasing lexicographic order of $(\Sigma,-\Psi)$, since every move increases this pair.

\section{The construction}
\label{sec:construction}

We now build, from a quantified Boolean formula, a line position that Alice wins exactly when the formula is true. This section and the next three prove the following theorem, the first half of the reduction.

\begin{theorem}[Formulas to Line Kalah]
\label{thm:formula-to-line}
From a quantified Boolean formula of the form described in \Cref{sec:source} one can compute, in polynomial time, a line position whose targets are all saturated and whose counts are at most its width, such that under either capture convention Alice wins it if and only if the formula is true.
\end{theorem}

This section gives the construction and its intended play. \Cref{sec:bands} proves deadline lemmas for a single stage. \Cref{sec:false} gives Bob a strategy that wins against every Alice strategy when the formula is false, and \Cref{sec:true} gives Alice a strategy that wins against every Bob strategy when the formula is true. The proof of \Cref{thm:formula-to-line} is at the end of \Cref{sec:true}.

\subsection{The formula and the assignment game}
\label{sec:source}

We reduce from closed prenex formulas
\[
 \mathcal Q_0x_0\,\mathcal Q_1x_1\cdots\mathcal Q_{m-1}x_{m-1}\;
 \varphi_0\wedge\varphi_1\wedge\cdots\wedge\varphi_{C-1},
\]
where each $\mathcal Q_i$ is $\exists$ or $\forall$, $m\ge1$, $C\ge1$, $\mathcal Q_0=\exists$, and each clause $\varphi_j$ is a nonempty disjunction of literals $x_i$ or $\neg x_i$. Deciding the truth of such formulas is $\PSPACE$-complete~\cite{StockmeyerMeyer1973}, and the conjunctive normal form is the standard one (problem LO11 in~\cite{GareyJohnson1979}). The other conditions cost nothing. An unused existential variable can be added at the front, and a formula with no variable, no clause or an empty clause is trivially true or false. Let $\rho\ge1$ be the largest number of occurrences of any single variable in the clauses (or $1$ if no variable occurs).

Put $n=m+2C-2$. The construction has $n$ \emph{stages}, called writers, numbered $0,\ldots,n-1$, followed by a final stage called the reader. Writer $i<m$ is the \emph{genuine} writer of $x_i$, and writers $m,\ldots,n-1$ are \emph{counter} writers. Each writer produces a bit $\epsilon_i$. A stage is \emph{universal} if it is the writer of a universally quantified variable or a counter writer, and \emph{existential} otherwise. Stage $0$ is existential. Let $n_\forall$ be the number of universal stages.

\begin{definition}[Assignment game]
\label{def:assignment-game}
In the \emph{assignment game}, the bits $\epsilon_0,\ldots,\epsilon_{n-1}$ are chosen in this order, by Alice at existential stages and by Bob at universal stages, each player seeing all earlier bits. Let $\nu=\epsilon_m+\cdots+\epsilon_{n-1}$ be the number of counter bits equal to one. Alice wins if $\nu$ is odd, or if $\nu=2j$ and the clause $\varphi_j$ is true under $x_i=\epsilon_i$ for $i<m$.
\end{definition}

\begin{lemma}[Assignment game]
\label{lem:assignment-game}
Alice wins the assignment game if and only if the formula is true.
\end{lemma}
\begin{proof}
Every even value $\nu=2j$ with $0\le\nu\le2C-2$ names a clause. If the formula is true, Alice chooses $\epsilon_0,\ldots,\epsilon_{m-1}$ by a winning strategy of the formula. The resulting assignment satisfies every clause, so she wins for every value of $\nu$. If the formula is false, Bob chooses the universal bits of $\epsilon_0,\ldots,\epsilon_{m-1}$ by a winning strategy of the falsifier, picks a clause $\varphi_j$ that is false under the resulting assignment, and sets exactly $2j\le 2C-2$ counter bits to one.
\end{proof}

\subsection{The pieces}

Each stage has a \emph{band} of six targets, consisting of a left triple $\mathcal A=\{a-2,a-1,a\}$ and a right triple $\mathcal B=\{b-2,b-1,b\}$, with $a<b$. Bob owns structural piles of three kinds.
\begin{itemize}
\item Each writer $i$ has four \emph{timers}. The \emph{left timer} $L_i$ and the \emph{left backup} $L'_i$ threaten its left triple, and the \emph{right timer} $R_i$ and the \emph{right backup} $R'_i$ threaten its right triple. The reader has a left timer $L_\star$ and a right timer $R_\star$. We use the same letters for the columns of these piles. Each timer is named for the triple it threatens. From left to right, a writer's timers occur in the order $R'_i<L'_i<R_i<L_i$. The timer $L_i$ thus lies to the right of $R_i$ and holds more stones, since its sow must reach the left triple. Both guards of a genuine writer hold the same number of stones, so a clearing at $L_i$ ends farther up the line than a clearing at $R_i$, by $L_i-R_i$ columns, and the records of $x_i$ fit between the two. The reader's timers are adjacent, with $R_\star=L_\star+1$.
\item \emph{Clocks} $c_0,\ldots,c_{J-1}$ lie at the right end of the line and land at a common column $g$ in the intended play. A clock firing adds one stone to every timer, so it moves every landing column one step closer to its band.
\item Each universal writer $i$ has two \emph{selectors} $S^R_i$ and $S^L_i$, with which Bob chooses the bit of that writer.
\end{itemize}
Alice owns the following special piles. All her other pits hold so many stones that they are saturated.
\begin{itemize}
\item Each writer $i$ has two \emph{guards}, Alice's pits opposite $R_i$ and $L_i$, and the reader has two guards, Alice's pits opposite $L_\star$ and $R_\star$. A guard blocks Alice's own capture of the opposite timer (\Cref{lem:moves}(\labelcref{mv:capture})), so she must first \emph{clear} the guard by selecting it.
\item Each writer $i$ has two \emph{weapons} $w^R_i$ and $w^L_i$, piles whose count is exactly the distance to $R_i$ or $L_i$ at the intended time. Firing a weapon after clearing the guard captures the timer. A \emph{shot} is this calibrated selection, whose last stone lands at its designated timer. An arbitrary selection of a weapon need not be a shot. At a universal writer, the selectors $S^R_i$ and $S^L_i$ can capture $w^R_i$ and $w^L_i$ respectively.
\item Each occurrence of a variable in a clause has a \emph{record} pile. Clearing a guard adds a stone to a fixed interval of records, and this is how writers store bits. There are also \emph{dummy records} $d_\eta$, one for each odd $\eta$ with $1\le\eta\le2C-3$, which let the reader accept when the number $\nu$ of counter bits equal to one is odd (\Cref{def:assignment-game}).
\end{itemize}

\paragraph{A choice and its record.} Consider a genuine writer $i$ in the intended play, with Bob replying by clock firings. Alice first clears one guard, then fires the corresponding weapon, and then saves the six targets. The records of $x_i$ lie strictly between the two possible landing columns of the clearing. Clearing the guard at $R_i$ stops below these records, whereas clearing the guard at $L_i$ crosses them and adds one stone to each. Either shot stops at its timer, below the records. Each of the six saves adds one stone to every record. Thus the choice has the following effect on each record of $x_i$.
\begin{center}
\begin{tabular}{@{}clc@{}}
\toprule
Bit & Preparation & Increase in the record count\\
\midrule
$0$ & clear at $R_i$, shoot from $w^R_i$ & $6$\\
$1$ & clear at $L_i$, shoot from $w^L_i$ & $7$\\
\bottomrule
\end{tabular}
\end{center}
The two choices leave the record counts one stone apart, so selecting a record later lands one column farther to the right in the second case. The reader will test for an exact landing. The coordinates of \Cref{app:coordinates} place the records between the clearing endpoints and calibrate their initial counts for that test. \Cref{lem:standard} proves the resulting counts through all writers.

\subsection{Layout}
\label{sec:layout}

\Cref{fig:ov-stages}(a) shows the order of the line. The exact columns and counts are given in \Cref{app:coordinates}, together with a table of notation (\Cref{tab:notation}), the routine calculations about them, and a lemma that lists which moves cross which piles. Here we describe what the proofs use. Three parameters are linear in $n$ and in the largest number $\rho$ of occurrences of a variable. The number $H=16n+32$ bounds the number of Bob moves and every leftover Bob pile until every target is saved, $W=3H+9$ is the height $b_i-a_i$ of a band and the gap between consecutive bands, and $\gamma=H+\rho+8$ is the distance $L_i-R_i$ between the two main timers of a writer. The width $D$ of the line is $O(n^2+n\rho)$.

The bands come first. The reader's band, with tops $a_\star<b_\star$, is the lowest, followed by the bands of writers $n-1,\ldots,0$, where writer $i$ has tops $a_i<b_i$. Then come the weapons $w^R_i<w^L_i$ of all writers, the common clock landing column $g$, and the timer blocks $R'_i<L'_i<R_i<L_i$ of writers $n-1,\ldots,0$. Above them lie the records and dummy records, the reader timers $L_\star$ and $R_\star=L_\star+1$, the selectors, and the $J=8n+7-n_\forall$ clocks $c_{J-1}<\cdots<c_0$. The targets and every Alice pit without a role hold $D$ stones, so they are saturated.

The counts of Bob's structural piles fix their initial landing columns. These are $a_i+8i+2$ and $b_i+8i+2$ for the writer timers $L_i$ and $R_i$, $a_i+8i+5$ and $b_i+8i+5$ for the backups, $a_\star+8n+2$ and $b_\star+8n+1$ for the reader timers, $w^R_i+8i-1$ and $w^L_i+8i-1$ for the selectors, and $g+j$ for the clock $c_j$. In the intended play, every Bob move before writer $i$ starts is a clock or a selector, there are $8i$ such moves, and each of them adds one stone to every timer of writer $i$, so writer $i$ starts with its timers landing at $a_i+2$, $b_i+2$, $a_i+5$ and $b_i+5$. Each weapon of writer $i$ then holds exactly its distance to its timer. The counts of Alice's guards determine where her clearings land, and the initial counts of the records are calibrated for the reader's exact-count test.

\paragraph{Layout facts.} The proofs also use the following facts, which \Cref{lem:layout} checks by direct calculation. Every count is an integer between $0$ and $D$, every pile with a role is nonempty, and every Bob pile lands at a column of at least $1$. The position can be computed in time polynomial in the size of the formula. Consecutive band tops differ by at least $W$. Weapons of different writers are more than $2H$ apart, and every weapon lies more than $4H$ above $b_0$ and more than $4H$ below $g$. Every timer lies more than $H$ above $g+J$, and every record and dummy record lies more than $H$ below $L_\star$. In particular the interval $(g+J,R_\star]$ contains every timer column, and so every guard, and every record and dummy record.

\subsection{Intended play}
\label{sec:records}

At writer $i$ the intended play has eight Alice moves.

\begin{definition}[Intended branch]
\label{def:standard-branch}
The \emph{intended branch} of writer $i$ with bit $0$ consists of clearing the guard opposite $R_i$, shooting $w^R_i$ at $R_i$, and saving the six targets of the band. The intended branch with bit $1$ uses $L_i$ and $w^L_i$ instead. The \emph{intended order} of the saves is $a_i,a_i-1,a_i-2,b_i,b_i-1,b_i-2$ for bit $0$ and $b_i,b_i-1,b_i-2,a_i,a_i-1,a_i-2$ for bit $1$.
\end{definition}

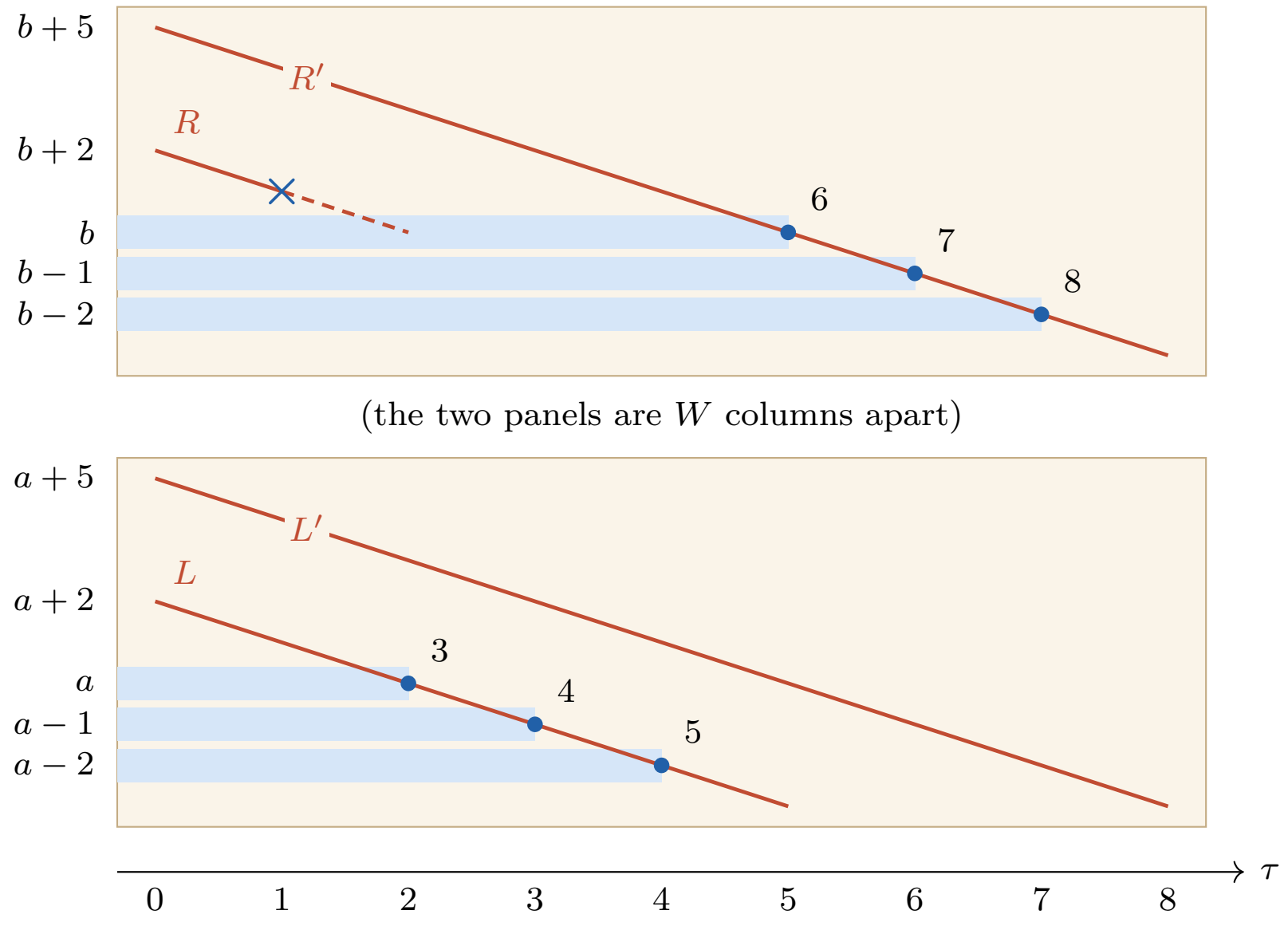
\begin{figure}[t]
\centering
\begin{tikzpicture}[x=1.05cm,y=0.34cm,
  lab/.style={font=\scriptsize},
  tim/.style={line width=0.9pt,draw=ovbobc},
  sav/.style={circle,fill=ovalice,inner sep=1.3pt}]
\foreach \base/\t in {0/a,11/b}{
  \draw[fill=ovboardfill!60,draw=ovboardline] (-0.3,\base-3.5) rectangle (8.3,\base+5.5);
  \foreach \d/\s in {0/,-1/-1,-2/-2,2/+2,5/+5}
    \node[lab,anchor=east] at (-0.35,\base+\d) {$\t\s$};
}
\foreach \y/\h in {0/2,-1/3,-2/4,11/5,10/6,9/7}
  \fill[ovalicefill] (-0.3,\y-0.4) rectangle (\h,\y+0.4);
\draw[tim] (0,2) node[lab,above right,text=ovbobc]{$L$} -- (5,-3);
\draw[tim] (0,5) -- node[lab,pos=0.15,fill=ovboardfill!60,inner sep=1pt,text=ovbobc]{$L'$} (8,-3);
\draw[tim] (0,13) node[lab,above right,text=ovbobc]{$R$} -- (1,12);
\draw[tim,densely dashed] (1,12) -- (2,11);
\node[text=ovalice] at (1,12) {$\boldsymbol\times$};
\draw[tim] (0,16) -- node[lab,pos=0.15,fill=ovboardfill!60,inner sep=1pt,text=ovbobc]{$R'$} (8,8);
\foreach \x/\y/\k in {2/0/3,3/-1/4,4/-2/5,5/11/6,6/10/7,7/9/8}
  \node[sav,label={[lab]above right:\k}] at (\x,\y) {};
\node[lab] at (4,6.5) {(the two panels are $W$ columns apart)};
\draw[->] (-0.3,-4.6) -- (8.6,-4.6) node[lab,right]{$\tau$};
\foreach \h in {0,...,8} \node[lab,below] at (\h,-4.6) {$\h$};
\end{tikzpicture}
\caption{Writer $i$ with bit $0$ in the intended play, with $a=a_i$ and $b=b_i$. Columns increase upward, so the right triple is in the top panel. The horizontal axis $\tau$ counts Bob's moves of the writer, each of which lowers every timer's landing column by one. A blue bar is a target while it is pending, and the red lines are Bob's timers. Alice clears the guard at $R$ when $\tau=0$ and captures $R$ when $\tau=1$ (the cross). The dashed segment shows where $R$ would have reached $b$. Each dot is a save, labelled with Alice's move number, made exactly when a timer reaches that target. The left backup $L'$ only crosses targets that are already saved.}
\label{fig:writer-timeline}
\end{figure}

\Cref{tab:writer} shows the intended orders, and \Cref{fig:writer-timeline} shows why the timing is tight. After Alice captures $R_i$, the timer $L_i$ makes $a_i$, $a_i-1$ and $a_i-2$ due on her moves $3$, $4$ and $5$, and the backup $R'_i$ makes $b_i$, $b_i-1$ and $b_i-2$ due on her moves $6$, $7$ and $8$. Without the capture, $L_i$ and $R_i$ would make $a_i$ and $b_i$ due on the same move. Capturing $L_i$ instead exchanges the roles of the two triples, with the backup $L'_i$.

\begin{table}[ht]
\centering
\begin{tabular}{c ll}
\toprule
Alice's move & Bit $0$ & Bit $1$\\
\midrule
1 & Clear the guard at $R_i$ & Clear the guard at $L_i$\\
2 & Shoot $w^R_i$ at $R_i$ & Shoot $w^L_i$ at $L_i$\\
3,4,5 & Save $a_i,a_i-1,a_i-2$ & Save $b_i,b_i-1,b_i-2$\\
6,7,8 & Save $b_i,b_i-1,b_i-2$ & Save $a_i,a_i-1,a_i-2$\\
\bottomrule
\end{tabular}
\caption{The two intended branches of writer $i$ in the intended order. Bob moves once between consecutive Alice moves. Clearing the guard writes the bit into the records, and the shot removes one deadline.}
\label{tab:writer}
\end{table}

The next lemma collects what Alice's piles look like after she has played intended branches. Its proof follows every pile through the intended branches with the exact counts of \Cref{app:coordinates}, and is given there. Its second hypothesis holds in both directions of the proof. In the intended play Bob captures only weapons, and in \Cref{sec:true} his captures lie near $g$, the weapons or the bands, or above $R_\star$. All of these lie outside the interval $(g+J,R_\star]$, which contains every guard, record and dummy record (\Cref{sec:layout}).

\begin{lemma}[Intended play]
\label{lem:standard}
Consider a time at which Alice's moves so far are, writer by writer in the order $0,1,\ldots$, the moves of intended branches, each beginning with its clearing and its shot and followed by its six saves in any order, where the last branch may be incomplete. Suppose also that by then Bob has captured at no column in $(g+J,R_\star]$.
\begin{enumerate}
\item\label{sp:clear} \textbf{Clearings.} Every clearing of a writer's guard so far landed in a pit of $(g+J,R_\star]$ that Alice had never selected and that held at least $D$ stones before the move, so it captured nothing.
\item\label{sp:counts} \textbf{Weapons and guards.} Suppose Alice has played exactly the branches of writers $0,\ldots,i-1$, all complete, where $0\le i\le n$. If $i<n$, an uncaptured $w^R_i$ lands exactly at $R_i$, an uncaptured $w^L_i$ lands exactly at $L_i$, and both guards of writer $i$ are nonempty. The reader's guards hold at least $D$ stones.
\item\label{sp:timers} \textbf{Piles that reach a timer.} Under the condition of item~\labelcref{sp:counts} with $i<n$, no Alice pile at a column below $R_i$ other than $w^R_i$ lands exactly at $R_i$, and $w^R_i$ does if it is uncaptured. The same holds for $L_i$ and $w^L_i$. No Alice pile lands at $R_i-1$ or at $L_i-1$.
\item\label{sp:records} \textbf{Records.} Suppose all $n$ branches are complete, let $\epsilon_i$ be the bit of the branch of writer $i$, and put $\nu=\epsilon_m+\cdots+\epsilon_{n-1}$. Consider an occurrence of $x_i$ in a clause $\varphi_{j'}$, and put $\delta=1$ if the occurrence is the literal $x_i$ and $\delta=0$ if it is $\neg x_i$. Its record lands at $R_\star+\nu-2j'-\delta+\epsilon_i$. The dummy record $d_\eta$ lands at $R_\star+\nu-\eta$. Consequently some record or dummy record lands exactly at $R_\star$ if and only if $\nu$ is odd, or $\nu=2j$ and the clause $\varphi_j$ is true under $x_i=\epsilon_i$ $(i<m)$.
\item\label{sp:inputs} \textbf{Inputs.} If all $n$ branches are complete, every Alice pile at a column below $R_\star$ that lands exactly at $R_\star$ is a record or a dummy record.
\end{enumerate}
\end{lemma}

\section{Deadlines for one stage}
\label{sec:bands}

This section gives the deadline arguments for one stage. Recall that the six targets of a stage form its \emph{band}, a left triple and a right triple. The results are stated for a general line position, so that the same lemmas serve the writers, the reader, and the situations where Bob deviates from the intended play. Each lemma gives Alice a \emph{policy}, a rule that tells her which target to save after each Bob move. We state the lemmas here with the ideas behind them, and give the proofs in \Cref{app:deadline-proofs}.

The idea is simple. A triple is safe as long as every Bob pile that can move lands above its highest pending target, and the margin by which it does is the \emph{slack} of the triple. A Bob move lowers the slack by at most one, and Alice restores it by saving the highest pending target (\Cref{lem:deadlines}). So Alice can finish one triple first if the other has slack for the moves this takes, and the firing of a left timer covers the right triple and so removes its deadline (\Cref{lem:first}). \Cref{lem:upper-fired} handles a band whose right timer has fired early. The last two subsections combine these policies for a writer and for the reader, which begin by preparing a capture (\Cref{thm:writer,thm:reader-accept}). We fix the positive integer $H$ of \Cref{def:quiet}.

\subsection{The band setting}

A band setting fixes one band, the timers that threaten it, and a set of \emph{omitted} piles that Bob is assumed not to fire. In \Cref{sec:true} the omitted piles are the timers of later stages, and a firing of one of them is handled there as a separate event. The standing assumptions below say that every residue pile is small, that all other structural piles lie and land far above the band, and that Alice answers every Bob move that is not quiet with a move of her policy.

\begin{definition}[Band setting]
\label{def:band-setting}
A \emph{band setting} consists of a play of Line Kalah and a time $t_0$ in it when Bob is to move, together with the following data.
\begin{itemize}
\item Integers $a\ge4$ and $b$ with $b-a>2H+8$, such that the six columns of $\mathcal A=\{a-2,a-1,a\}$ and $\mathcal B=\{b-2,b-1,b\}$ are targets. These are the \emph{left} and the \emph{right triple}. Every target of the band that is pending at $t_0$ is saturated.
\item A set $T$ of unfired structural piles, called the \emph{timers} of the band.
\item A set $\Omega$ of unfired structural piles disjoint from $T$, called the \emph{omitted} piles.
\end{itemize}
A timer that fires or is captured leaves $T$, as for $\Omega$. The \emph{standing assumptions} are the following.
\begin{enumerate}
\item[(S1)] Every residue pile holds at most $H$ stones.
\item[(S2)] No omitted pile fires.
\item[(S3)] Every unfired structural pile lies at a column greater than $b+2H+10$, and those outside $T\cup\Omega$ have landing columns greater than $b+H+10$.
\item[(S4)] Call a Bob move \emph{quiet} if it is $(b+H+6)$-quiet in the sense of \Cref{def:quiet}. After a quiet Bob move, Alice either follows her policy or makes a move that changes no Bob pile and saves no target of the band. After every other Bob move she follows her policy.
\end{enumerate}
A policy \emph{saves the band} if, whenever the standing assumptions hold from $t_0$ up to some time, Bob captures no pending target of $\mathcal A\cup\mathcal B$ up to that time, and each move that Alice makes by the policy saves a different pending target of the band, except for any guard clearing and shot explicitly prescribed as preparation.
\end{definition}

The proofs of these results argue one move at a time, so they show that no pending target is captured as long as the standing assumptions hold. This is what \Cref{sec:true} needs, because there a stage may be abandoned, after which (S2) can fail.

The \emph{highest pending target} of a triple is the largest column of the triple that is still pending. If none is pending, we take it to be $a-3$ for $\mathcal A$ and $b-3$ for $\mathcal B$. We write $\hat a$ and $\hat b$ for the highest pending targets of $\mathcal A$ and $\mathcal B$. Every policy below starts with the following rule.

\begin{itemize}
\item[] \textbf{Cover rule.} If a target of the band is exposed, save it.
\end{itemize}

\begin{definition}[Slack]
\label{def:slack}
Put $F_{\mathrm{left}}=\min\{E_\Omega,a+4\}$, and for $\Lambda\subseteq T$ put $F_{\mathrm{right}}(\Lambda)=\min\{E_{\Lambda\cup\Omega},b+4\}$. The \emph{slack} of the left triple is $F_{\mathrm{left}}-\hat a$, and the \emph{slack of the right triple outside $\Lambda$} is $F_{\mathrm{right}}(\Lambda)-\hat b$.
\end{definition}

The right slack is measured outside $\Lambda$ because, in the construction, the piles of $\Lambda$ are timers of the left triple, and when they fire they cover the right triple instead of threatening it. The clipping at $a+4$ and $b+4$ only discards margin that is never used.

\paragraph{Single moves near a band.} \Cref{lem:band-facts} in \Cref{app:deadline-proofs} collects what one move can do to a band. A quiet move changes nothing that the band depends on. A move that empties a pit opposite a right target cannot affect the left triple. A Bob move lowers a slack by at most one, and while a slack is positive no Bob move can capture a pending target of that triple. A pending target below the frontier is never covered. Finally, the firing of a pile that lands between the two triples covers every pending right target, and covered targets stay safe under the cover rule.

\subsection{Deadlines}

The next lemma is the basic deadline argument for one triple. Saving the highest pending target after every Bob move that is not quiet keeps the slack of the triple at least one until the triple is saved. For the right triple the lemma allows timers of a set $\Lambda$ to remain, because they threaten the left triple, and their firing covers the right triple instead of threatening it.

\begin{lemma}[Deadlines]
\label{lem:deadlines}
In a band setting, consider a Bob turn $t$, and suppose that from $t$ on Alice applies the cover rule and answers quiet Bob moves only with moves that change no Bob pile.
\begin{enumerate}
\item\label{dl:lower} Suppose the left triple has slack at least one at $t$, and from $t$ on, after every Bob move that is not quiet and does not expose a target of the band, Alice saves the highest pending left target while one remains. Then no pending left target is captured, covered or exposed, and while a left target is pending the left triple has slack at least one at every Bob turn.
\item\label{dl:upper} Let $\Lambda\subseteq T$ be a set of piles none of which has fired by time $t$. Suppose the right triple has slack at least one outside $\Lambda$ at $t$, and from $t$ until a pile of $\Lambda$ fires, after every Bob move that is not quiet and does not expose a target of the band, Alice saves the highest pending right target while one remains. Then no pending right target is captured before a pile of $\Lambda$ fires. Moreover, while no pile of $\Lambda$ has fired and a right target is pending, no pending band target is covered or exposed, and the right triple has slack at least one outside $\Lambda$ at every Bob turn.
\end{enumerate}
\end{lemma}
\subsection{One triple first}

In the next lemma, $\Lambda\subseteq T$ is a set of timers each of which lands at most at $a+H+4$ at time $t_0$. In the construction, $\Lambda$ consists of the unfired timers that threaten the left triple. By \Cref{lem:moves}(\labelcref{mv:frontier}), a pile of $\Lambda$ still lands at most at $a+H+4<b-2$ when it fires, and it lies above $b$ by (S3), so by \Cref{lem:band-facts}(\labelcref{bf:lower}) its firing captures no right target and covers every pending right target. We use two policies.
\begin{itemize}
\item The \emph{right-first} policy applies the cover rule, and otherwise saves the highest pending right target, or the highest pending left target if no right target is pending.
\item The \emph{left-first} policy applies the cover rule, and otherwise saves the highest pending left target, or the highest pending right target if no left target is pending.
\end{itemize}

In words, Alice may postpone one triple while its slack exceeds the number of saves that the other triple still needs.

\begin{lemma}[One triple first]
\label{lem:first}
In a band setting, let $k_{\mathcal A}$ and $k_{\mathcal B}$ be the numbers of pending left and right targets at time $t_0$.
\begin{enumerate}
\item\label{first:upper} If at $t_0$ the left triple has slack at least $k_{\mathcal B}+1$, and either every pending right target is covered or no pile of $\Lambda$ has fired and the right triple has slack at least one outside $\Lambda$, then the right-first policy saves the band.
\item\label{first:lower} If at $t_0$ the left triple has slack at least one, and either every pending right target is covered or no pile of $\Lambda$ has fired and the right triple has slack at least $k_{\mathcal A}+1$ outside $\Lambda$, then the left-first policy saves the band.
\end{enumerate}
\end{lemma}
\subsection{When a right timer has fired}

The last policy handles the case where Bob fires a pile that threatens the right triple at a moment when it lands above that triple. Its residue then threatens the right triple on a new schedule, while the left timers keep their own schedule. Alice answers each Bob move according to which of the two schedules it can advance. A move from above the left timers can add stones to them, and any other move cannot.

\begin{lemma}[After a right timer fires]
\label{lem:upper-fired}
In a band setting, let $T=\Lambda\cup T_{\mathrm{right}}$, where $\Lambda$ is a nonempty set of timers that land at most at $a+H+4$, as in \Cref{lem:first}, and $T_{\mathrm{right}}$ is either empty or consists of a single timer $R'$ at a column below every pile of $\Lambda$. Let $y_0$ be the lowest column of a pile of $\Lambda$. Suppose that at time $t_0$ both triples have pending targets, no band target is covered, the left triple has slack at least one, the right triple has slack at least one outside $T$, and, if $T_{\mathrm{right}}=\{R'\}$, $e(R')\ge\hat b+2+k$, where $k$ is the number of pending left targets. Then the following policy saves the band.
\begin{itemize}
\item Apply the cover rule.
\item Once a pile of $\Lambda$ has fired, follow the left-first policy.
\item Before that, after a Bob move from a column above $y_0$, save the highest pending left target, or the highest pending right target if no left target is pending. After any other Bob move, including a pass, save the highest pending right target, or the highest pending left target if no right target is pending.
\end{itemize}
\end{lemma}

\subsection{Prepared writers}
\label{sec:gadgets}

A writer and the reader each begin with a \emph{preparation}, in which Alice clears the guard of a timer and on her next move shoots it. This subsection and the next show that, under hypotheses we verify later for the construction, a prepared writer or reader saves its band against every Bob move. The proofs, in \Cref{app:deadline-proofs}, reduce every case to one of the policies above. Here a policy \emph{saves the band} in the sense of \Cref{def:band-setting}, except that the time $t_0$ is now a time when Alice is to move. Its prescribed clearing and, when needed, shot are preparation moves. The distinct-save requirement applies to the continuation after them.

\begin{definition}[Writer setting]
\label{def:writer-setting}
A \emph{writer setting} consists of a play, a time $t_0$ when Alice is to move, integers $a,b$, a set $T=\{L,R,L',R'\}$ of unfired structural piles, and a set $\Omega$ of unfired structural piles disjoint from $T$, such that the conditions of \Cref{def:band-setting} on $a$, $b$, $T$ and $\Omega$ hold, together with the following conditions.
\begin{enumerate}
\item[(W1)] $R'<L'<\min\{L,R\}$.
\item[(W2)] $a+2\le e(L)\le a+H+1$, $b+2\le e(R)\le b+H+1$, $a+5\le e(L')\le a+H+4$ and $b+5\le e(R')\le b+H+4$.
\item[(W3)] Every Bob pile outside $T\cup\Omega$ has landing column greater than $b+2H+10$.
\item[(W4)] All six targets of the band are pending, and none is covered.
\end{enumerate}
A \emph{prepared choice} is a timer $V\in\{L,R\}$ and a column $w<V$ with $w>b+2H+10$, such that $A_V>0$ and $A_w=V-w$, clearing the guard at $V$ captures nothing, and Bob's reply to that clearing either fires $V$ or leaves Alice's pile at $w$ in place.
\end{definition}

Clearing the guard at $V$ adds stones only above $V$, so it does not change Alice's pile at $w$. The pile at $w$ is the \emph{weapon} of the prepared choice.

The idea is as follows. After the clearing, Bob's reply either fires the timer $V$ whose capture Alice is preparing or leaves her weapon in place. In the first case the firing itself removes a deadline. A fired left timer lands between the triples and covers the right triple, and a fired right timer lands above the band, so that \Cref{lem:upper-fired} applies. In the second case the shot captures $V$. Either way one of the two timers no longer threatens the band, the deadlines leave enough slack for the remaining saves, and one of the policies above saves all six targets.

\begin{theorem}[Writer]
\label{thm:writer}
In a writer setting with a prepared choice $(V,w)$, the following policy saves the band. Alice clears the guard at $V$.
\begin{itemize}
\item If Bob's reply fires $V$, Alice saves $a$, and then follows the left-first policy of \Cref{lem:first} if $V=L$, and the policy of \Cref{lem:upper-fired} with $\Lambda=\{L,L'\}$ and $T_{\mathrm{right}}=\{R'\}$ if $V=R$.
\item Otherwise Alice shoots $w$ at $V$, which captures $V$, and then follows the right-first policy if $V=L$ and the left-first policy if $V=R$. In both cases $\Lambda$ is the set of unfired piles among $L$ and $L'$.
\end{itemize}
In the continuation, quiet Bob moves are answered as in (S4). The local policy uses at most eight Alice moves, and at most seven if Bob fires $V$. These bounds exclude the interruptions allowed by (S4), which save no target of this band and are charged separately.
\end{theorem}
\subsection{The prepared reader}

\begin{definition}[Reader setting]
\label{def:reader-setting}
A \emph{reader setting} consists of a play, a time $t_0$ when Alice is to move, integers $a,b$, a set $T=\{L,R\}$ of unfired structural piles, and a set $\Omega$ of unfired structural piles disjoint from $T$, such that the conditions of \Cref{def:band-setting} on $a$, $b$, $T$ and $\Omega$ hold, together with the following conditions.
\begin{enumerate}
\item[(R1)] $R=L+1$.
\item[(R2)] $a+2\le e(L)\le a+H+1$ and $b+1\le e(R)\le b+H+1$.
\item[(R3)] Every Bob pile outside $T\cup\Omega$ has landing column greater than $b+2H+10$.
\item[(R4)] All six targets of the band are pending, and none is covered.
\end{enumerate}
An \emph{input} is a column $x<R$ with $A_x=R-x$ such that $A_R>0$, clearing the guard at $R$ captures nothing, and Bob's reply to that clearing either fires $R$ or leaves Alice's pile at $x$ in place.
\end{definition}

The reader works in the same way, with an input in the role of the weapon. If Bob fires $R$ in reply to the clearing, it lands above the band, and its first stone falls on $L$, which lies directly below it. Alice then saves the top target and continues with the policy of \Cref{lem:upper-fired}. Otherwise the shot from the input captures $R$, and the left-first policy saves the band.

\begin{theorem}[Reader with an input]
\label{thm:reader-accept}
In a reader setting with an input $x$, the following policy saves the band. Alice clears the guard at $R$.
\begin{itemize}
\item If Bob's reply fires $R$, Alice saves $b$ and then follows the policy of \Cref{lem:upper-fired} with $\Lambda=\{L\}$ and $T_{\mathrm{right}}$ empty.
\item Otherwise Alice shoots $x$ at $R$, which captures $R$, and then follows the left-first policy of \Cref{lem:first} with $\Lambda$ the set of unfired piles among $\{L\}$.
\end{itemize}
In the continuation, quiet Bob moves are answered as in (S4). The local policy uses at most eight Alice moves, excluding the interruptions allowed by (S4), which save no target of this band and are charged separately.
\end{theorem}

\section{Bob's strategy for false formulas}
\label{sec:false}

In this section the formula is false, and we show that Bob wins the line position. Bob plays the intended moves, and the deadlines then force Alice to play the intended computation, and the reader rejects.

\subsection{The strategy}

Throughout this section, $H$, the coordinates and the counts are those of \Cref{sec:construction} and \Cref{app:coordinates}. We read a bit from Alice's play. Writer $i$ has bit $\epsilon_i=0$ if Alice captured $R_i$ and $\epsilon_i=1$ if she captured $L_i$.

\begin{definition}[Intended strategy]
\label{def:canonical}
Fix a strategy $\beta$ for Bob in the assignment game. Bob's \emph{intended strategy} chooses his move as follows.
\begin{enumerate}
\item If a pile at the column of a writer timer or a reader timer can capture a pending target on this move, select it.
\item Otherwise, if this is Bob's move number $8i$ for a universal writer $i$, let $\epsilon_i$ be the bit that $\beta$ chooses given the bits read so far, and select $S^R_i$ if $\epsilon_i=1$ and $S^L_i$ if $\epsilon_i=0$. (If that selector has already been captured, use rule~3 instead.)
\item Otherwise select the highest unfired clock.
\end{enumerate}
\end{definition}

The selector $S^R_i$ captures the weapon $w^R_i$, which leaves Alice only the branch with bit $1$, and $S^L_i$ leaves only bit $0$.

\subsection{Calibration}

We call a play \emph{intended up to writer $i$} if Bob has followed the intended strategy, and Alice has played, at each writer $i'<i$ in turn, the eight moves of an intended branch in the intended order (\Cref{def:standard-branch}). \emph{Intended up to the reader} means intended up to writer $n$.

\begin{lemma}[Calibration]
\label{lem:calibration}
In a play that is intended up to writer $i$, at the start of writer $i$ (or of the reader, if $i=n$), the following hold.
\begin{enumerate}
\item\label{cal:moves} Bob has made $8i$ moves, of which one at each universal writer $i'\le i$ was a selector and all others were clocks. Every clock landed at $g$. Alice has selected no pit other than guards, weapons and targets of the writers before $i$. Bob's only captures are, for each universal writer $i'\le i$, the weapon $w^R_{i'}$ or $w^L_{i'}$ captured by the selector on his move $8i'$, according to rule~2. In particular, the hypotheses of \Cref{lem:standard} hold.
\item\label{cal:clocks} At least $7(n-i)+8$ clocks are unfired if $i<n$, and at least $7$ if $i=n$. Consequently, however Alice plays from then on, some clock is unfired before each of Bob's eight moves of writer $i$ if $i<n$, and before each of his first five moves of the reader if $i=n$.
\item\label{cal:timers} Every timer of a writer $i'$ with $i\le i'<n$, and both reader timers, are unfired, and their landing columns are their initial values minus $8i$. If $i<n$, in particular, $e(L_i)=a_i+2$, $e(R_i)=b_i+2$, $e(L'_i)=a_i+5$, $e(R'_i)=b_i+5$.
\item\label{cal:bands} Every target of a writer $i'$ with $i\le i'<n$ and of the reader is pending, and every Bob pit at a target column is empty.
\item\label{cal:records} If $i\ge1$, every record and dummy record is covered.
\end{enumerate}
\end{lemma}

The lemma follows by induction on $i$, and its proof, like the other proofs of this section, is in \Cref{app:false-proofs}.

\subsection{Alice's forced moves}

At a writer, a target becomes urgent when a timer is about to land on it. We say that a pending target is \emph{due} on Alice's $\tau$-th move of writer $i$ if, after Bob's $(\tau-1)$-th move of the writer, an uncaptured timer of writer $i$ lands on it. \Cref{tab:due} lists these dates when Bob follows the intended strategy.

\begin{table}[ht]
\centering
\begin{tabular}{l cccccc}
\toprule
Alice's move & $3$ & $4$ & $5$ & $6$ & $7$ & $8$\\
\midrule
due from $L_i$   & $a$ & $a-1$ & $a-2$ & & & \\
due from $R_i$   & $b$ & $b-1$ & $b-2$ & & & \\
due from $L'_i$ & & & & $a$ & $a-1$ & $a-2$\\
due from $R'_i$ & & & & $b$ & $b-1$ & $b-2$\\
\bottomrule
\end{tabular}
\caption{The targets of writer $i$ that each of its timers makes due, by Alice's move number, when every Bob move of the writer fires a clock or a selector ($a=a_i$, $b=b_i$). A target is due only while it is pending and the timer is uncaptured. Nothing is due on Alice's first two moves.}
\label{tab:due}
\end{table}

\begin{lemma}[Forced writer]
\label{lem:forced-writer}
Suppose a play is intended up to writer $i$, and Bob keeps playing the intended strategy. If Bob captures no pending target during writer $i$, then Alice's eight moves at writer $i$ are an intended branch in the intended order whose weapon was not captured. In particular, at a universal writer her branch is the one that Bob's selector left.
\end{lemma}

The proof counts Alice's moves. Capturing $L_i$ or $R_i$ takes two of her moves, a clearing and then a shot, with no save in between, because a save adds a stone to the guard. A capture on her fourth move or later comes too late, because until then two targets are due on each of her moves $3$, $4$ and $5$. A capture on her third move is impossible, because the target that is due on that move must be saved before the clearing, and that save adds a stone to every weapon. So she clears a guard on her first move and shoots on her second, and after that the six due dates leave her no spare move.

\begin{lemma}[Reader rejection]
\label{lem:reader-reject}
Suppose the play is intended up to the reader, and Bob keeps playing the intended strategy. If at the start of the reader no Alice pile at a column below $R_\star$ lands exactly at $R_\star$, then Bob captures a pending target during the reader.
\end{lemma}

At the reader the same counting forces Alice to clear the guard at $R_\star$ on her first move and to capture $R_\star$ on her second, which needs a pile that already landed exactly at $R_\star$.

\begin{theorem}[False formulas]
\label{thm:false}
If the formula is false, Bob wins the line position of the construction.
\end{theorem}
\begin{proof}
By \Cref{lem:assignment-game}, Bob has a winning strategy $\beta$ in the assignment game, and he plays the intended strategy with $\beta$. If Bob captures a pending target during some writer, he wins. Otherwise, by \Cref{lem:forced-writer} and induction, the play is intended up to the reader. At each universal writer Alice's bit is the one chosen by $\beta$, and at each existential writer it is Alice's choice, so the bits are a play of the assignment game against $\beta$, which Alice loses. By \Cref{lem:calibration}, the hypotheses of \Cref{lem:standard} hold at the start of the reader, so by \Cref{lem:standard}(\labelcref{sp:records}) no record or dummy record lands at $R_\star$, and by \Cref{lem:standard}(\labelcref{sp:inputs}) no Alice pile below $R_\star$ lands at $R_\star$. \Cref{lem:reader-reject} shows that Bob captures a pending target.
\end{proof}

\section{Alice's strategy for true formulas}
\label{sec:true}

In this section the formula is true, and we give Alice a strategy that defeats every Bob strategy, even if Bob may also pass on any turn when his row is nonempty. Passes only give Bob more options, and the reduction to Kalah in \Cref{sec:empty-stores} needs them. As in \Cref{sec:quiet}, a pass counts as a quiet Bob move, so every count of Bob moves below includes his passes. Throughout, $H$, the coordinates and the counts are those of \Cref{sec:construction} and \Cref{app:coordinates}, and $\alpha$ is a winning strategy for Alice in the assignment game (\Cref{lem:assignment-game}).

The plan is as follows. Alice works on one stage at a time. Each stage is started by an \emph{entry}, after which Alice follows one of the policies of \Cref{sec:bands}. \Cref{lem:entries} shows that each entry puts her into the hypotheses of the policy she then follows, and \Cref{lem:active} shows that the standing assumptions of that policy hold while the stage is active. Two further lemmas collect the counting facts that these proofs use.

\subsection{Stages}

We number the stages $0,\ldots,n$, where stage $s<n$ is writer $s$ and stage $n$ is the reader. Stage $s$ has the band with tops $a_s,b_s$, where $a_n=a_\star$ and $b_n=b_\star$, and its \emph{timers} $T_s$ are $L_s,R_s,L'_s,R'_s$ for a writer and $L_\star,R_\star$ for the reader. Bands of later stages lie lower on the line (\Cref{lem:layout}). For a stage $s$ let $\Omega_s$ be the set of unfired timers of all stages after $s$.

\subsection{The strategy}

\begin{definition}[Alice's strategy]
\label{def:alice-strategy}
Alice keeps an \emph{active stage} $s$, initially $0$, and a \emph{mode}, initially \emph{computing}. On her first move she performs the computing-mode entry into writer $0$ described below. She switches permanently to the \emph{escaping} mode at the first Bob move that is a pass, a residue move, the firing of a timer of a stage after the active one (a \emph{jump}), or the firing of the timer $V$ immediately after Alice cleared the guard at $V$. On every subsequent turn she applies the first of the following rules that applies.
\begin{enumerate}
\item[(J)] \emph{Jump.} If Bob's last move fired a timer of a stage $s'>s$, she sets $s=s'$ and plays the \emph{jump entry} of stage $s'$ (below).
\item[(C)] \emph{Cover rule.} If a target is exposed, she saves it.
\item[(P)] \emph{Stage policy.} If every target of stage $s$ is saved and $s<n$, she sets $s=s+1$ and plays the \emph{entry} of the new stage. If every target of stage $n$ is saved, she saves any pending target, or passes if there is none. Otherwise she makes the next move of the policy that she is following at stage $s$.
\end{enumerate}
The entries and the policies they start are as follows.
\begin{itemize}
\item \emph{Writer entry, computing mode.} Let $j$ be the writer. If exactly one of its weapons still holds exactly the distance to its timer, she uses that one. If both do, she uses the bit chosen by $\alpha$ at an existential writer and bit $0$ at a universal writer. Bit $0$ means $V=R_j$ with weapon $w^R_j$, and bit $1$ means $V=L_j$ with weapon $w^L_j$. She follows the policy of \Cref{thm:writer} with this prepared choice.
\item \emph{Writer entry, escaping mode.} If $e(L_j)\ge a_j+3$, she saves $b_j$ and follows the right-first policy of \Cref{lem:first}. Otherwise she saves $a_j$ and follows the left-first policy. In both cases $\Lambda$ is the set of unfired piles among $L_j$ and $L'_j$.
\item \emph{Reader entry.} If $e(L_\star)\ge a_\star+3$, she saves $b_\star$ and follows the right-first policy with $\Lambda$ the set of unfired piles among $\{L_\star\}$. Otherwise she follows the policy of \Cref{thm:reader-accept}, with input the lowest Alice pile below $R_\star$ that lands exactly at $R_\star$.
\item \emph{Jump entry.} If the firing timer is $L_{s'}$, $L'_{s'}$ or $L_\star$, she saves $a_{s'}$ and follows the left-first policy. If it is $R_{s'}$ or $R'_{s'}$, she saves $b_{s'}$ and follows the right-first policy with $\Lambda$ the set of unfired piles among $L_{s'}$ and $L'_{s'}$. If it is $R_\star$, she saves $b_\star$ and follows the policy of \Cref{lem:upper-fired} with $\Lambda=\{L_\star\}$.
\end{itemize}
\end{definition}

In the policies, the band is that of the active stage, $T$ is the set of its unfired timers and $\Omega=\Omega_s$. Every move prescribed by the strategy before all targets are saved is a save, a clearing, or a shot. \Cref{tab:entries} summarizes the entries.

\begin{table}[ht]
\centering
\begin{tabular}{lll l}
\toprule
Entry & First move & Then & Shown in\\
\midrule
Writer, computing mode & clear the chosen guard & \Cref{thm:writer}
  & \Cref{lem:entries}(\labelcref{en:computing})\\
Writer, escaping, $e(L_j)\ge a_j+3$ & save $b_j$ & right first
  & \Cref{lem:entries}(\labelcref{en:escaping})\\
Writer, escaping, otherwise & save $a_j$ & left first
  & \Cref{lem:entries}(\labelcref{en:escaping})\\
Jump by $L$, $L'$ or $L_\star$ & save $a_{s'}$ & left first
  & \Cref{lem:entries}(\labelcref{en:jump})\\
Jump by $R$ or $R'$ & save $b_{s'}$ & right first
  & \Cref{lem:entries}(\labelcref{en:jump})\\
Jump by $R_\star$ & save $b_\star$ & \Cref{lem:upper-fired}
  & \Cref{lem:entries}(\labelcref{en:jump})\\
Reader, $e(L_\star)\ge a_\star+3$ & save $b_\star$ & right first
  & \Cref{lem:entries}(\labelcref{en:reader})\\
Reader, otherwise & clear the guard at $R_\star$ & \Cref{thm:reader-accept}
  & \Cref{lem:entries}(\labelcref{en:reader})\\
\bottomrule
\end{tabular}
\caption{The entries of Alice's strategy.}
\label{tab:entries}
\end{table}

\subsection{The induction}

The next four lemmas are proved together by induction over the moves of the prescribed strategy. Whenever a proof counts earlier saves and preparations or uses an earlier stage policy, it uses only the legal prefix already established by the induction. At a new entry, the bookkeeping and computing-mode facts for that prefix give the hypotheses checked in \Cref{lem:entries}, which also proves the first prescribed move legal. Between entries, \Cref{lem:active} verifies the standing assumptions, so the local policy results ensure legal moves and prevent captures of pending targets. A jump ends the old policy and is checked as a new entry. The initial entry into writer $0$, with no preceding moves, is the base case. Thus these lemmas do not assume that later entries or moves are legal. Their proofs are in \Cref{app:true-proofs}.

\Cref{lem:bookkeeping} bounds the length of the play and shows that the stages after the active one are untouched. \Cref{lem:active} shows that the targets of earlier stages stay protected and that the active stage satisfies the standing assumptions of \Cref{sec:bands}. \Cref{lem:computing} records what the play looks like while Alice is in computing mode, and in particular which weapons Bob can capture. \Cref{lem:entries} is the heart of the argument. At each entry it counts the stones that the timers of the new stage have received, and at the reader it uses the adjacency of the two reader timers.

\begin{lemma}[Bookkeeping]
\label{lem:bookkeeping}
\leavevmode
\begin{enumerate}
\item\label{bk:short} \textbf{Short play.} Until every target is saved, Alice makes at most $8n+8$ moves, Bob makes at most $8n+8\le H-1$ moves, and every residue pile holds at most $H$ stones.
\item\label{bk:untouched} \textbf{Later stages are untouched.} Let $j$ be a stage. A firing of a timer of a stage $j$ or later while the active stage is smaller than $j$ is a jump, so if stage $j$ becomes active by an advance, Bob has fired no timer of a stage $j$ or later before. At every time before every target is saved at which Bob has fired no timer of a stage $j$ or later, every Bob pile other than an unfired timer of a stage $j$ or later has landing column greater than $b_j+2H+10$, every Bob pit at a column at most $b_j+2H+10$ is empty, and no target of a stage $j$ or later has been captured. In particular these targets are uncovered.
\item\label{bk:charging} \textbf{Charging.} Charge each preparation move to its stage and each save to the stage of its target. Before all targets are saved, every Alice move is charged to the active stage or to an earlier one, and each stage receives at most eight charges. Consequently, before stage $j$ becomes active Alice has made at most $8j$ moves, a jump to stage $j$ is Bob's move number $8j$ or earlier, and at Alice's first move at stage $j$ after an advance, Bob has made at most $8j$ moves.
\end{enumerate}
\end{lemma}

\begin{lemma}[The active stage]
\label{lem:active}
Let $s$ be the active stage.
\begin{enumerate}
\item\label{act:earlier} At every time when Alice is to move, every pending target of a stage before $s$ is covered or exposed. A Bob move that empties the pit opposite such a target is quiet in the band setting of stage $s$.
\item\label{act:standing} While $s$ is active and Alice applies its policy, the standing assumptions (S1), (S3) and (S4) of \Cref{def:band-setting} hold for its band, with $T$ the unfired timers of stage $s$ and $\Omega=\Omega_s$, and (S2) holds until Bob fires a pile of $\Omega_s$, which is a jump and ends the activity of $s$.
\end{enumerate}
\end{lemma}

\begin{lemma}[Computing mode]
\label{lem:computing}
Suppose Alice is in computing mode, during writer play or at reader entry before her first reader move. Then the following hold.
\begin{enumerate}
\item\label{cm:history} Every writer before the active stage was entered in computing mode and made a clearing, a shot and six saves, in this order. When a writer $j$ became active, Bob had made exactly $8j$ moves, each firing a clock, a selector or a timer of an earlier writer. Bob has captured only at columns at most $g+J$, so the hypotheses of \Cref{lem:standard} hold. Alice's clearings and saves capture nothing.
\item\label{cm:selector} A weapon of writer $i$ can be captured only by its own selector, and only on Bob's move number $8i$. At most one weapon of each writer is captured, and none at an existential writer.
\item\label{cm:covered} If some Bob move came from a column below the corresponding selector of a weapon of a universal writer, then that weapon is covered from then on.
\end{enumerate}
Moreover, suppose Alice is in computing mode at writer $i$ and has just cleared the guard at $V$. If Bob replies with his first residue move, the reply does not capture Alice's weapon $w$ for $V$.
\end{lemma}

\begin{lemma}[Entries]
\label{lem:entries}
\leavevmode
\begin{enumerate}
\item\label{en:computing} \textbf{Writer, computing mode.} Suppose $j=0$ at the initial entry, or writer $j$ becomes active by an advance while Alice is in computing mode. At her first move there the writer setting of \Cref{def:writer-setting} holds with $T=T_j$ and $\Omega=\Omega_j$, the landing columns of the timers are exactly $a_j+2$, $b_j+2$, $a_j+5$ and $b_j+5$, and the choice of \Cref{def:alice-strategy} is a prepared choice.
\item\label{en:escaping} \textbf{Writer, escaping mode.} If writer $j$ becomes active by an advance while Alice is in escaping mode, then after her entry move the hypotheses of \Cref{lem:first}(\labelcref{first:upper}) hold if $e(L_j)\ge a_j+3$, and those of \Cref{lem:first}(\labelcref{first:lower}) hold otherwise.
\item\label{en:jump} \textbf{Jumps.} If Bob fires a timer $X$ of a stage $s'$ after the active stage, then the firing captures no target, covers every pending target of every stage before $s'$, and exposes nothing. After Alice's jump entry move the hypotheses of the policy she then follows hold.
\item\label{en:reader} \textbf{The reader.} Suppose the reader becomes active by an advance, and put $\Delta_L=e(L_\star)-(a_\star+2)$ and $\Delta_R=e(R_\star)-(b_\star+1)$ at Alice's first move there. Then $\Delta_L\ge\Delta_R\ge0$. If $\Delta_L\ge1$, then after Alice saves $b_\star$ the hypotheses of \Cref{lem:first}(\labelcref{first:upper}) hold with $\Lambda=\{L_\star\}$. If $\Delta_L=0$, then Alice's moves before the reader are, writer by writer, the moves of intended branches, and their bits form a play of the assignment game in which Alice's bits follow $\alpha$. In this case the hypotheses of \Cref{lem:standard} hold, every record and dummy record is covered, and the reader setting of \Cref{def:reader-setting} holds with $\Omega=\emptyset$. Moreover, some Alice pile lands exactly at $R_\star$, and every such pile is an input.
\end{enumerate}
\end{lemma}

\subsection{The theorem}

\begin{theorem}[True formulas]
\label{thm:true}
If the formula is true, Alice wins the line position of the construction. Her strategy of \Cref{def:alice-strategy} prevents every capture of a pending target even if Bob may also pass.
\end{theorem}
\begin{proof}
Alice follows \Cref{def:alice-strategy}. We complete the simultaneous induction described above by showing that Bob never captures a pending target. At each entry, \Cref{lem:entries} makes the next prescribed move legal, and the applicable local policy does the same between entries. Consider a Bob move, and let $s$ be the active stage before it. A pending target of a stage after $s$ is not captured by \Cref{lem:bookkeeping}(\labelcref{bk:untouched}), and if the move is a jump, it captures no target by \Cref{lem:entries}(\labelcref{en:jump}). A pending target of a stage before $s$ is covered or exposed at Alice's turn (\Cref{lem:active}(\labelcref{act:earlier})). If it is exposed, rule~(C) saves it, so at Bob's turn it is covered, and \Cref{lem:moves}(\labelcref{mv:covered}) shows that the move does not capture it.

For the targets of stage $s$ we use the result that Alice is following. When stage $s$ became active, its hypotheses held, by \Cref{lem:entries}(\labelcref{en:computing}) and \Cref{thm:writer} for a writer in computing mode, by \Cref{lem:entries}(\labelcref{en:escaping}) for a writer in escaping mode, by \Cref{lem:entries}(\labelcref{en:reader}) with \Cref{lem:first} or \Cref{thm:reader-accept} for the reader, and by \Cref{lem:entries}(\labelcref{en:jump}) after a jump. Its standing assumptions hold while $s$ is active, up to the move that ends its activity (\Cref{lem:active}(\labelcref{act:standing})). Alice deviates from the band policy only to save an exposed target of an earlier stage, which by \Cref{lem:active}(\labelcref{act:earlier}) happens only after a quiet Bob move, as (S4) allows. Rule~(C) never preempts a preparation, because a writer in computing mode and a reader in the case $\Delta_L=0$ are entered only when no target of an earlier stage is pending, and the targets of the active stage are uncovered when the preparation starts. Hence the result Alice is following shows that no pending target of stage $s$ is captured while $s$ is active.

Once every target is saved, Bob cannot capture a pending target. So no pending target is ever captured, whether or not Bob may pass. In Line Kalah itself Bob may not pass. Each of his moves then lowers the nonnegative integer $\sum_x xB_x$, because his stones only move to smaller columns or leave the line, and Alice never adds stones to his row. So the play is finite and ends when his row is empty, and Alice wins.
\end{proof}

The arguments of \Cref{sec:false} and of this section do not depend on the capture convention.

\begin{proposition}[Capture robustness]
\label{prop:capture-robustness}
For the line position constructed from a formula in \Cref{sec:coordinates}, Alice wins if and only if the formula is true, under either capture convention. Alice's strategy of \Cref{def:alice-strategy} and Bob's intended strategy for a false formula work under both conventions.
\end{proposition}
The proof, in \Cref{app:true-proofs}, recalls why the convention does not affect these arguments.

\begin{proof}[Proof of \Cref{thm:formula-to-line}]
The line position of \Cref{sec:coordinates} can be computed in polynomial time, its targets are saturated, and its counts are at most its width, by \Cref{lem:layout}. By \Cref{prop:capture-robustness}, Alice wins it under either capture convention exactly when the formula is true.
\end{proof}

The reduction to Kalah in \Cref{sec:empty-stores} uses these results in the following form.

\begin{corollary}[Bob passes]
\label{cor:bob-passes}
For the line position constructed from a true formula, allow Bob also to pass on any turn when his row is nonempty. Under either capture convention, Alice's strategy of \Cref{def:alice-strategy} ensures that no pending target is captured. Until Bob's row empties or every target is saved, each of her moves selects a nonempty pit, and one of these events occurs within $8n+8$ Alice moves.
\end{corollary}
\begin{proof}
No pending target is captured by \Cref{thm:true}, under either capture convention by \Cref{prop:capture-robustness}. Before every target is saved, each prescribed move is a save, a clearing or a shot, so it selects a nonempty pit, and Alice makes at most $8n+8$ such moves (\Cref{lem:bookkeeping}(\labelcref{bk:short})).
\end{proof}

\section{From Line Kalah to Kalah}
\label{sec:empty-stores}

We now embed the formula line positions in Kalah, with all stones in pits and both stores empty. Moves outside the line then act as passes. An Alice pass is allowed in Line Kalah. A Bob pass is not, but \Cref{cor:bob-passes} shows that Alice's strategy still works when Bob may pass, and that it makes a useful selection on every turn until all targets are saved or Bob's row empties.

The score can be arranged by a strict majority. If there are $p$ targets, put a payload of $\Pi$ stones on each, and give Bob $(p-1)\Pi$ additional stones in a separate pit. Choose $\Pi$ larger than the entire original line mass. Alice wins by keeping all $p$ payloads. Bob wins by taking one payload and collecting his additional pile. The next lemmas make both assertions valid under unrestricted play.

\subsection{The board}

\begin{definition}[Empty-store board]
\label{def:empty-store-board}
Let $(D,A,B,P)$ be a line position with saturated targets, $p=|P|\ge1$, and a nonempty Bob row. Put
\[
\begin{aligned}
 M_B&=\sum_{x=1}^D B_x,& M_0&=\sum_{x=1}^D(A_x+B_x),\\
 \Pi&=M_0+1,& Z&=(p-1)\Pi,\\
 M&=M_0+p\Pi+Z=2p\Pi-1,\\
 q&=Z+M_B+1,& h&=q+M_B+1,\\
 \Phi&=qZ+\sum_{x=1}^D(h+x)B_x,&
 N&=h+D+(\Phi+1)M+1.
\end{aligned}
\]
The \emph{empty-store board} has width $N$, both stores empty, and Alice to move. At column $h+x$, Alice holds $A_x$ stones plus $\Pi$ if $x\in P$, and Bob holds $B_x$ stones, for $1\le x\le D$. Bob also holds $Z$ stones at column $q$. All other pits are empty. The $\Pi$ extra stones at each target are its \emph{payload}, and the $Z$ stones initially at Bob's pit $q$ are the \emph{padding stones}. If $Z=0$, that pit is empty. The columns $h+1,\ldots,h+D$ are the \emph{line segment}, and $1,\ldots,h$ are the \emph{buffer}.
\end{definition}

The total number of stones is the odd number $M=2p\Pi-1$, so $p\Pi$ stones are a strict majority. The column $q$ lies far enough below the segment that Bob can leave its pile untouched throughout the forcing play. The long empty stretch above the segment prevents Alice from sowing into her store.

\subsection{Isolation and payloads}

\begin{lemma}[Isolation]
\label{lem:empty-isolation}
In every play from an empty-store board, under either capture convention and with or without the bonus rule, the following hold.
\begin{enumerate}
\item Bob has no pit stone above column $h+D$. If a Bob sow goes past his store, it puts stones only into Alice's buffer pits and does not return to Bob's row.
\item No Alice sow reaches her store, and Alice receives no bonus turn. In fact every Alice sow has terminal column at most $N-1$.
\item Bob makes at most $\Phi$ moves, and Alice makes at most $\Phi+1$ moves.
\end{enumerate}
\end{lemma}

Bob's potential $\sum_x xB_x$ drops with each of his moves, which bounds the number of moves, and the empty stretch above the segment is longer than all of Alice's sows together. The proofs of the three lemmas of this subsection and the next are in \Cref{app:kalah-proofs}.

\begin{lemma}[Ownership of a payload]
\label{lem:empty-payload}
Consider a target in an empty-store board. If Bob captures it before Alice first selects it, its original payload ends in Bob's store. If Alice selects it first, its original payload ends in Alice's store. If neither event occurs before termination, its original payload also ends in Alice's store. Later refills and captures of the same pit do not change these conclusions.
\end{lemma}

\subsection{Collecting stones near a store}

The following elementary fact explains why Bob can leave the padding alone until it is needed. It applies to any Kalah position satisfying its hypotheses.

\begin{lemma}[Nearest-store collection]
\label{lem:collection}
Consider a continuation of a Kalah play in which Alice never sows into her store. At its start suppose that Bob's pit at each column $x$ holds at most $x$ stones, and that Alice has no pit stone at a column below some $u\in\{1,\ldots,N+1\}$. If Bob always selects his nonempty pit of least column, then he never sows past his store, Alice's pits below $u$ remain empty, and every stone initially in Bob's pits below $u$ ends in his store. The conclusion holds under either capture convention, with or without the bonus rule.
\end{lemma}

The reason is that Bob's lowest nonempty pit never holds more stones than its column, so his sows never go past his store.

\subsection{The reduction}

Throughout this subsection we fix a formula of the form described in \Cref{sec:source}, its line position of \Cref{sec:coordinates}, and the empty-store board of that line position. The targets of the line position are saturated and its Bob row is nonempty by \Cref{lem:layout}, so the board is defined. The two lemmas below hold under either capture convention, with or without the bonus rule.

\begin{lemma}[True formulas in Kalah]
\label{lem:kalah-true}
If the formula is true, Alice wins the empty-store board.
\end{lemma}
\begin{proof}
Until all targets are saved or Bob has no stone left in the segment, Alice follows the strategy of \Cref{def:alice-strategy}, as in \Cref{cor:bob-passes}. Each of her prescribed moves selects a pit in the segment. She never selects a buffer pit during this part of the play. The segment counts agree with the line counts, Bob's exactly and Alice's except for the payloads of pending targets. A Bob segment move has the same effect there as its line move. A pending target is saturated, so selecting it gives the same deposits within the segment with or without its payload. Every other Alice segment move also has the same effect there. These statements preserve the agreement up to a first capture of a pending target.

Every Bob move outside the segment starts below it, sows to the left in his row, and, if it goes past his store, changes only Alice's buffer by \Cref{lem:empty-isolation}. Such a move leaves the segment unchanged. A segment move cannot earn a bonus, since its column is at least $h+1$ and Bob's entire pit mass is less than $h$. To account for possible bonus turns, consider the Bob moves between two consecutive Alice turns. All but the last, if any, are buffer bonus moves. Erase these from the simulated history. The last move is a segment move, which gives one ordinary line move, or a buffer move, which gives one Bob pass. If the game ends during such a sequence, there is no further Alice turn to simulate. This gives exactly the alternating line play required by \Cref{cor:bob-passes}.

That corollary supplies legal nonpassing Alice moves and rules out a first pending-target capture. If all targets are saved, all $p$ payloads belong to Alice by \Cref{lem:empty-payload}. If Bob's segment becomes empty first, no later Bob stone can return to it, so no pending target can ever be captured. Again every payload belongs to Alice, whatever legal moves she makes afterwards. The same conclusion holds if the Kalah game ends earlier. Alice therefore receives at least $p\Pi$ of the $2p\Pi-1$ stones and wins.
\end{proof}

\begin{lemma}[False formulas in Kalah]
\label{lem:kalah-false}
If the formula is false, Bob wins the empty-store board.
\end{lemma}
\begin{proof}
Bob follows the winning line strategy of \Cref{thm:false}, selecting only segment pits until a pending target is captured. Until then, no Bob sow touches the padding or reaches his store. Only original line stones can lie in his segment pits, so each such move holds at most $M_B$ stones and lands at a column of at least
\[
 h+1-M_B=q+2.
\]
Alice's buffer stays empty. Every legal Alice segment move has its corresponding line effect, and every selection above the segment represents an allowed Alice pass. By \Cref{lem:empty-isolation} she cannot sow around the board. Thus Bob forces a pending-target capture just as in Line Kalah. Kalah cannot end earlier by emptying Alice's row, since a remaining pending target has a positive pile. If all targets had been selected, or Bob's segment had emptied, without such a capture, the corresponding line play would contradict his winning strategy.

At the first pending-target capture, Bob receives its $\Pi$ payload stones. His padding at $q$ is untouched, and every other Bob pit still holding a stone lies at a column at least $q+2$. Each Bob pit holds at most $M_B+Z<q$ stones, so in particular it holds at most its column. Alice still has no stone below $h+1$. If play continues, Bob applies \Cref{lem:collection} with $u=h+1$. The hypotheses on Alice's sows hold by \Cref{lem:empty-isolation}. All $Z$ padding stones therefore end in Bob's store. If the target capture itself ends the game, the padding is swept to his store immediately. In either case he receives at least
\[
 \Pi+Z=p\Pi
\]
stones and wins.
\end{proof}

\begin{theorem}[Reduction with empty stores]
\label{thm:empty-transfer}
There is a polynomial-time map from the quantified formulas of \Cref{sec:source} to unary Kalah positions with both stores empty and Alice to move, such that Alice wins if and only if the formula is true. The output has an odd total number of stones. The same positions establish this equivalence under either capture convention, with or without the bonus rule.
\end{theorem}
\begin{proof}
Map the formula to the empty-store board of its line position. By \Cref{lem:kalah-true,lem:kalah-false}, Alice wins exactly when the formula is true, and the total number of stones is the odd number $M=2p\Pi-1$.

Finally, the map has polynomial size. The formula line position has polynomial width $D$, and all its counts are at most $D$ by \Cref{lem:layout}. Hence $p\le D$ and $M_B\le M_0\le2D^2$. Every parameter in \Cref{def:empty-store-board} is obtained by a fixed number of sums and products of these quantities. In particular $M$, $\Phi$, and $N$ are polynomially bounded in $D$. Writing the $2N$ pit counts in unary takes polynomial time. The total mass is $M=2p\Pi-1$. The construction depends on neither convention, the strategy arguments above treat bonus turns explicitly, and the line strategies work under either capture convention by \Cref{prop:capture-robustness}.
\end{proof}

\subsection{Proof of the main theorem}

\begin{proof}[Proof of \Cref{thm:main}]
Membership is \Cref{thm:membership}, including when both stores are initially empty and Alice is to move. Hardness for that restriction follows from \Cref{thm:empty-transfer} and the $\PSPACE$-hardness of the quantified formulas of \Cref{sec:source}. The same output positions have both stores empty and Alice to move, and they work under all four rule choices. This also proves hardness when arbitrary initial stores and either player to move are allowed.
\end{proof}

\section{Consequences and open questions}
\label{sec:consequences}

\paragraph{Line Kalah.} By \Cref{thm:formula-to-line}, the Line Kalah problem is $\PSPACE$-hard under either capture convention. It also belongs to $\PSPACE$. Each Bob move lowers the nonnegative integer $\sum_x xB_x$, since Alice never adds stones to his row, and the players alternate, so a play has at most $2\sum_x xB_x+1$ moves and the game tree can be searched in polynomial space. Hence the Line Kalah problem is $\PSPACE$-complete.

Line Kalah remains $\mathsf{PSPACE}$-complete when counts are written in binary. On a line of width $D$, a pile of $D$ or more stones sows once into every pit ahead of it, and the rest fall off the line. We can therefore cap every initial count at $D$: subsequent deposits preserve this equivalence, and a capture empties the entire pile. The resulting unary instance has size $O(D^2)$.

\paragraph{Binary counts.} Our upper bound needs unary counts. With counts written in binary, a position can hold exponentially many stones, and we show the following. The proof is in \Cref{app:extensions}.

\begin{proposition}[Binary counts]
\label{prop:binary-counts}
With counts written in binary, deciding the winner of a given Kalah position belongs to $\mathsf{EXPTIME}$ and is $\PSPACE$-hard under either capture convention, with or without the bonus rule. Hardness holds even with both stores initially empty and Alice to move.
\end{proposition}

\paragraph{Open questions.} We list four questions that our results leave open.
\begin{itemize}
\item \emph{Uniform openings.} Our hard positions are far from uniform. The winner of the uniform opening $\Kalah(N,c)$, with $N$ and $c$ written in unary, is a sparse language in $\PSPACE$. Whether it can be decided in polynomial time is open.
\item \emph{Kalah without captures.} If captures are omitted, \Cref{lem:polynomial-play} still applies, so the unary problem is in $\PSPACE$. Our lower bound uses captures both to implement choices and to make the loss of a target decisive, and it does not show hardness for this variant. Simple-Kalah~\cite{MaeiKiyaHanakaOno2018} also omits captures but ends only when both rows are empty.
\item \emph{Binary counts.} With counts written in binary the problem is $\PSPACE$-hard and in $\mathsf{EXPTIME}$ (\Cref{prop:binary-counts}). Our bound on play length can be exponential in the binary input length. Whether the winner can still be decided in polynomial space is open.
\item \emph{Oware.} Oware differs from Kalah in its sowing, capture and feeding rules. Its play can cycle, so a complexity question must also fix a repetition rule. The complexity of Oware with a variable number of pits and unary counts is open.
\end{itemize}

\section*{AI disclosure}
The authors used large language models (OpenAI GPT-6.0 Astra, GPT-6.1 Sol/Terra, and Anthropic Claude Opus 4.8 and 5.5) as research and writing assistants during the preparation of this manuscript. Specifically, these tools were used to explore candidate reductions and intermediate game variants (including Line Kalah), assist with drafting and restructuring the exposition and terminology, and generate figures. All mathematical constructions, reductions, and proofs were rigorously verified, refined, and validated by the authors, who take full responsibility for the content of this paper.

\bibliographystyle{plain}
\bibliography{references}

\addtocontents{toc}{\protect\setcounter{tocdepth}{1}}
\appendix
\section{The coordinates}
\label[appendix]{app:coordinates}

This appendix gives the exact columns and counts of the construction of \Cref{sec:construction}, proves the layout facts that the other proofs use, and proves \Cref{lem:standard}.

\subsection{Columns and counts}
\label{sec:coordinates}

\begin{table}[t]
\centering
\small
\begin{tabular}{@{}l p{0.68\textwidth}@{}}
\toprule
Symbol & Meaning\\
\midrule
$m$, $C$, $\rho$ & number of variables, number of clauses, and largest number
  of occurrences of one variable\\
$n=m+2C-2$ & number of writers, namely one genuine writer per variable and $2C-2$
  counter writers\\
$n_\forall$ & number of universal writers\\
$\epsilon_i$, $\nu$ & bit of writer $i$, and number of counter bits equal to one\\
$\delta$ & $1$ for an occurrence $x_i$, and $0$ for an occurrence
  $\neg x_i$\\
$H=16n+32$ & bounds Bob's move count and every residue pile during the
  prescribed strategy prefix through the last target save\\
$\gamma=H+\rho+8$ & equals $L_i-R_i$ and $w^L_i-w^R_i$, and leaves room for the
  records of one variable\\
$\sigma=8\gamma+40$ & distance between the timer blocks of consecutive
  writers\\
$W=3H+9$ & height $b_i-a_i$ of a band, and the gap between consecutive
  writer bands\\
$J=8n+7-n_\forall$ & number of clocks\\
\midrule
$a_i$, $b_i$ & tops of the left and right triples of writer $i$\\
$a_\star$, $b_\star$ & tops of the left and right triples of the reader\\
$w^R_i$, $w^L_i$, $k_0$ & weapons of writer $i$, each holding $k_0$ stones
  initially\\
$g$ & common clock landing column in the intended play\\
$R'_i$, $L'_i$, $R_i$, $L_i$ & right backup, left backup, right timer
  and left timer of writer $i$, with the guards opposite $R_i$ and $L_i$\\
$K_0$ & initial count of a guard of a genuine writer\\
$f_0$ & column where a counter writer's clearing at $R_i$ lands\\
$r_{i,\ell}$, $P_{i,\ell}$ & column and initial count of the record of the
  $\ell$-th occurrence of $x_i$\\
$Q=L_0+K_0$ & lies above every record\\
$d_\eta$ & dummy records, for odd $\eta\le2C-3$\\
$Q^+$ & the largest of $Q$ and the dummy records\\
$f_1$ & column where a counter writer's clearing at $L_i$ lands\\
$L_\star$, $R_\star$ & left and right timers of the reader, with $R_\star=L_\star+1$\\
$S^R_i$, $S^L_i$ & selectors of a universal writer $i$, which reach $w^R_i$
  and $w^L_i$ on Bob's move $8i$\\
$c_j$ & clocks, $0\le j<J$, where $c_j$ initially lands at $g+j$\\
$D$ & width of the line\\
\bottomrule
\end{tabular}
\caption{Notation of the construction. The second group lists the columns of \Cref{sec:coordinates} in the order of the line (\Cref{fig:ov-stages}(a)), with the initial counts next to their piles.}
\label{tab:notation}
\end{table}

The following numbers depend only on $n$ and $\rho$.
\[
 H=16n+32,\qquad \gamma=H+\rho+8,\qquad \sigma=8\gamma+40,\qquad
 W=3H+9.
\]
The bands come first, the reader's lowest and writer $0$'s highest.
\[
\begin{aligned}
 a_\star&=4,& b_\star&=4+W,\\
 a_i&=b_\star+W+16n+50+2(n-1-i)W,& b_i&=a_i+W\qquad(0\le i<n).
\end{aligned}
\]
The weapons and the clock landing column are
\[
\begin{aligned}
 w_0&=b_0+4H+100+(\sigma+6)(n-1), & w^R_i&=w_0-(\sigma+6)i,\\
 g&=w_0+\gamma+4H+100, & w^L_i&=w^R_i+\gamma.
\end{aligned}
\]
The writer timers are
\[
 k_0=g+\sigma n+100n+100,\quad R_i=w_0+k_0-\sigma i,\quad L_i=R_i+\gamma,
 \quad L'_i=R_i-\gamma-4,\quad R'_i=R_i-2\gamma-8.
\]
For the records put $K_0=L_0+\sigma n+100n+100$. If $x_i$ occurs in the clauses, number its occurrences $\ell=1,2,\ldots$ and place the record of its $\ell$-th occurrence at
\[
 r_{i,\ell}=R_i+K_0+7i+1+\ell .
\]
Put $Q=L_0+K_0$. For each odd $\eta$ with $1\le\eta\le 2C-3$ place a dummy record at $d_\eta=Q+10+(\eta-1)/2$, and let $Q^+$ be the largest of $Q$ and these columns. Then
\[
 f_0=L_0+20n+20,\qquad f_1=Q^++20n+20,\qquad L_\star=f_1+100n+100,\qquad
 R_\star=L_\star+1.
\]
For each universal writer $i$ (so $i\ge1$) the selectors are
\[
 S^R_i=R_\star+100n+100+10(n-1-i),\qquad S^L_i=S^R_i+3,
\]
and with $J=8n+7-n_\forall$ the clocks and the width are
\[
 c_j=R_\star+110n+100+3(J-j)\quad(0\le j<J),\qquad D=c_0+100n+100.
\]

\paragraph{Alice's counts.} Every Alice pit holds $D$ stones, except the following. Each weapon holds $k_0$. The guards of a genuine writer $i<m$ hold $K_0$ each. The guards of a counter writer $i\ge m$ hold
\[
 A_{R_i}=f_0-R_i-7i,\qquad A_{L_i}=f_1-L_i-7i.
\]
If the $\ell$-th occurrence of $x_i$ is in clause $\varphi_j$, the record $r_{i,\ell}$ holds
\[
 P_{i,\ell}=R_\star-r_{i,\ell}-6n-i-2j-\delta,\qquad
 \delta=\begin{cases}1&\text{if the occurrence is }x_i,\\
 0&\text{if it is }\neg x_i.\end{cases}
\]
The dummy record $d_\eta$ holds $R_\star-d_\eta-6n-\eta$. The targets are the $6n+6$ columns of the bands, and they hold $D$ stones.

\paragraph{Bob's counts.} Bob's pit at $g$ holds one stone, and his structural piles hold
\[
\begin{aligned}
 B_{L_i}&=L_i-a_i-(8i+2), & B_{R_i}&=R_i-b_i-(8i+2),\\
 B_{L'_i}&=L'_i-a_i-(8i+5), & B_{R'_i}&=R'_i-b_i-(8i+5),\\
 B_{L_\star}&=L_\star-a_\star-(8n+2), & B_{R_\star}&=R_\star-b_\star-(8n+1),\\
 B_{S^R_i}&=S^R_i-w^R_i-(8i-1), & B_{S^L_i}&=S^L_i-w^L_i-(8i-1),\\
 B_{c_j}&=c_j-g-j.
\end{aligned}
\]
All other Bob pits are empty. The structural piles are the timers, the selectors and the clocks. The single stone at $g$ is not structural.

So, as stated in \Cref{sec:layout}, the initial landing columns are $a_i+8i+2$ and $b_i+8i+2$ for the writer timers $L_i$ and $R_i$, $a_i+8i+5$ and $b_i+8i+5$ for the backups, $a_\star+8n+2$ and $b_\star+8n+1$ for the reader timers, $w^R_i+8i-1$ and $w^L_i+8i-1$ for the selectors, and $g+j$ for the clock $c_j$. In the intended play, every Bob move before writer $i$ starts is a clock or a selector, there are $8i$ such moves, and each of them adds one stone to every timer of writer $i$, so writer $i$ starts with its timers landing at $a_i+2$, $b_i+2$, $a_i+5$ and $b_i+5$. \Cref{tab:notation} collects the notation.

\subsection{Layout facts}

The next lemma lists the facts about the coordinates that the proofs use. Each item is a direct calculation from the definitions, and the calculations are in \Cref{app:arithmetic}.

\begin{lemma}[Layout]
\label{lem:layout}
\leavevmode
\begin{enumerate}
\item\label{lay:order} \textbf{Order.} The targets, weapons, timers, records, dummy records, selectors, clocks, and columns $g,f_0,f_1$ are pairwise distinct and lie in $\{1,\ldots,D\}$. They appear in the order of \Cref{fig:ov-stages}(a).
  \[
   a_\star-2<b_\star<a_{n-1}-2<b_{n-1}<\cdots<a_0-2<b_0<w^R_{n-1}<w^L_{n-1}
   <\cdots<w^R_0<w^L_0<g,
  \]
then the timer blocks $R'_i<L'_i<R_i<L_i$ for $i=n-1,\ldots,0$ in this order, then $f_0$, the records, $Q^+$, $f_1$, $L_\star<R_\star$, the selectors, and the clocks $c_{J-1}<\cdots<c_0<D$. Auxiliary names may denote the same column. In particular, $w_0=w^R_0$, and $Q^+$ is the largest of $Q$ and the dummy record columns.
\item\label{lay:gaps} \textbf{Gaps.} $W>3H+8$. Consecutive band tops differ by at least $W$. $\gamma>H+\rho+1$. Two weapons of different writers are more than $2H$ apart. Every weapon lies more than $4H$ below $g$ and more than $4H$ above $b_0$. Every timer lies more than $H$ above $g+J$. Every record and dummy record lies more than $H$ below $L_\star$.
\item\label{lay:size} \textbf{Size.} $D=O(n^2+n\rho)$, and the line position can be computed in time polynomial in the size of the formula.
\item\label{lay:counts} \textbf{Counts.} Every count is a nonnegative integer at most $D$, every pile with a role is nonempty, every target is saturated, and every Bob pile has landing column at least~$1$.
\item\label{lay:records} \textbf{Records.} The records of $x_i$ lie in $(R_i+K_0+7i,\;L_i+K_0+7i)$. If $i<i'$, every record of $x_{i'}$ lies below $R_i+K_0+7i$, and every record of $x_i$ lies above $L_{i'}+K_0+7i'$. All records lie between $f_0$ and $Q$, and the dummy records lie between $Q$ and $f_1$.
\end{enumerate}
\end{lemma}

\subsection{Arithmetic for the layout}
\label{app:arithmetic}

This subsection proves \Cref{lem:layout} by direct calculation from the definitions in \Cref{sec:coordinates}. Recall $H=16n+32$, $\gamma=H+\rho+8$, $\sigma=8\gamma+40$ and $W=3H+9$, and that $R_i=w_0+k_0-\sigma i$.

\paragraph{Order.} We go through the chain of \Cref{lem:layout}(\labelcref{lay:order}) from left to right.
\begin{itemize}
\item $a_\star-2=2\ge1$ and $b_\star=a_\star+W$.
\item $a_{n-1}-2=b_\star+W+16n+48>b_\star$, and for $1\le i<n$, $a_{i-1}-2=a_i+2W-2>a_i+W=b_i$.
\item $w^R_{n-1}=w_0-(\sigma+6)(n-1)=b_0+4H+100>b_0$.
\item $w^L_i-w^R_i=\gamma>0$ and $w^R_{i-1}-w^L_i=\sigma+6-\gamma=7\gamma+46>0$.
\item $g-w^L_0=4H+100$.
\item $R'_{n-1}-g=w_0+k_0-\sigma(n-1)-2\gamma-8-g=w_0+\sigma+100n+92-2\gamma$, which exceeds $H+J$ because $w_0>b_0>W$.
\item Within block $i$, $L'_i-R'_i=\gamma+4$, $R_i-L'_i=\gamma+4$, $L_i-R_i=\gamma$. Between blocks, $R'_i-L_{i+1}=\sigma-3\gamma-8 =5\gamma+32>0$.
\item $f_0-L_0=20n+20$. The lowest record is at least $R_{n-1}+K_0+2>K_0=L_0+\sigma n+100n+100>f_0$.
\item The records of $x_i$ lie in $[R_i+K_0+7i+2,\;R_i+K_0+7i+1+\rho]$, and the records of $x_{i+1}$ lie below those of $x_i$ because $\sigma>\rho+6$. The highest record is at most $R_0+K_0+1+\rho<R_0+\gamma+K_0=L_0+K_0=Q$.
\item $d_\eta=Q+10+(\eta-1)/2$, so the dummy records are distinct and lie in $[Q+10,Q+C+8]$, so $Q^+\le Q+C+8$.
\item $f_1-Q^+=20n+20$, $L_\star-f_1=100n+100$, $R_\star=L_\star+1$.
\item $S^R_i-R_\star=100n+100+10(n-1-i)\ge100n+100$, $S^L_i-S^R_i=3$, and $S^R_{i-1}-S^L_i=7$.
\item The highest selector is at most $R_\star+110n+83$, and $c_{J-1}=R_\star+110n+103$, consecutive clocks differ by $3$, and $D-c_0=100n+100$.
\end{itemize}
All differences are positive, so the operational sites are distinct and in the stated order, and they lie in $\{1,\ldots,D\}$. This does not assert distinctness of auxiliary names for the same column, such as $w_0=w^R_0$ or $Q^+$ when it is a dummy record column.

\paragraph{Gaps.} $W=3H+9>3H+8$. Consecutive band tops differ by $W$, except $a_{n-1}-b_\star=W+16n+50$. $\gamma-(H+\rho+1)=7$. Weapons of different writers are at least $7\gamma+46>2H$ apart. The weapons lie between $w^R_{n-1}=b_0+4H+100$ and $w^L_0=g-4H-100$. The lowest timer $R'_{n-1}$ lies more than $H$ above $g+J$, as computed above. The highest record or dummy record is at most $Q^+$, and $L_\star-Q^+=120n+120>H$.

\paragraph{Size.} $H=O(n)$, $W=O(n)$ and $\gamma,\sigma=O(n+\rho)$. Then $b_0=O(nW)=O(n^2)$, and each later coordinate adds $O(n\sigma+H)$ to an earlier one or to the sum of two earlier ones. Since $R_0=w_0+k_0$ and $k_0=g+O(n\sigma)$, all coordinates are $O(n^2+n\rho)$, and so is $D$. The position lists $2D$ counts, each at most $D$, so its unary encoding has size $O(D^2)$, and each count is computed from its formula in \Cref{sec:coordinates} in polynomial time.

\paragraph{Counts.}
\begin{itemize}
\item Weapons hold $k_0$ and genuine guards hold $K_0$, and both are positive and less than $D$.
\item A counter guard at $R_i$ holds $f_0-R_i-7i\ge f_0-R_0-7n=\gamma+13n+20>0$, using $R_0=L_0-\gamma$. A counter guard at $L_i$ holds $f_1-L_i-7i\ge f_1-L_0-7n$. Since $f_1>Q=L_0+K_0$, we have $f_1-L_0-7n>K_0-7n>0$.
\item A record holds $P_{i,\ell}=R_\star-r_{i,\ell}-6n-i-2j-\delta$. Here $R_\star-r_{i,\ell}\ge R_\star-Q^+=120n+121$, and $6n+i+2j+\delta\le 6n+(m-1)+2(C-1)+1=6n+(n-m)+m=7n$, so $P_{i,\ell}\ge113n+121>0$. A dummy record holds $R_\star-d_\eta-6n-\eta\ge 120n+121-6n-2C>0$.
\item Every Alice pit not listed above holds $D$. In particular, every target $x$ holds $D\ge D-x+1$ and is saturated.
\item Each structural Bob pile holds its column minus its landing column, and the landing columns $a_i+8i+2$, $b_i+8i+2$, $a_i+8i+5$, $b_i+8i+5$, $a_\star+8n+2$, $b_\star+8n+1$, $w^R_i+8i-1$, $w^L_i+8i-1$ and $g+j$ are positive and smaller than the corresponding columns, by the order above and because $8n+5<W$ and $J<H$. So every structural Bob count is positive, at most $D$, and has landing column at least $1$. The remaining Bob counts are zero except for $B_g=1$.
\end{itemize}
Every count is therefore a nonnegative integer at most $D$, and every designated pile is nonempty.

\paragraph{Records.} The records of $x_i$ lie in $[R_i+K_0+7i+2,\;R_i+K_0+7i+1+\rho]$, which is inside $(R_i+K_0+7i,\;L_i+K_0+7i)$ because $\rho+1<\gamma=L_i-R_i$. For $i<i'$,
\[
 R_i+K_0+7i-(R_{i'}+K_0+7i'+1+\rho)=(\sigma-7)(i'-i)-1-\rho>0,
\]
so the records of $x_{i'}$ lie below $R_i+K_0+7i$, and
\[
 R_i+K_0+7i+2-(L_{i'}+K_0+7i')=(\sigma-7)(i'-i)+2-\gamma>0,
\]
so the records of $x_i$ lie above $L_{i'}+K_0+7i'$. The records lie between $f_0$ and $Q$, and the dummy records between $Q$ and $f_1$, by the order computations.

\subsection{What each move crosses}

Recall that a move crosses the columns strictly between its source and its last stone (\Cref{def:landing}). A Bob move adds a stone to each Bob pit that it crosses and covers Alice's pit there (\Cref{lem:moves}(\labelcref{mv:sow})). The following lemma says which moves cross which piles. It only uses bounds on landing columns that hold during the first $H$ Bob moves of any play.

\begin{lemma}[Crossings]
\label{lem:crossings}
Consider any play in which Bob has made at most $H$ moves.
\begin{enumerate}
\item\label{cross:landing} Every unfired writer timer of writer $i$ has landing column within $H$ of $a_i$ (left timer and backup) or of $b_i$ (right timer and backup), and the same holds for the reader timers with $a_\star,b_\star$. Every unfired selector of writer $i$ lands within $H$ of its weapon, and every unfired clock lands in $[g-H,\,g+J]$.
\item\label{cross:clock} A clock or selector firing crosses every unfired timer and every record, and lands above $b_0+3H$.
\item\label{cross:writer} The firing of a timer of writer $i$ crosses every unfired timer of every writer $i'>i$, the column $g$, every weapon, and every target of every writer $i'<i$. It does not cross the reader timers or any record, and it lands more than $W-2H$ above every band below band $i$.
\item\label{cross:reader} The firing of a reader timer crosses every record, every writer timer, every weapon, and every writer target.
\item\label{cross:lower} When a left timer or left backup of any stage fires, it crosses the right triple of its own band.
\item\label{cross:save} A save adds one stone to every Alice pit to its right, in particular to every weapon, guard and record.
\item\label{cross:alice} Clearing a guard of writer $i$ adds stones only to Alice pits above $R_i$, and it crosses no weapon and no guard of a writer $i'>i$. A shot of writer $i$ goes from a weapon to $R_i$ or $L_i$. It crosses both guards of every writer $i'>i$, and no weapon of a writer $i'>i$ and no record.
\item\label{cross:guard} Suppose every writer clears at most one guard and shoots at most once, and a writer that does both shoots at the timer opposite the guard it cleared. Then a guard that is never cleared receives at most one stone from the clearings and shots of each writer.
\end{enumerate}
\end{lemma}
\begin{proof}
An unfired structural pile only gains stones, at most one per Bob move, so its landing column decreases by at most $H$ from its initial value. The initial landing columns are listed above, and writer timers start at most $8n+5<H$ above their band tops, so item~\labelcref{cross:landing} follows. The clock $c_j$ starts at $g+j$ with $j<J$.

\Crefrange{cross:clock}{cross:reader} compare a source column with a landing column. Clocks and selectors lie above every timer and record, and by item~\labelcref{cross:landing} they land at or above $w^R_{n-1}-H>b_0+3H$, below every timer (\Cref{lem:layout}). A timer of writer $i$ lies in block $i$, above the blocks of writers $i'>i$, above $g$ and the weapons, and below the records and the reader timers. It lands within $H$ of band $i$, below the bands of writers $i'<i$ (which are at least $W>H$ higher) and more than $W-2H$ above the next lower band. The reader timers lie above every record and every writer timer, and land within $H$ of the reader band, which is the lowest. For item~\labelcref{cross:lower}, a left timer or backup lies above $b+2H$ and lands within $H$ of $a$, which is below $b-2$ because $b-a=W>H+2$. Item~\labelcref{cross:save} holds because targets are saturated. For item~\labelcref{cross:alice}, the guards of writer $i$ lie in block $i$, above every weapon and above the blocks of the writers $i'>i$, and Alice sows to the right. A weapon of writer $i$ lies above the weapons of the writers $i'>i$ and below $g$, and its shot ends in block $i$, so it crosses the blocks of the writers $i'>i$ and stops below every record.

For item~\labelcref{cross:guard}, take a guard of writer $i$ that is never cleared. A writer $i'<i$ clears a guard in block $i'$, above it, so only the shot of $i'$ can add a stone to it. A writer $i'>i$ shoots into block $i'$, below it, so only the clearing of $i'$ can. If writer $i$ makes both moves, then either it clears the guard at $R_i$ and shoots at $R_i$, in which case the shot ends below $L_i$, or it clears the guard at $L_i$ and shoots at $L_i$, in which case the clearing adds stones only above $L_i$. In both cases at most one of its two moves adds a stone to its other guard.
\end{proof}

\subsection{Intended play}

\begin{proof}[Proof of \Cref{lem:standard}]
Alice's counts change only through her own moves and Bob's captures, and Bob has captured no pit in $(g+J,R_\star]$. That interval contains every guard, record and dummy record, and the columns $f_0$, $f_1$, $R_i+K_0+7i$ and $L_i+K_0+7i$ (\Cref{lem:layout}). Alice selects only targets, weapons and guards of writers, and these lie below $f_0$.

\emph{Item~\labelcref{sp:clear}.} A guard of writer $i$ gains one stone from each save and from the shot of each writer $i'<i$, and nothing from the clearings of those writers (\Cref{lem:crossings}), and writer $i$ clears before its other moves. So when writer $i$ clears a guard, the guard holds its initial count plus $7i$, namely $K_0+7i$ if $i<m$, and $f_0-R_i$ or $f_1-L_i$ if $i\ge m$. Adding the column of the guard, the clearing lands at $R_i+K_0+7i$ or $L_i+K_0+7i$, or at $f_0$ or $f_1$. By \Cref{lem:layout}(\labelcref{lay:records}), the records of $x_i$ lie strictly between $R_i+K_0+7i$ and $L_i+K_0+7i$, the records of later variables lie below $R_i+K_0+7i$, and those of earlier variables lie above $L_i+K_0+7i$. By \Cref{lem:layout}(\labelcref{lay:order}), the columns $f_0$ and $f_1$ and the dummy records are distinct from these. So the landing pit is not a target, weapon, guard or record. Alice has never selected it and Bob has not captured it, so it holds at least $D$ stones, and the clearing captures nothing.

\emph{Item~\labelcref{sp:counts}.} A weapon of a writer $i'$ with $i\le i'<n$ receives one stone from each save, and nothing from the shots of earlier writers, which start above it, or from clearings, which add stones only above timers (\Cref{lem:crossings}). So an uncaptured one holds $k_0+6i$ stones, and the identities $R_i-w^R_i=L_i-w^L_i=k_0+6i$ give the landing columns when $i<n$. In that case the guards of writer $i$ hold $K_0+7i$ stones, or $f_0-R_i$ and $f_1-L_i$, by the computation in the proof of item~\labelcref{sp:clear}, and these counts are positive because the initial counts are (\Cref{lem:layout}(\labelcref{lay:counts})). The reader's guards lie in $(g+J,R_\star]$ and were never selected, so they hold at least $D$ stones.

\emph{Item~\labelcref{sp:timers}.} Assume the condition of item~\labelcref{sp:counts} with $i<n$. We go through Alice's pits below $L_i$, since a pit above $L_i$ cannot land at or below $L_i$. Each Alice move adds at most one stone to a given pit, and Alice has made $8i\le8n$ moves.
\begin{itemize}
\item A pit that Alice never selected and Bob never captured, other than a weapon, guard or record, holds at least $D$ stones, so its last stone leaves the line. This includes the pits at $R_i-1$ and $L_i-1$, which lie in $(g+J,R_\star]$.
\item A pit below $L_i$ that Alice selected or Bob captured is not a cleared guard, since those lie in the blocks of the writers before $i$, above $L_i$. So it is a saved target, a used or captured weapon, or a pit that Bob captured, all at columns at most $g+J$. Since it was emptied, such a pit has received at most $8n$ stones, so it lands at most at $g+J+8n$, more than $H$ below every timer column.
\item Consider an unused weapon that Bob has not captured. Those of writer $i$ land at $R_i$ and $L_i$ (item~\labelcref{sp:counts}). One of a later writer $i'$ holds $k_0+6i$ stones (proof of item~\labelcref{sp:counts}) and lands at $R_{i'}-6(i'-i)$ or $L_{i'}-6(i'-i)$, which is at most $L_{i'}\le R_i-\sigma+\gamma<R_i-1$. One of an earlier writer $i'$ has received the $6i$ saves, so it holds at least $k_0+6i$ stones and lands at or above $R_{i'}+6(i-i')$ or $L_{i'}+6(i-i')$, above $L_i$ because $R_{i'}\ge R_i+\sigma$.
\item The guards of a later writer $i'$ have received only the $6i$ saves and the $i$ shots of the writers before $i$ (\Cref{lem:crossings}). So they hold $K_0+7i$ stones (genuine) or $f_0-R_{i'}-7(i'-i)$ and $f_1-L_{i'}-7(i'-i)$ (counter), and they land at $R_{i'}+K_0+7i$ or $L_{i'}+K_0+7i$, above $K_0>L_0$, or at least at $f_0-7n>L_0$. The guard at $R_i$ lands at $R_i+K_0+7i$ or at $f_0$, above $L_i$, by the proof of item~\labelcref{sp:clear}.
\item The records and dummy records lie above $L_0$, so above $L_i$.
\end{itemize}
So the only pile below $R_i$ that lands at $R_i$ is an uncaptured $w^R_i$, the only pile below $L_i$ that lands at $L_i$ is an uncaptured $w^L_i$, and no pile lands at $R_i-1$ or $L_i-1$.

\emph{Item~\labelcref{sp:records}.} Records and dummy records are above every target, so each of the $6n$ saves adds one stone to each of them. By the landing columns in the proof of item~\labelcref{sp:clear} and \Cref{lem:layout}(\labelcref{lay:records}), clearing the guard at $L_{i'}$ crosses the records of $x_{i'}$ and of every $x_i$ with $i>i'$, and clearing the guard at $R_{i'}$ crosses only the records of $x_i$ with $i>i'$. So the records of $x_i$ receive one stone from each of the $i$ genuine writers before it, and one more if $\epsilon_i=1$. A counter clearing at $R_{i'}$ lands at $f_0$, below every record, and one at $L_{i'}$ lands at $f_1$, above every record and dummy record. So the counter writers add $\nu$ to every record and dummy record, while genuine clearings land below the dummy records. Shots land at timers, below every record. So the record $r_{i,\ell}$ of an occurrence in clause $\varphi_{j'}$ holds $P_{i,\ell}+6n+i+\epsilon_i+\nu$ stones and lands at $r_{i,\ell}+P_{i,\ell}+6n+i+\epsilon_i+\nu=R_\star+\nu-2j'-\delta+\epsilon_i$, and the dummy record $d_\eta$ holds $R_\star-d_\eta-\eta+\nu$ stones and lands at $R_\star+\nu-\eta$.

If $\nu$ is odd, the dummy record $d_\nu$ lands at $R_\star$. If $\nu=2j$ is even, no dummy record lands at $R_\star$, and the record of an occurrence in clause $\varphi_{j'}$ lands at $R_\star+2(j-j')-\delta+\epsilon_i$. Since $-\delta+\epsilon_i\in\{-1,0,1\}$, this equals $R_\star$ only if $j'=j$ and $\epsilon_i=\delta$, that is, exactly when the occurrence is a true literal of $\varphi_j$.

\emph{Item~\labelcref{sp:inputs}.} Now all $n$ branches are complete, and Alice has made $8n$ moves. We go through Alice's pits below $R_\star$ other than the records and dummy records.
\begin{itemize}
\item A pit that Alice never selected and Bob never captured, other than a writer's weapon or guard, a record or a dummy record, holds at least $D$ stones, and so do the reader's guards (item~\labelcref{sp:counts}), so their last stones leave the line.
\item A pit that Alice selected or Bob captured is a saved target, a used or captured weapon, a cleared guard, or a pit that Bob captured at a column at most $g+J$, so it lies at most at $L_0$. Since it was emptied it has received at most $8n$ stones, so it lands at most at $L_0+8n<f_0$.
\item An unused weapon that Bob has not captured holds $k_0$ stones, plus one from each of the $6n$ saves and at most one from each of the $n$ shots, so it lands at most at $L_0+7n<f_0$.
\item A writer's guard that was not cleared holds its initial count plus one stone from each of the $6n$ saves and, by \Cref{lem:crossings}(\labelcref{cross:guard}), at most one from the clearing and shot of each writer. So a genuine one lands at most at $L_0+K_0+7n=Q+7n$ and a counter one at most at $f_1+7n$, both below $R_\star=f_1+100n+101$.
\end{itemize}
So only records and dummy records can land at $R_\star$. The comparisons in both walks use the gaps of \Cref{lem:layout}.
\end{proof}

\section{Proofs of the deadline lemmas}
\label[appendix]{app:deadline-proofs}

This appendix proves the lemmas and theorems of \Cref{sec:bands}. Their statements, and the definitions and standing assumptions they use, are given there.

We first state the facts about single moves that the proofs use.

\begin{lemma}[Single moves near a band]
\label{lem:band-facts}
In a band setting, the following hold as long as the standing assumptions do.
\begin{enumerate}
\item\label{bf:quiet} A quiet Bob move leaves $F_{\mathrm{left}}$ and every $F_{\mathrm{right}}(\Lambda)$ unchanged, captures no band target, and leaves every covered band target covered.
\item\label{bf:upper} A Bob move that empties the pit opposite a target of $\mathcal B$ leaves $F_{\mathrm{left}}$ unchanged, captures no target of $\mathcal A$, and leaves every other covered target covered.
\item\label{bf:slack} If, when Bob is to move, the left triple has slack at least one, then Bob's move captures no pending left target. If the right triple has slack at least one outside $\Lambda$ and Bob's move fires no pile of $\Lambda$, then it captures no pending right target. Every Bob move lowers the left slack by at most one, a Bob move that fires no pile of $\Lambda$ lowers the right slack outside $\Lambda$ by at most one, and an Alice move lowers neither.
\item\label{bf:below} If, when Bob is to move, a set $\mathcal X$ of pending band targets lies below $E_Y$ for a set $Y$ of unfired structural piles, then no target of $\mathcal X$ is covered, and Bob's move exposes none.
\item\label{bf:lower} A Bob move that fires a pile at a column greater than $b$ whose landing column $y$ satisfies $y<b-2$ captures no target of $\mathcal B$, and afterwards every pending target of $\mathcal B$ is covered.
\item\label{bf:covered} If at some Bob turn every pending target of $\mathcal B$ is covered, and from then on Alice applies the cover rule after every Bob move that is not quiet and makes only moves that change no Bob pile, then no pending target of $\mathcal B$ is captured.
\end{enumerate}
\end{lemma}

\begin{proof}[Proof of \Cref{lem:band-facts}]
\emph{Item~\labelcref{bf:quiet}.} A quiet move selects a column $z$ with $z-2H>b+H+6$. By (S1) and \Cref{lem:quiet} with $\lambda=b+H+6$, it leaves $\min\{E_Y,b+H+7\}$ unchanged for every set $Y$ of unfired structural piles, hence also $F_{\mathrm{left}}$ and $F_{\mathrm{right}}(\Lambda)$, whose clipping levels are smaller. It captures only at columns of at least $z-H>b$ and changes only pits at columns of at least $z-H>b$.

\emph{Item~\labelcref{bf:upper}.} The selected pile lies at a column $z\ge b-2$ and is residue, since every unfired structural pile lies above $b+2H+10$ by (S3). So the move is $(a+3)$-quiet, because $z-2H\ge b-2-2H>a+6$ and no unfired structural pile lies in $[z-H,z-1]$. By \Cref{lem:quiet} with $\lambda=a+3$ it leaves $\min\{E_\Omega,a+4\}$ unchanged and captures only at a column of at least $z-H>a$. The last claim is \Cref{lem:moves}(\labelcref{mv:covered}).

\emph{Item~\labelcref{bf:slack}.} Let $Y$ be $\Omega$ for the left triple and $\Lambda\cup\Omega$ for the right one, and $F$ the corresponding clipped frontier. A capture at a pending target $z$ needs a Bob pile with landing column $z$. The piles of $\Omega$ do not move by (S2), those of $\Lambda$ do not move by assumption, and every other pile lands at or above $E_Y\ge F$, which is larger than every pending target of the triple. The second claim is \Cref{lem:moves}(\labelcref{mv:frontier}), since $\hat a$ and $\hat b$ never increase.

\emph{Item~\labelcref{bf:below}.} A covered target $z$ of $\mathcal X$ has a nonempty Bob pit at $z$. That pile is not in $Y$, because unfired structural piles lie above $b+2H+10$ by (S3), and its landing column is at most $z-1<E_Y$, which contradicts the definition of $E_Y$. A target that is not covered cannot become exposed.

\emph{Item~\labelcref{bf:lower}.} By \Cref{lem:moves}(\labelcref{mv:capture}) the move captures at most at column $y$, which is not in $\mathcal B$. The columns of $\mathcal B$ lie strictly between $\max\{y,0\}$ and the column of the pile, because $b-2>y$ and $b-2\ge1$, so \Cref{lem:moves}(\labelcref{mv:sow}) shows that they are covered.

\emph{Item~\labelcref{bf:covered}.} By \Cref{lem:moves}(\labelcref{mv:covered}), a covered target is not captured, and it stays covered until Bob empties the pit opposite it. A quiet move does not do this (item~\labelcref{bf:quiet}), so the move that does it is not quiet. The target is then exposed, and Alice saves it before Bob moves again.
\end{proof}

\begin{proof}[Proof of \Cref{lem:deadlines}]
\emph{Item~\labelcref{dl:lower}.} We show by induction that the left triple has slack at least one at every Bob turn from $t$ on while a left target is pending. Then no pending left target is covered, by \Cref{lem:band-facts}(\labelcref{bf:below}) with $Y=\Omega$, so Bob's move exposes none, and it captures none by \Cref{lem:band-facts}(\labelcref{bf:slack}). If the move is quiet, or empties the pit opposite a right target, it leaves $F_{\mathrm{left}}$ unchanged (\Cref{lem:band-facts}(\labelcref{bf:quiet}),(\labelcref{bf:upper})), and Alice's answer, which changes no Bob pile, does not lower the slack. Any other move exposes nothing, since a left target was not covered and a right target is exposed only when its pit is emptied. It lowers the slack by at most one (\Cref{lem:band-facts}(\labelcref{bf:slack})), and Alice saves $\hat a$, which lowers $\hat a$ by at least one.

\emph{Item~\labelcref{dl:upper}.} We show by induction that the right triple has slack at least one outside $\Lambda$ at every Bob turn while no pile of $\Lambda$ has fired and a right target is pending. Then $\hat b\ge b-2$, and every pending band target lies below $F_{\mathrm{right}}(\Lambda)\le E_{\Lambda\cup\Omega}$. By \Cref{lem:band-facts}(\labelcref{bf:below}) no pending band target is covered, so Bob's move exposes none and the cover rule does not apply. If the move is quiet, it leaves $F_{\mathrm{right}}(\Lambda)$ unchanged (\Cref{lem:band-facts}(\labelcref{bf:quiet})), and Alice's answer does not lower the slack. If it fires no pile of $\Lambda$, it captures no pending right target and lowers the slack by at most one (\Cref{lem:band-facts}(\labelcref{bf:slack})), and Alice saves $\hat b$.
\end{proof}

\begin{proof}[Proof of \Cref{lem:first}]
Let $k$ be $k_{\mathcal B}$ for item~\labelcref{first:upper} and $k_{\mathcal A}$ for item~\labelcref{first:lower}, and let $t_1$ be the Bob turn just after Alice's $k$-th policy move. By (S4) every Bob move that is not quiet is followed by a policy move, so at most $k$ such moves are made from $t_0$ to $t_1$. Quiet moves change neither slack (\Cref{lem:band-facts}(\labelcref{bf:quiet})), and by (S4) Alice's answers to them change no Bob pile, since her policy moves are saves.

\emph{Item~\labelcref{first:upper}, the left triple.} By \Cref{lem:band-facts}(\labelcref{bf:slack}) the slack of the left triple drops by at most $k$ from $t_0$ to $t_1$, so it is at least one at every Bob turn in that period. So no pending left target is covered or exposed (\Cref{lem:band-facts}(\labelcref{bf:below})), or captured (\Cref{lem:band-facts}(\labelcref{bf:slack})), before $t_1$. Consequently each of the $k$ policy moves before $t_1$ saves a right target, by the cover rule or as the highest pending one, and the right triple is saved by $t_1$. From $t_1$ on the policy applies the cover rule and otherwise saves the highest pending left target, so \Cref{lem:deadlines}(\labelcref{dl:lower}) protects the left triple.

\emph{Item~\labelcref{first:upper}, the right triple.} If every pending right target is covered at $t_0$, \Cref{lem:band-facts}(\labelcref{bf:covered}) protects them. Otherwise no pile of $\Lambda$ has fired, and \Cref{lem:deadlines}(\labelcref{dl:upper}) protects the right targets until a pile of $\Lambda$ fires. That firing captures no right target and covers the pending ones, and \Cref{lem:band-facts}(\labelcref{bf:covered}) protects them from then on.

\emph{Item~\labelcref{first:lower}, the left triple.} The policy applies the cover rule and otherwise saves the highest pending left target, so \Cref{lem:deadlines}(\labelcref{dl:lower}) protects the left targets and shows that none is exposed while pending.

\emph{Item~\labelcref{first:lower}, the right triple.} If every pending right target is covered at $t_0$, \Cref{lem:band-facts}(\labelcref{bf:covered}) protects them. Otherwise no pile of $\Lambda$ has fired. If a pile of $\Lambda$ fires at any point, it covers every pending right target, and \Cref{lem:band-facts}(\labelcref{bf:covered}) protects them from then on. So it is enough to consider the time before such a firing. By \Cref{lem:band-facts}(\labelcref{bf:slack}) the slack of the right triple outside $\Lambda$ drops by at most $k$ from $t_0$ to $t_1$, so it is at least one at every Bob turn in that period. So no pending right target is covered or exposed (\Cref{lem:band-facts}(\labelcref{bf:below})), or captured (\Cref{lem:band-facts}(\labelcref{bf:slack})), before $t_1$, and each of the $k$ policy moves before $t_1$ saves a left target. So the left triple is saved by $t_1$, and from then on the policy applies the cover rule and otherwise saves the highest pending right target, so \Cref{lem:deadlines}(\labelcref{dl:upper}) applies.
\end{proof}

\begin{proof}[Proof of \Cref{lem:upper-fired}]
Call a Bob move from a column above $y_0$ a \emph{left-triple move}, and any other Bob move, including a pass, a \emph{right-triple move}. The name indicates which triple the last clause of the policy gives priority. Before a pile of $\Lambda$ fires, every save that Alice makes in the band saves the highest pending target of a triple, because the cover rule does not apply then (shown below), and by (S4) her other answers to quiet moves save no band target. The piles of $\Lambda$ keep landing at most at $a+H+4$ (\Cref{lem:moves}(\labelcref{mv:frontier})).

We show that the following hold until a pile of $\Lambda$ fires or one of the triples has no pending target, and also at the Bob turn when this first happens.
\begin{enumerate}
\item[(i)] The left triple has slack at least one.
\item[(ii)] The right triple has slack at least one outside $T$.
\item[(iii)] If $R'$ is present and unfired, $e(R')\ge\hat b+2+k$, where $k$ is the number of pending left targets.
\end{enumerate}
These hold at $t_0$.

\emph{Safety.} Suppose (i)--(iii) hold at a Bob turn at which both triples have pending targets. By \Cref{lem:band-facts}(\labelcref{bf:slack}), Bob's move captures no pending left target, and if it fires no pile of $T$ it captures no pending right target. If it fires $R'$, it lands at $e(R')\ge\hat b+3$, above every pending right target, and if it fires a pile of $\Lambda$, it lands at most at $a+H+4<b-2$. So it captures no pending band target. Every pending band target lies below $F_{\mathrm{right}}(T)\le E_{T\cup\Omega}$, so by \Cref{lem:band-facts}(\labelcref{bf:below}) no pending band target is covered. So none is exposed, and the cover rule does not apply.

\emph{Right-triple moves.} A right-triple move that is not quiet selects a column below $y_0$, or fires $R'$, which also lies below $y_0$, so it adds no stones to the piles of $\Lambda$. Every Bob pile outside $\Omega\cup \Lambda$ lands at or above $b-1$ before the move, by (ii) and (iii), and so at or above $b-2>a+4$ after it (\Cref{lem:moves}(\labelcref{mv:frontier})). Hence the move leaves $F_{\mathrm{left}}$ unchanged. If it does not fire $R'$, it lowers the right slack outside $T$ by at most one (\Cref{lem:band-facts}(\labelcref{bf:slack})). If it fires $R'$, then $R'$ leaves $T$. It landed at $e(R')\ge\hat b+3$, so the piles it creates land at or above $\hat b+2$, and the landing column of every other pile outside $T\cup\Omega$ dropped by at most one. So afterwards $F_{\mathrm{right}}(T)\ge\hat b$ for the new $T$. In both cases Alice saves $\hat b$, which restores (ii). The move adds at most one stone to $R'$, and the save lowers $\hat b$ by at least one, so (iii) still holds, and (i) is unchanged.

\emph{Left-triple moves.} A left-triple move that is not quiet selects a pile at a column above $y_0$ other than a pile of $\Lambda$, and $y_0$ lies above $b+2H+10$ by (S3). If the pile is residue, it holds at most $H$ stones by (S1), so it only changes pits above $y_0-H>b+H+10$, and the residue piles there keep landing columns above $b+10$. If it is an unfired structural pile outside $T\cup\Omega$, it lands above $b+H+10$ by (S3), so it only changes pits above $b+H+10$. The residue piles there keep landing columns above $b+10$, and the unfired structural piles outside $T\cup\Omega$ keep landing columns above $b+H+10$. Either way no affected pile outside $T\cup\Omega$ ends with a landing column of $b+4$ or less. Unaffected piles keep their landing columns, so the right slack outside $T$ does not decrease. The move lowers the left slack by at most one (\Cref{lem:band-facts}(\labelcref{bf:slack})) and adds at most one stone to $R'$. Alice saves $\hat a$, which restores (i) and lowers $k$ by one, so (iii) still holds.

\emph{Quiet moves.} A quiet move changes no unfired structural pile (\Cref{lem:quiet}) and neither slack (\Cref{lem:band-facts}(\labelcref{bf:quiet})). Alice's answer either saves the highest pending target of a triple, which only raises a slack and keeps (iii), or changes no Bob pile and saves no band target.

\emph{Endings.} If the right triple is saved first, the left triple has slack at least one by (i), every later policy move saves the highest pending left target while one remains, and \Cref{lem:deadlines}(\labelcref{dl:lower}) protects the left triple, also after a pile of $\Lambda$ fires.

If the left triple is saved first, then $k=0$, so by (iii) an unfired $R'$ lands at or above $\hat b+2$, and with (ii) the right triple has slack at least one outside $\Lambda$. Every later policy move saves the highest pending right target, and \Cref{lem:deadlines}(\labelcref{dl:upper}) protects the right triple until a pile of $\Lambda$ fires. That firing covers the remaining right targets (\Cref{lem:band-facts}(\labelcref{bf:lower})), and \Cref{lem:band-facts}(\labelcref{bf:covered}) finishes them.

If a pile of $\Lambda$ fires while both triples are pending, then by (i) it lands above $\hat a$, since it is not omitted, and it lands at most at $a+H+4<b-2$. So it captures no pending band target and covers every pending right target (\Cref{lem:band-facts}(\labelcref{bf:lower})). It lowers the left slack by at most one and exposes nothing, so Alice saves $\hat a$ and the left slack is again at least one. At the next Bob turn \Cref{lem:first}(\labelcref{first:lower}) applies, because every pending right target is covered, and its left-first policy is the policy stated above.
\end{proof}

\begin{proof}[Proof of \Cref{thm:writer}]
Clearing the guard changes no Bob pile. Bob's reply to the clearing captures no target of the band, because the targets are not covered, the timers land above $a$ and outside the triples by (W2), and all other piles that may move land above $b+2H+10$ by (W3) and (S2).

\emph{Bob fires $V=L$.} By (W2), $L$ lands strictly between the triples, so it captures nothing in the band, and it covers the three right targets (\Cref{lem:band-facts}(\labelcref{bf:lower})). Its new piles have landing columns at least $a+1$, and every other pile not in $\Omega$ lost at most one, so $E_\Omega\ge a+1$. Alice saves $a$. At the next Bob turn the pending left targets are $a-2$ and $a-1$, so the left triple has slack at least $(a+1)-(a-1)=2$, and every pending right target is covered. So \Cref{lem:first}(\labelcref{first:lower}) applies.

\emph{Bob fires $V=R$.} By (W2), $R$ lands above $b$, so it captures and covers nothing in the band. Its new piles have landing columns at least $b+1$, and every other pile outside $T\cup\Omega$ still lands above $b+2H+9$. The timers $L$, $L'$, $R'$ have lost at most one each, so $e(L)\ge a+1$, $e(L')\ge a+4$ and $e(R')\ge b+4$, and their upper bounds from (W2) still hold. Alice saves $a$. At the next Bob turn every pile not in $\Omega$ lands at or above $a+1$, so the left triple has slack at least $(a+1)-(a-1)=2$. Every pile outside $\{L,L',R'\}\cup\Omega$ lands at or above $b+1$, so the right triple has slack at least one outside $\{L,L',R'\}$. Finally, $e(R')\ge b+4=\hat b+2+2$, with two pending left targets. By (W1), $R'$ lies below $L'$ and $L$. So the hypotheses of \Cref{lem:upper-fired} hold with $\Lambda=\{L,L'\}$ and $T_{\mathrm{right}}=\{R'\}$.

\emph{Bob does not fire $V$.} Then $V$ is still an unfired pile with $B_V>0$, Alice's pit at $V$ is still empty, and $A_w=V-w$ by the definition of a prepared choice. So the shot lands its last stone in the empty pit $V$ and captures $V$ (\Cref{lem:moves}(\labelcref{mv:capture})). It changes no other Bob pile. Let $t_s$ be the time just after the shot. At that time Bob has made one move since $t_0$, and no target has been saved.

If $V=L$, we apply \Cref{lem:first}(\labelcref{first:upper}) at $t_s$, with three pending right targets. At $t_0$ every pile outside $\Omega\cup\{L\}$ lands at or above $a+5$, by (W2) and (W3). The reply did not fire $L$, so $E_{\Omega\cup\{L\}}\ge a+4$ after it (\Cref{lem:moves}(\labelcref{mv:frontier})), and once $L$ is captured this is $E_\Omega$. So the left triple has slack at least $(a+4)-a=4$. If $L'$ has not fired, every pile outside $\Omega\cup\{L,L'\}$ landed at or above $b+2$ at $t_0$, so the right triple has slack at least $(b+1)-b=1$ outside $\Lambda=\{L'\}$ at $t_s$. If $L'$ fired as the reply, then $\Lambda$ is empty and the right targets are covered (\Cref{lem:band-facts}(\labelcref{bf:lower})). The bound $e(L')\le a+H+4$ comes from (W2) and \Cref{lem:moves}(\labelcref{mv:frontier}).

If $V=R$, we apply \Cref{lem:first}(\labelcref{first:lower}) at $t_s$, with three pending left targets. At $t_0$ every pile outside $\Omega\cup\{R\}$ lands at or above $a+2$, so after the reply and the capture of $R$ we have $E_\Omega\ge a+1$, and the left triple has slack at least one. If neither $L$ nor $L'$ has fired, every pile outside $\Omega\cup\{R,L,L'\}$ landed at or above $b+5$ at $t_0$, so the right triple has slack at least $(b+4)-b=4$ outside $\Lambda$ at $t_s$. If one of them fired as the reply, it covered the right triple (\Cref{lem:band-facts}(\labelcref{bf:lower})), because it lies above $b$ and lands at most at $a+H+4<b-2$.

In every case the continuation saves each of the six targets with one policy move, so Alice makes two preparation moves and six saves, or one preparation move and six saves when Bob fires $V$.
\end{proof}

\begin{proof}[Proof of \Cref{thm:reader-accept}]
Clearing the guard at $R$ changes no Bob pile. Bob's reply to the clearing captures no target of the band. By (R2), $L$ lands strictly between the triples and $R$ lands above them, and every other pile that may move lands above $b+2H+10$ by (R3) and (S2).

\emph{Bob fires $R$.} It lands at $e(R)\ge b+1$, above the band, so it captures and covers nothing in the band. Since $R=L+1$, its first stone falls on $L$, so $e(L)\ge a+1$, and still $e(L)\le a+H+1$. Its new piles have landing columns at least $b$, and every other pile outside $T\cup\Omega$ still lands above $b+2H+9$. Alice saves $b$. At the next Bob turn every pile not in $\Omega$ lands at or above $a+1$, so the left triple has slack at least one. Every pile outside $\{L\}\cup\Omega$ lands at or above $b$, while the highest pending right target is $b-1$, so the right triple has slack at least one outside $\{L\}$. So the hypotheses of \Cref{lem:upper-fired} hold with $\Lambda=\{L\}$ and $T_{\mathrm{right}}$ empty.

\emph{Bob does not fire $R$.} Then $R$ is still an unfired pile with $B_R>0$, Alice's pit at $R$ is still empty, and $A_x=R-x$ by the definition of an input. So the shot from $x$ lands its last stone in the empty pit $R$ and captures $R$ (\Cref{lem:moves}(\labelcref{mv:capture})), and it changes no other Bob pile. Let $t_s$ be the time just after it. We apply \Cref{lem:first}(\labelcref{first:lower}) at $t_s$, with three pending left targets. At $t_0$ every pile outside $\Omega\cup\{R\}$ lands at or above $a+2$, so $E_\Omega\ge a+1$ at $t_s$, and the left triple has slack at least one. If $L$ has not fired, every pile outside $\Omega\cup\{L,R\}$ landed above $b+2H+10$ at $t_0$, so the right triple has slack $(b+4)-b=4$ outside $\{L\}$ at $t_s$. If $L$ fired as the reply, it covered the right triple by \Cref{lem:band-facts}(\labelcref{bf:lower}), since it lies above $b$ and lands at most at $a+H+1<b-2$. In both cases Alice makes one preparation move, at most one shot, and one save per target.
\end{proof}

\section{Proofs for false formulas}
\label[appendix]{app:false-proofs}

This appendix proves the lemmas of \Cref{sec:false}.

\begin{proof}[Proof of \Cref{lem:calibration}]
We first derive item~\labelcref{cal:clocks} from item~\labelcref{cal:moves}. Let $u_i$ be the number of universal writers $i'\le i$. Bob has fired $8i-u_i$ clocks, and Alice has captured none, since that needs a move that selects the pit opposite the clock. With $J=8n+7-n_\forall$, this leaves $8(n-i)+7-(n_\forall-u_i)$ unfired clocks. The universal writers not counted by $u_i$ are among the $n-1-i$ writers after $i$, so this number is at least $7(n-i)+8$ if $i<n$, and it is $7$ if $i=n$. Before Bob's $\tau$-th move of the stage, Bob has fired at most $\tau-1$ clocks during it, and Alice, who has made $\tau$ moves of the stage, has captured at most $\lfloor \tau/2\rfloor$ clocks, since each capture of a clock needs a move that selects the pit opposite it and a move that lands there. Empty capture does not shorten this preparation, since it removes only a last singleton from a pit that was empty before it arrived, and so cannot clear the positive guard of a clock. So at least $15-7-4\ge1$ clocks are unfired for $\tau\le8$ at a writer, and at least $7-(\tau-1)-\lfloor \tau/2\rfloor\ge1$ for $\tau\le5$ at the reader.

The other items are proved by induction on $i$. For $i=0$ they hold by construction. Assume them at the start of writer $i'$, and suppose Alice plays an intended branch in the intended order at writer $i'$. Write $a=a_{i'}$ and $b=b_{i'}$.

\emph{Bob's moves.} Bob's rule~1 never applies during the writer. The timers of earlier writers land within $H$ of their own bands (\Cref{lem:crossings}), whose targets are all saved, and other bands are more than $W-2H$ away. The timers of later writers and of the reader land at no target, by item~\labelcref{cal:timers}, since at most eight Bob moves of the writer are made. Their left timers land between the two triples of their bands, and their right timers above both. The timers of writer $i'$ land at $a+2-\tau$, $b+2-\tau$, $a+5-\tau$ and $b+5-\tau$ after Bob's $\tau$-th move of the writer, and in the intended order Alice saves each target of the band before the Bob move on which an uncaptured timer would land on it (\Cref{fig:writer-timeline}). The pits of captured timers, of this writer and of earlier ones, refill only with stones from clock and selector firings, at most $8n+8<H$ of them, so the piles there land more than $H$ above every band. Alice's moves are a clearing, a shot and six saves. By \Cref{lem:standard}(\labelcref{sp:clear}), whose hypotheses hold by item~\labelcref{cal:moves}, the clearing captures nothing, the shot captures its timer, and saves capture nothing. So she selects only a guard, a weapon and targets of writer $i'$, and her only capture is of a timer. By item~\labelcref{cal:clocks}, rule~3 always finds an unfired clock. Since no clock has been captured, the highest unfired clock has received one stone from each higher clock, so it lands at $g$, where Bob's pit is nonempty, and it captures nothing.

\emph{Timers and bands.} A clock or selector firing crosses every unfired timer (\Cref{lem:crossings}), so each of Bob's moves adds one stone to each timer in item~\labelcref{cal:timers}, which proves that item. Bob's moves land at $g$ or at weapons, so no Bob move reaches a band, and Alice saves only targets of writer $i'$, which proves item~\labelcref{cal:bands}.

\emph{Selectors.} Consider a universal writer $i''$ and its two selectors. The clocks lie above every selector, and the selectors of earlier writers lie above those of writer $i''$ and land at their own weapons, below every selector. So each of Bob's first $8i''-1$ moves crosses both selectors of writer $i''$, and before his move $8i''$ each of them lands exactly at its weapon. The weapon is not covered, because clocks land at $g$ and the selectors of earlier writers land at higher weapons. The weapon pile is also nonempty. So the selector chosen by rule~2 captures its weapon on move $8i''$. This proves item~\labelcref{cal:moves}.

\emph{Records.} The first clock firing crosses every record and dummy record and covers it (\Cref{lem:crossings}). Bob never selects a pit at a record, and Alice captures only timers, so these pits stay nonempty.
\end{proof}

\begin{proof}[Proof of \Cref{lem:forced-writer}]
We assume that Bob captures no pending target during the writer, so his rule~1 never applies. Rule~2 applies at most on his eighth move of the writer, and by \Cref{lem:calibration}(\labelcref{cal:clocks}) rule~3 always finds a clock. So each of his moves of the writer fires a clock or a selector and adds one stone to every unfired timer (\Cref{lem:crossings}). With the landing columns of \Cref{lem:calibration}(\labelcref{cal:timers}), this gives the dates of \Cref{tab:due}. The Bob pits opposite the band are empty, because no Bob move reaches the band. So if Alice does not save a target on the move on which it is due, Bob captures it by rule~1 on his next move, because she cannot add stones to Bob's row, and she cannot empty a target without saving it.

Under either capture convention, the removal of a last singleton cannot clear a positive guard, remove a positive Bob timer, or save a pending target without selecting it. Thus the same due dates apply even to arbitrary Alice moves. Capturing $L_i$ or $R_i$ takes two moves of Alice, one that clears its guard and a later one, the shot, whose last stone lands there. No save may come in between, because a save adds a stone to the guard. Let $\tau$ be the move on which Alice first captures $L_i$ or $R_i$, with $\tau=\infty$ if she never does. We count the moves that she needs before move $\tau$.
\begin{itemize}
\item If $\tau\ge6$, the six targets due on moves $3$, $4$ and $5$ must all be saved by move $5$, which is six saves in five moves.
\item If $\tau\in\{4,5\}$, say with $R_i$ captured, Alice must save the $2(\tau-3)$ targets due on moves $3,\ldots,\tau-1$ and the target that $L_i$ makes due on move $\tau$, and she must clear the guard. These are $2\tau-4$ moves among her first $\tau-1$, which is too many.
\item If $\tau=3$, say with $R_i$ captured, then $a$ is due on move $3$, so moves $1$ and $2$ are the save of $a$ followed by the clearing. By \Cref{lem:standard}(\labelcref{sp:timers}), whose hypotheses hold by \Cref{lem:calibration}, at the start of the writer only $w^R_i$ lands at $R_i$ and no pile lands at $R_i-1$. The save of $a$ adds one stone to every pile above $a$, so afterwards no pile below $R_i$ lands at $R_i$. Bob's clocks change no Alice pile, and the clearing adds stones only above $R_i$. So the shot on move $3$ is impossible. Likewise for $L_i$, with $b$ in place of $a$.
\item $\tau=1$ is impossible, because the guards are nonempty at the start of the writer (\Cref{lem:standard}(\labelcref{sp:counts})).
\end{itemize}
So $\tau=2$. Alice clears a guard on move $1$ and captures the opposite timer on move $2$. By \Cref{lem:standard}(\labelcref{sp:timers}) the only pile that lands there is the dedicated weapon, so it was not captured by a selector, and by \Cref{lem:standard}(\labelcref{sp:clear}) the clearing captures nothing. A guard is emptied only by selecting it, so the clearing empties no other guard.

After the capture, say of $R_i$, \Cref{tab:due} leaves one target due on each of the moves $3$ to $8$, namely $a$, $a-1$, $a-2$ from $L_i$ and $b$, $b-1$, $b-2$ from $R'_i$, provided Alice saves each on time, and the targets that $L'_i$ reaches are saved by then. Removing $L_i$ or $R'_i$ instead of saving would take two moves while a target is due. So moves $3$ to $8$ are the saves in the intended order. Capturing $L_i$ instead exchanges the roles of the triples, with the backup $L'_i$.
\end{proof}

\begin{proof}[Proof of \Cref{lem:reader-reject}]
By \Cref{lem:calibration} with $i=n$, the reader timers land at $a_\star+2$ and $b_\star+1$, and the reader targets are pending and uncovered. By \Cref{lem:standard}(\labelcref{sp:counts}), the reader's guards hold at least $D$ stones. Write $a=a_\star$, $b=b_\star$. By \Cref{lem:calibration}(\labelcref{cal:clocks}), Bob's first three moves of the reader are clock firings unless rule~1 applies, and each adds one stone to both reader timers, so after his $\tau$-th move they land at $a+2-\tau$ and $b+1-\tau$.

Neither timer can be captured on Alice's first move, because its guard is nonempty. To survive Bob's second move, Alice must save $b$ with one of her first two moves, unless she captures $R_\star$ on her second move. Suppose she does not capture $R_\star$ on move two. Then one of her first two moves saves $b$, and the other neither captures a timer nor saves two targets. If the other move cleared a guard, then after Bob's second move both $a$ and $b-1$ are threatened, and her third move can deal with only one of them. If the other move saved $a$ or $b-1$, her third move must save the remaining one of these, no guard has been cleared, and after Bob's third move both $a-1$ and $b-2$ are threatened. Otherwise both $a$ and $b-1$ are threatened after Bob's second move and she has cleared no guard. In every case Bob captures a target.

So Alice must clear the guard at $R_\star$ on move one and capture $R_\star$ on move two. Clearing that guard changes only pits above $R_\star$, and Bob's clock changes no Alice pile. So the capture needs an Alice pile below $R_\star$ that landed exactly at $R_\star$ at the start of the reader.
\end{proof}

\section{Proofs for true formulas}
\label[appendix]{app:true-proofs}

This appendix proves the lemmas of \Cref{sec:true}, as part of the simultaneous induction described there, and \Cref{prop:capture-robustness}.

\begin{proof}[Proof of \Cref{lem:bookkeeping}]
\emph{Item~\labelcref{bk:short}.} Each prescribed move is a save of a pending target, a clearing or a shot, and Alice clears at most one guard and shoots at most once per stage. There are $6n+6$ targets and $n+1$ stages, so Alice makes at most $8n+8$ moves before every target is saved, and Bob moves once after each of them. A Bob move adds at most one stone to each pit other than the one it selects, and Alice never adds stones to Bob's row. The only initial Bob pile that is not structural is the single stone at $g$. So the pit of a residue pile either held that stone and has never been emptied, or was empty at some time after which the current pile accumulated. In both cases the pile holds at most $\tau+1$ stones after $\tau$ Bob moves, and $8n+9\le H$. Removing a last singleton under the empty-capture rule can only lower this bound.

\emph{Item~\labelcref{bk:untouched}.} A firing of a timer of a stage $j$ or later while the active stage is smaller is a jump, and rule~(J) then makes that stage active, which gives the first claim. For the second, let $\Omega_{-1}$ be the set of all timers. The timers of stages $j$ and later that are still on the board are unfired, since Alice's captures only remove piles, so the frontier $E_{\Omega_{j-1}}$ loses at most one per Bob move by \Cref{lem:moves}(\labelcref{mv:frontier}). Initially the piles outside $\Omega_{j-1}$ are the timers of stages before $j$, the clocks, the selectors and the stone at $g$, and they land at or above $a_{j-1}+2$ if $j\ge1$, and at or above $w^R_{n-1}$ if $j=0$. After at most $H-1$ Bob moves (item~\labelcref{bk:short}) they land above $a_{j-1}+2-H\ge b_j+W+2-H> b_j+2H+10$, respectively above $b_0+3H$, by \Cref{lem:layout}. Unfired timers lie above the weapons. A nonempty Bob pit at a column at most $b_j+2H+10$ would be a pile outside $\Omega_{j-1}$ with a smaller landing column, which is impossible. The same holds at every earlier time, and a capture at a column needs a pile landing there, so no target of a stage $j$ or later has been captured.

\emph{Item~\labelcref{bk:charging}.} Preparations happen only at the active stage. A save by the stage policy saves a target of the active stage, a save by rule~(P) after the reader saves a target of an earlier stage, and a save by the cover rule saves an exposed target. A target of a stage after the active one is never covered, by item~\labelcref{bk:untouched}, because a firing of a timer of its stage or a later one would have made that stage active. So it cannot be exposed. A stage receives at most two preparation charges and six save charges. Before stage $j$ is active, all charges go to the $j$ stages before it. Bob's moves alternate with Alice's, Alice moving first, which gives the last two claims.
\end{proof}

\begin{proof}[Proof of \Cref{lem:active}]
\emph{Item~\labelcref{act:earlier}.} A stage stops being active either because all its targets are saved, or through a jump to a later stage $s'$. The jumping timer crosses every target of every stage before $s'$ (\Cref{lem:crossings}), so after the jump every pending target of those stages is covered. By \Cref{lem:moves}(\labelcref{mv:covered}), such a target stays covered until Bob empties the pit opposite it. It is then exposed, and rule~(C) saves it at Alice's next turn, since rule~(J) applies only after a firing, which exposes nothing. The covers are strict-interior deposits of the jumping timer. An empty capture removes only a newly arrived last singleton and cannot remove an existing cover.

Let $z$ be such a target. Then $z\ge a_{s-1}-2\ge b_s+W-2$, so $z-2H>b_s+H+6$. The pit opposite $z$ holds residue, and no unfired structural pile lies at a column in $[z-H,z-1]$, since all of them lie above the weapons. So the move is $(b_s+H+6)$-quiet.

\emph{Item~\labelcref{act:standing}.} (S1) is \Cref{lem:bookkeeping}(\labelcref{bk:short}). (S2) holds while $s$ is active, because a firing of a pile of $\Omega_s$ is a jump. For (S3), every unfired structural pile is a clock, a selector or a timer, and all of them lie above the weapons, which lie more than $4H$ above $b_0\ge b_s$. The ones outside $T\cup\Omega_s$ are the clocks, the selectors and the timers of earlier stages. By \Cref{lem:crossings}, clocks land at or above $g-H$, selectors within $H$ of their weapons, and every timer of an earlier stage $s'$ has landing column at least $a_{s'}-H\ge b_s+W-H$. All of these are above $b_s+H+10$. For (S4), after a quiet Bob move, Alice either saves an exposed target of an earlier stage, which changes no Bob pile and saves no target of the band, or follows the policy. After any other Bob move that does not end the activity of $s$, no target of an earlier stage is exposed (item~\labelcref{act:earlier}), no target of a later stage is exposed, since those are uncovered (\Cref{lem:bookkeeping}(\labelcref{bk:untouched})), and rule~(J) does not apply, so she follows the policy.
\end{proof}

\begin{proof}[Proof of \Cref{lem:computing}]
In computing mode there has been no pass, no residue move and no jump. Writer $0$ began with its initial entry, and each subsequent stage became active by an advance, after all targets of the previous stage were saved. No target of an earlier stage is pending. Every Bob move so far fired an unfired structural pile, namely a clock, a selector, or a timer of the active writer or an earlier one. An exposure needs Bob to empty the pit opposite a target, which holds residue, because every structural pile lies above the bands. So no target has been exposed, and Alice has made no cover-rule save.

\emph{Item~\labelcref{cm:history}.} A writer entered in computing mode follows \Cref{thm:writer}. Bob did not fire $V$ right after the clearing, so Alice shot and then made six policy saves. So every earlier writer made a clearing, a shot and six saves, and Bob made exactly $8j$ moves before a writer $j$ became active. Bob's moves land within $H$ of $g$, of a weapon, or of a band (\Cref{lem:crossings}), and so do his captures, all at columns at most $g+J$. Alice's moves so far are intended branches, writer by writer. So the hypotheses of \Cref{lem:standard} hold, and by its item~\labelcref{sp:clear} her clearings capture nothing, and neither do her saves.

\emph{\Cref{cm:selector,cm:covered}.} By \Cref{lem:crossings}, clocks land at or above $g-H$, timers land within $H$ of their bands, and a selector lands within $H$ of its own weapon. By \Cref{lem:layout}, weapons of different writers are more than $2H$ apart, and the two weapons of one writer are $\gamma>H$ apart. All weapons lie more than $4H$ below $g$ and more than $4H$ above the bands. So only a weapon's own selector, if it has one, can land at that weapon. In particular, no weapon of an existential writer can be captured. For a universal writer, let $S$ be the own selector of a weapon $w$.

Every Bob move from a column above $S$ crosses $S$, because it lands below every selector. Let $\mu$ be the number of Bob moves before move number $\tau$ that came from columns below $S$. Then before move $\tau$, $S$ has gained $\tau-1-\mu$ stones and lands at $w+8i-\tau+\mu$, where $i$ is its writer. A move from below $S$ is a selector of a later writer, the other selector of the same writer when $S=S^L_i$, or a timer. Each of them starts above $w$ and lands below $w$, so it puts a stone on Bob's pit at $w$. That pit is never emptied afterwards, because Bob selects no residue, and Alice's captures are at the columns where her shots end, which are timer columns, since her clearings and saves capture nothing (item~\labelcref{cm:history}). This proves item~\labelcref{cm:covered}. So a capture of $w$ needs $\mu=0$ and $\tau=8i$, which is the first claim of item~\labelcref{cm:selector}.

If $S^R_i$ captures $w^R_i$ on move $8i$, then $S^R_i$ lies below $S^L_i$, so every later move of $S^L_i$ has $\mu\ge1$ and $w^L_i$ is covered. If $S^L_i$ captures $w^L_i$ on move $8i$, then a later firing of $S^R_i$ on a move $\tau>8i$ either has $\mu\ge1$, so that $w^R_i$ is covered, or lands at $w^R_i+8i-\tau<w^R_i$ and so captures nothing at $w^R_i$. This proves the rest of item~\labelcref{cm:selector}.

\emph{The first residue move.} By item~\labelcref{cm:history}, Bob has made exactly $8i$ moves before the clearing, so the reply is his move $8i+1$. If Bob's pit at $w$ is nonempty, the reply cannot capture at $w$ (\Cref{lem:moves}(\labelcref{mv:capture})). Otherwise no earlier Bob move crossed $w$, since its strict-interior deposit would have covered $w$, and that cover could be emptied only by a Bob residue move or an Alice capture there, whereas Bob has made no residue move and Alice has captured only timers. We show that no residue pile lands at $w$, so that the reply cannot capture there.

Residue piles above $w+H$ hold at most $H$ stones (\Cref{lem:bookkeeping}(\labelcref{bk:short})), so they land above $w$. Residue piles in $(w,w+H]$ were created by Bob moves that landed in $(w,w+H]$ without crossing $w$. By \Cref{lem:crossings}, clocks land at or above $g-H>w+H$, timers land within $H$ of their bands, far below $w$, and a selector lands within $H$ of its own weapon, while the weapons of other writers are more than $2H$ away from $w$. So the only candidates are the selectors of writer $i$, fired on Bob's moves $\tau\le8i$. At an existential writer there are no such selectors, so $(w,w+H]$ contains no residue. This also covers writer $0$, before which Bob has made no move.

Now let writer $i$ be universal, and let $S$ be the own selector of $w$. By item~\labelcref{cm:covered}, no earlier Bob move came from below $S$. The weapon $w$ holds exactly the positive distance to $V$ (\Cref{def:alice-strategy}), whereas a captured weapon of this writer was captured on Bob's move $8i$ (item~\labelcref{cm:selector}) and is still empty. So $w$ has not been captured, and $S$ did not fire on move $8i$, since with no earlier move from below $S$, it would have landed at $w$ and captured it. The other selector of the pair lands at or above its own weapon on moves $\tau\le8i$. For $w=w^R_i$, $S^L_i$ lands at or above $w^L_i=w+\gamma>w+H$, and for $w=w^L_i$, $S^R_i$ lands at most at $w^R_i+8i<w^L_i$ and would have crossed $w$. So only $S$ itself remains, fired on a move $\tau<8i$ with landing column $w+d$, $d=8i-\tau\ge1$. Bob's pit at $w+d$ was empty and Alice's pit there holds at least $D$ stones, so $S$ captured there and its last stone was removed. Every pit in $(w+d,w+H]$ then holds exactly one stone from $S$, and lands at or above $w+d>w$.
\end{proof}

\begin{proof}[Proof of \Cref{lem:entries}]
\emph{Item~\labelcref{en:computing}.} Write $a=a_j$, $b=b_j$, with $j=0$ for the initial entry. At that initial entry Bob has made no move, and the four timers have their initial landing columns $a_0+2$, $b_0+2$, $a_0+5$ and $b_0+5$. At an entry by advance, \Cref{lem:computing}(\labelcref{cm:history}) gives exactly $8j$ preceding Bob moves, each firing a clock, a selector or a timer of an earlier writer. Each crosses every timer of writer $j$ (\Cref{lem:crossings}), so their landing columns are the initial ones minus $8j$, giving the same four stated values. The conditions of \Cref{def:band-setting} on $a$, $b$, $T$ and $\Omega$, (W1), and the upper bounds in (W2) come from \Cref{lem:layout,lem:crossings}, and (W3) and (W4) come from \Cref{lem:bookkeeping}(\labelcref{bk:untouched}).

The hypotheses of \Cref{lem:standard} hold for the empty history at the initial entry, and otherwise by \Cref{lem:computing}(\labelcref{cm:history}). By its item~\labelcref{sp:counts}, an uncaptured weapon of writer $j$ lands exactly at its timer, and the guards of writer $j$ are nonempty. A captured weapon of writer $j$ was captured on Bob's move $8j$ (\Cref{lem:computing}(\labelcref{cm:selector})), just before the writer started, and is empty. So a weapon of writer $j$ holds exactly the distance to its timer if and only if it is uncaptured. By \Cref{lem:computing}(\labelcref{cm:selector}), at least one is, and both are at an existential writer. By \Cref{lem:standard}(\labelcref{sp:clear}), applied just after the clearing, the clearing captures nothing. Finally, Bob's reply to the clearing either fires $V$ or does not capture the weapon, because a pass captures nothing, a selector cannot capture on move $8j+1$ (\Cref{lem:computing}(\labelcref{cm:selector})), a residue reply spares the weapon by the last claim of \Cref{lem:computing}, and every other move lands within $H$ of $g$, of a band, or of its own weapon (\Cref{lem:crossings}).

\emph{Item~\labelcref{en:escaping}.} Write $a=a_j$, $b=b_j$. By \Cref{lem:bookkeeping}(\labelcref{bk:untouched}) the timers of writer $j$ are unfired, the band is pending and uncovered, and every pile outside $T_j\cup\Omega_j$ lands above $b+2H+10$. By \Cref{lem:bookkeeping}(\labelcref{bk:charging}), Bob has made $\tau\le8j$ moves, and each adds at most one stone to each timer, so
\[
 e(L_j)\ge a+2,\quad e(R_j)\ge b+2,\quad e(L'_j)\ge a+5,\quad
 e(R'_j)\ge b+5.
\]
We claim that $e(L_j)\ge a+3$ or $e(R_j)\ge b+3$. Consider the event that started the escaping mode. If it was a residue move, that pile held at most $H<\gamma=L_j-R_j$ stones, so it added a stone to at most one of $L_j$ and $R_j$, and the other has landing column at least $8j+2-(\tau-1)\ge3$ above its top. If it was a pass, it added a stone to neither, so both have landing columns at least $8j+2-(\tau-1)\ge3$ above their tops. If it was a jump, to some stage before $j$, that stage was charged only saves, at most six. If it was the firing of a prepared timer, that writer was charged at most seven moves. In both cases $\tau\le8j-1$, and both $L_j$ and $R_j$ land at least three above their tops. The upper bounds $e(L_j)\le a+H+1$ and $e(L'_j)\le a+H+4$ come from \Cref{lem:crossings}.

If $e(L_j)\ge a+3$, Alice saves $b$. At the next Bob turn every pile outside $\Omega_j$ lands at or above $a+3$, so the left triple has slack at least $3$, one more than the number of pending right targets. Every pile outside $\Lambda\cup\Omega_j$ lands at or above $b+2$, so the right triple has slack at least one outside $\Lambda$. These are the hypotheses of \Cref{lem:first}(\labelcref{first:upper}). Otherwise $e(L_j)=a+2$ and $e(R_j)\ge b+3$, and Alice saves $a$. At the next Bob turn every pile outside $\Omega_j$ lands at or above $a+2$, so the left triple has slack at least $3$. Every pile outside $\Lambda\cup\Omega_j$ lands at or above $b+3$, so the right triple has slack at least $3$ outside $\Lambda$, one more than the number of pending left targets. These are the hypotheses of \Cref{lem:first}(\labelcref{first:lower}).

\emph{Item~\labelcref{en:jump}.} Write $a=a_{s'}$, $b=b_{s'}$. By \Cref{lem:bookkeeping}(\labelcref{bk:charging}) the firing is Bob's move number at most $8s'$, so at most $8s'-1$ moves precede it, and by \Cref{lem:bookkeeping}(\labelcref{bk:untouched}) the timers of stage $s'$ are unfired. So just before the firing, if $s'<n$,
\[
 e(L_{s'})\ge a+3,\quad e(R_{s'})\ge b+3,\quad e(L'_{s'})\ge a+6,\quad
 e(R'_{s'})\ge b+6,
\]
while for the reader $e(L_\star)\ge a+3$ and $e(R_\star)\ge b+2$. With the upper bounds of \Cref{lem:crossings}, a left timer or left backup lands strictly between the two triples of its band, at least $a+3$ and at most $a+H+4<b-2$, and a right timer or right backup lands above both, at least $b+2$. Every other band is more than $W-2H$ away from these landing columns. So $X$ captures no target. By \Cref{lem:crossings} it crosses every target of every stage before $s'$, covering the pending ones. A firing empties no pit opposite a target, so it exposes nothing. Every pile outside $T_{s'}\cup\Omega_{s'}$ landed above $b+2H+10$ before the firing (\Cref{lem:bookkeeping}(\labelcref{bk:untouched})).

If $X$ is $L_{s'}$, $L'_{s'}$ or $L_\star$, it lands below $b-2$ and covers the right triple (\Cref{lem:crossings}). Its new piles land at or above $a+2$, and the remaining timers of the stage still land at or above $a+3$. Alice saves $a$, and at the next Bob turn the left triple has slack at least $(a+2)-(a-1)=3$ and every pending right target is covered, so \Cref{lem:first}(\labelcref{first:lower}) applies.

If $X$ is $R_{s'}$ or $R'_{s'}$, it lands at or above $b+3$ and creates piles landing at or above $b+2$. It does not cross $L_{s'}$, which lies above $R_{s'}$, and it adds at most one stone to $L'_{s'}$. So $e(L_{s'})\ge a+3$ and $e(L'_{s'})\ge a+5$, and every other pile outside $\Omega_{s'}$ lands at or above $b+2$. Alice saves $b$. At the next Bob turn the left triple has slack at least $3$, one more than the number of pending right targets, and the right triple has slack at least $(b+2)-(b-1)=3$ outside $\Lambda$. These are the hypotheses of \Cref{lem:first}(\labelcref{first:upper}).

If $X=R_\star$, it adds one stone to $L_\star=R_\star-1$, so $a+2\le e(L_\star)\le a+H+1$, and its new piles land at or above $b+1$. Alice saves $b$. At the next Bob turn the left triple has slack at least $(a+2)-a=2$, the right triple has slack at least $(b+1)-(b-1)=2$ outside $\{L_\star\}$, and no band target is covered, so the hypotheses of \Cref{lem:upper-fired} hold with $\Lambda=\{L_\star\}$ and $T_{\mathrm{right}}$ empty.

\emph{Item~\labelcref{en:reader}.} Write $a=a_\star$, $b=b_\star$. By \Cref{lem:bookkeeping}(\labelcref{bk:untouched}) both reader timers are unfired. Let $\tau\le8n$ be the number of Bob moves so far (\Cref{lem:bookkeeping}(\labelcref{bk:charging})), and $I_L$, $I_R$ the numbers of stones that $L_\star$ and $R_\star$ have received. A move that adds a stone to $L_\star$ comes from a column above $L_\star$ other than $R_\star$, so it also crosses $R_\star=L_\star+1$. Both timers are still positive and unfired, so these increments are retained under either capture convention. Hence $I_L\le I_R\le\tau\le8n$, and $\Delta_L=8n-I_L\ge8n-I_R=\Delta_R\ge0$.

\emph{Case $\Delta_L\ge1$.} Then $e(L_\star)\ge a+3$ and $e(R_\star)\ge b+1$, and every pile outside $\{L_\star,R_\star\}$ lands above $b+2H+10$ (\Cref{lem:bookkeeping}(\labelcref{bk:untouched})). After Alice saves $b$, at the next Bob turn the left triple has slack at least $3$, one more than the number of pending right targets, and the right triple has slack at least $(b+1)-(b-1)=2$ outside $\{L_\star\}$. Finally, $e(L_\star)\le a+H+1$ by \Cref{lem:crossings}.

\emph{Case $\Delta_L=0$, the history.} Then $I_L=\tau=8n$. So Alice has made $8n$ moves, and every Bob move so far came from a column above $L_\star$. In particular Bob has not passed, since a pass adds no stone to $L_\star$. So no writer timer has fired, since writer timers lie below $L_\star$. So there was no jump to a writer and no firing of a prepared timer, and every residue move came from above $L_\star$. Since each writer is charged at most eight moves, each was charged exactly eight, so each made a clearing and a shot. A writer entered in escaping mode makes no preparation, so every writer was entered in computing mode. No target was ever exposed, since that needs a Bob move from a band column, below $L_\star$. So Alice played, writer by writer, the moves of \Cref{thm:writer} in the branch where Bob does not fire the chosen timer, namely a clearing, a shot and six saves, which are the moves of an intended branch. At an existential writer both weapons were available (\Cref{lem:computing}(\labelcref{cm:selector})) and the bit is $\alpha$'s, and at a universal writer the bit is the one Bob left or, if he left both, bit~$0$. So the bits are a play of the assignment game in which Alice follows $\alpha$.

\emph{Case $\Delta_L=0$, captures.} Bob's first move fired a clock or a selector, since at the start the only piles above $L_\star$ are clocks, selectors and $R_\star$, which has not fired. It crossed every Bob pit from $g+J+1$ up to $L_\star-1$ and made it nonempty by strict-interior deposits, which are unaffected by empty capture. We show by induction over the play that every capture by Bob is at a column at most $g+J$ or above $R_\star$, and that every capture by Alice is at a writer timer column. Then the pits from $g+J+1$ to $L_\star-1$ other than writer timer columns are never emptied, since emptying one needs a Bob move that selects it, from below $L_\star$, or an Alice capture there. Every Bob move fires a clock, which lands in $[g-H,g+J]$, or a selector, which lands within $H$ of its weapon (\Cref{lem:crossings}), or is a residue move from above $L_\star$. The pile of such a move holds at most $H$ stones (\Cref{lem:bookkeeping}(\labelcref{bk:short})) and so lands above $L_\star-H$, which is above every record and every writer timer column (\Cref{lem:layout}). Such a residue move lands above $R_\star$, or in one of the nonempty pits just described, or at $L_\star$ or $R_\star$, which hold unfired piles, and in the latter cases it captures nothing (\Cref{lem:moves}(\labelcref{mv:capture})). Alice's saves capture nothing, and her shots end at writer timer columns. By the induction hypothesis Bob has captured at no column in $(g+J,R_\star]$, so by \Cref{lem:standard}(\labelcref{sp:clear}), applied just after a clearing, the clearing captures nothing. In particular the hypotheses of \Cref{lem:standard} hold at the start of the reader, and every record and dummy record is covered.

\emph{Case $\Delta_L=0$, the reader.} By \Cref{lem:standard}(\labelcref{sp:records}) and \Cref{lem:assignment-game}, some record or dummy record lands exactly at $R_\star$, because $\alpha$ wins the assignment game. The reader setting holds. The conditions of \Cref{def:band-setting}, (R1) and the upper bounds in (R2) hold by \Cref{lem:layout,lem:crossings}, the exact landing columns by $\Delta_L=\Delta_R=0$, and (R3) and (R4) by \Cref{lem:bookkeeping}(\labelcref{bk:untouched}). By \Cref{lem:standard}(\labelcref{sp:counts}), the reader's guard at $R_\star$ holds at least $D$ stones, so clearing it sends its last stone off the line and captures nothing. Let $x$ be an Alice pile below $R_\star$ that lands at $R_\star$. By \Cref{lem:standard}(\labelcref{sp:inputs}), $x$ is a record or a dummy record, so Bob's pit at $x$ is nonempty, Bob's reply to the clearing cannot capture at $x$ (\Cref{lem:moves}(\labelcref{mv:covered})), and $x$ is an input.
\end{proof}

\begin{proof}[Proof of \Cref{prop:capture-robustness}]
\Cref{thm:false,thm:true} were proved under either capture convention, with the same coordinates and strategies. We recall why the convention does not affect the arguments for these positions.

By \Cref{lem:moves}(\labelcref{mv:empty}), an empty capture removes only the last newly deposited singleton. It changes neither the opposing row nor any deposit strictly between the source and landing columns, and it removes no pile that was already positive. In the intended play, every guard clearing lands in a pit holding at least $D$ stones (\Cref{lem:standard}(\labelcref{sp:clear})), every shot captures a positive timer, and every save leaves the line. Thus these Alice moves are identical under both conventions, as are their contributions to the record counts.

Against arbitrary Alice play, the clock availability, writer forcing and reader rejection arguments of \Cref{sec:false} already allow removal of the final singleton. A positive guard still requires a clearing before a shot, and the deposits on positive timers are retained. These facts give the same forced intended branches and the same rejection of a false clause.

Against arbitrary Bob play, \Cref{sec:true} applies the frontier and residue bounds to the actual position under the chosen convention. The covers used by the strategy are strict-interior deposits, and the preparation and reader-input arguments protect positive piles. The entry and policy lemmas therefore remain valid even when a final singleton disappears. Alice's strategy wins every true formula, and Bob's intended strategy wins every false formula, under either convention.
\end{proof}

\section{Proofs for the Kalah board}
\label[appendix]{app:kalah-proofs}

This appendix proves the lemmas of \Cref{sec:empty-stores} about isolation, payloads and collection.

\begin{proof}[Proof of \Cref{lem:empty-isolation}]
Consider a proposed first move at which Alice sows into her store or Bob sows past his store as far as Alice's column $h$. Before this move Alice has added no stone to Bob's row. Bob therefore has at most $M_B+Z$ stones in his pits, all at columns at most $h+D$. His potential $\sum_x xB_x$ initially equals $\Phi$ and decreases by at least one on every preceding Bob move. Indeed, selecting $c$ stones at column $x$ removes weight $xc$, every stone remaining in his row moves to a smaller column, and every stone leaving that row adds no weight. Alice's captures can only lower the potential. Thus at most $\Phi$ Bob moves precede the proposed move. Since Alice has had no bonus, her number of moves, including the proposed move if hers, is at most $\Phi+1$.

The proposed move cannot be Bob's. His entire pit mass satisfies $M_B+Z<h$, so a sow that goes past his store reaches only Alice columns below $h$ and cannot return to his own row.

For an Alice move, initially all her stones lie at columns at most $h+D$. Before the proposed move, Bob changes her row only below column $h$, and her earlier sows stayed in her row. An Alice sow holds at most $M$ stones, so a sow from a column at most $y$ ends at a column at most $y+M$. By induction on $t$, after $t-1$ Alice moves all her stones lie at columns at most $h+D+(t-1)M$, and her $t$-th sow, evaluated as if her row continued to the right, ends at a column at most $h+D+tM$. Since $t\le\Phi+1$, this column is at most
\[
 h+D+(\Phi+1)M=N-1.
\]
The sow cannot reach her store at column $N+1$. This excludes the first failure. The potential and terminal-column arguments consequently hold throughout the play and give all three claims. Either row may empty first, in which case the owner sweep ends the game.
\end{proof}

\begin{proof}[Proof of \Cref{lem:empty-payload}]
Before the first selection or capture, the original target pile remains nonempty and retains all its original stones. A capture by Bob therefore takes its entire payload to his store.

If Alice selects it first, regard its $\Pi$ payload stones as the last stones of the sow. This labeling changes no move. At its original line column $x$, the pit still contains at least its initial $D-x+1$ unlabelled stones, by saturation. These cover the remaining columns of the segment, so every payload stone is sown above $h+D$. A possible empty capture of the final singleton puts that stone in Alice's store. All other payload stones remain in her row above $h+D$. By \Cref{lem:empty-isolation}, Bob never has a pit there, and Alice never sows into her store or returns a stone from there to the segment. These stones therefore belong to Alice, either by later empty captures or by the final sweep.

If neither event occurs, the original target pile is still nonempty at termination. Alice's row has not emptied, so Bob's row has, and the owner sweep puts the payload in Alice's store. In every case the owner of each original payload stone is fixed before any later refill of the target.
\end{proof}

\begin{proof}[Proof of \Cref{lem:collection}]
The least nonempty Bob column $x$ holds $c\le x$ stones. Its sow stays in his row or ends at his store, so it never goes past that store. Every Bob pit below $x$ was empty before the move. The sow puts at most one stone in each, possibly removing its last singleton by a capture, and leaves every pit above $x$ unchanged. Thus every remaining Bob pit at a column $y$ still holds at most $y$ stones. Alice adds no stones to his row, since her sows do not reach her store, and her captures only empty his pits. This proves the invariant for every subsequent Bob move, including bonus moves.

No Bob sow gives Alice any stone. Since Alice sows to the right, her pits below $u$ remain empty. She therefore cannot capture a Bob pile below $u$, because doing so would require selecting a source at a still lower column. Stones already in Bob's pits below $u$ only move left, enter his store, or are transferred to it by his own captures or the owner sweep. Play is finite by \Cref{lem:polynomial-play}, so every one of those stones ends in Bob's store.
\end{proof}

\section{Score targets and binary counts}
\label[appendix]{app:extensions}

This appendix collects two extensions of the main theorem. The first shows that initial stores and score targets are interchangeable, and the second proves \Cref{prop:binary-counts} on counts written in binary.

\paragraph{Score targets.} One may also ask whether Alice can make her final store at least a given unary number $\theta$. With empty stores and the majority target $\theta=\lfloor M/2\rfloor+1$, where $M$ is the pit mass, this is the winner problem, so it is $\PSPACE$-hard by \Cref{thm:main}, and it belongs to $\PSPACE$ by \Cref{lem:polynomial-play} as in \Cref{thm:membership}. Initial stores and targets are interchangeable. Every stone ends in a store, and the stores do not affect any move. So if $\kappa$ is the number of pit stones that end in Alice's store, she wins from stores $\Sigma_A,\Sigma_B$ exactly when $\kappa\ge\lfloor(M+\Sigma_B-\Sigma_A)/2\rfloor+1$. Conversely, the target $\theta$ is represented by the stores $\Sigma_A=\max\{0,M+1-2\theta\}$ and $\Sigma_B=\max\{0,2\theta-M-1\}$. Then $\Sigma_B-\Sigma_A=2\theta-M-1$, so Alice wins exactly when $2\kappa>2\theta-1$, that is, when $\kappa\ge\theta$.

\begin{proof}[Proof of \Cref{prop:binary-counts}]
Let $M$ be the initial pit mass. The pits are listed explicitly, so $N$ and $\log(M+1)$ are at most the input length. The stores do not affect any move, so keep the initial stores as fixed offsets and track the pit counts, the two earned-store counts, and the player to move. There are at most
\[
 2\binom{M+2N+1}{2N+1}\le 2(M+1)^{2N+2}
\]
such positions, which is exponential in a polynomial of the input length. A successor is computable in time polynomial in the input length, even for a large pile. Divide its count by $2N+1$, the length of the sowing circuit, distribute the full circuits at once, and then distribute the remaining stones. The capture, final sweep, and player to move are then determined by binary arithmetic. Each position has at most $N$ successors. Every move increases the pair $(\Sigma,-\Psi)$ of \Cref{sec:membership} in the lexicographic order, so evaluating these positions in decreasing order of this pair takes exponential time. Hardness follows from \Cref{thm:main}, since converting its unary instances to binary takes polynomial time.
\end{proof}

\end{document}